\documentclass[aps,pra,twocolumn,showpacs,superscriptaddress]{iopart}
\usepackage{graphicx}
\usepackage{amssymb}

\input{epsf}
\begin{document}

\review{Few-body physics with
ultracold atomic and molecular systems in traps}

\author{D. Blume}
\address{Department of Physics and Astronomy,
Washington State University,
  Pullman, Washington 99164-2814, USA}

\date{\today}

\begin{abstract}
Few-body physics has played a prominent role in 
atomic, molecular and nuclear physics since the early days of quantum mechanics.
It is now possible---thanks
to tremendous progress in 
cooling, trapping, and manipulating ultracold samples---to
experimentally study few-body phenomena in trapped atomic and molecular systems
with unprecedented control.
This review summarizes recent studies of few-body phenomena
in trapped
atomic and molecular gases, with an emphasis on small trapped systems.
We start by introducing the free-space scattering properties and then
investigate what happens when two particles, bosons or fermions,
are placed in an external confinement. Next, various three-body systems
are treated
analytically in limiting cases.
Our current understanding of
larger two-component Fermi systems and Bose systems is reviewed,
and connections with the corresponding bulk systems are established.
Lastly, future prospects and challenges are discussed.
Throughout this review, commonalities with other systems such as 
nuclei or quantum dots are highlighted. 
\end{abstract}


\maketitle

\section{Introduction}
Although few-body physics plays an important role in all or nearly 
all branches of physics, few-body phenomena are often only 
covered briefly, if at all, in a typical undergraduate or 
graduate core curriculum. 
Only selected few-body problems
such as 
the celestial three-particle Kepler problem and 
the quantum mechanical treatment of the helium atom, which
becomes a three-body problem if the nucleus is treated as a
point-particle, can be found in modern textbooks.
One of the reasons 
is that
few-body problems do not, in general, reduce to an 
effective one-body problem. Instead, the dynamics typically depend
in non-trivial ways on all or nearly all degrees of freedom,
making the analytical and numerical  treatment of few-body systems
 in many cases challenging.
This review discusses few-body aspects of trapped atomic
and molecular Bose and Fermi gases, which are
ideal model systems with which to
study few-body physics. 
These systems
exhibit a variety of few-body phenomena that are amenable to
analytical treatments and that can, in certain
cases, even
be interpreted within an effective one-body framework despite the presence 
of strong interactions and multiple degrees of freedom.

Degenerate Bose and Fermi gases are nowadays being produced routinely
in laboratories all over the world. Most experimental set-ups 
work with alkali atoms 
(Li~\cite{brad95,trus01}, 
Na~\cite{davi95}, K~\cite{dema99,modu01}, 
Rb~\cite{ande95,corn00}, Cs~\cite{webe03}), 
which have one outer
valence electron. At ultracold temperatures, i.e., in the regime
where the de Broglie wave length $\lambda_{{\rm{dB}}}$
is 
larger than the 
range $r_0$ of the two-body interaction potential,
the interactions of the atomic gas are
dominated by isotropic $s$-wave collisions. The net effect of an
$s$-wave collision
can be parametrized through the 
$s$-wave scattering length $a_0(0)$, which, in many instances, is
the only microscopic parameter that enters into the description of
larger systems~\cite{stringari,pethick}. 
A well-known example of this type of description
is the mean-field Gross-Pitaevskii 
equation~\cite{ginz58,gros61,gros63,pita61,dalf98}, 
which depends on the 
product of the number of particles $N$ (or more precisely,
$N-1$~\cite{esry97}) and the $s$-wave scattering length
$a_0(0)$. The Gross-Pitaevskii equation has been proven to 
predict accurately a variety of aspects of dilute
bosonic gases, including a mechanical
instability for negative $a_0(0)$.
Besides alkalis, atoms
with a more complicated
valence electron structure such as
Ca~\cite{kraf09}, 
He$^*$~\cite{robe01,doss01}, 
Cr~\cite{grie05}, 
Yb~\cite{fuku07,fuku09}  and 
Sr~\cite{stel09,dees09}
have been
condensed. The atom-atom collisions between two Cr atoms, e.g., are 
anisotropic and long-ranged~\cite{bara08,laha08a}. 
These characteristics give rise to 
exciting beyond $s$-wave physics that
depends on the strength of the dipole moment
in addition to the
$s$-wave scattering length~\cite{bara08,laha08a}.

This review summarizes our current understanding of
trapped 
few-atom systems
 that interact through isotropic
short-range potentials.
The quantized energy spectrum
of the trapped system
is the result of the interplay of
the particle-particle interactions
and the external trapping geometry.
In fact, the external confinement can be viewed as
quantizing the scattering continuum 
that is characteristic for the free-space system. 
Experimentally, 
trapped few-particle systems can be realized by loading
so-called microtraps with just a few particles~\cite{serw11}.
Alternatively, arrays of few-particle systems can be
realized by
loading a deep optical lattice,
i.e., an optical lattice where tunneling between different lattice sites
is 
strongly suppressed and where nearest-neighbor
interactions can be neglected, with multiple atoms per
lattice site~\cite{grei02,koeh05,thal06}.
 
A distinct characteristic of cold atom systems 
is that the atom-atom interactions
can be tuned through the application
of an external magnetic field in the vicinity
of a Fano-Feshbach resonance~\cite{chin10}. 
A Fano-Feshbach resonance occurs when the two-body energy
of a closed channel molecule coincides with
the scattering threshold of the open channel.
Fano-Feshbach resonances in ultracold
atomic gases were first observed experimentally in 
1998~\cite{inou98,cour98,robe98}.
Their existence 
paves the way to perform
detailed comparisons between theory and experiment as a function
of the interaction strength.

Trapped few-atom systems---together with quantum dots,
nuclei, atomic and molecular clusters, and certain nano-particles---fall 
into the category
of mesoscopic systems. 
Quite generally, the 
study of mesoscopic systems 
provides a bridge between the microscopic and macroscopic
worlds. This review discusses several links
between small and large systems, 
and illustrates how the treatment of mesoscopic systems 
can provide an interpretation 
of the physics of large systems that is complementary to that
derived within effective many-body frameworks. 
In this context, two-component Fermi gases with
infinitely large $s$-wave scattering length provide
a particularly rich example~\cite{bake99,heis01,carl03,astr04c,gior08}. 
In this system, the 
underlying two-body interactions do not define a meaningful length
scale, thereby making the system scale invariant. 
This scale invariance
implies that the few-body system and the many-body system
share the same 
or related universal 
features~\cite{tan04,wern06,tan08a,tan08b,tan08c}. 
Moreover, the scale invariance implies
that atomic two-component Fermi gases
share certain characteristics
with nuclear and neutron matter,
despite the fact that the typical
energy and length scales
of these systems are vastly different.

A particularly interesting few-body phenomenon, 
which is discussed in Subsecs.~\ref{sec_efimov},
\ref{sec_fermigasunequalmasses} and \ref{sec_bosegasthreebodyparam}, 
is the Efimov
effect.
In the early 
1970s,
Vitaly
Efimov~\cite{efim70,efim71,efim73} 
considered a two-body system in free space with infinitely large
$s$-wave scattering length that supports exactly one zero-energy
bound state and no other weakly-bound states.
Efimov found the surprising and, at first sight,
counterintuitive result that
the corresponding three-body system, obtained by adding a third
particle and assuming pairwise interactions,
supports infinitely many 
weakly-bound three-body states.
Efimov's prediction, known as the 
three-body Efimov effect, 
stimulated much theoretical and experimental
efforts~\cite{braa06,braa07}, and
continues to be one of the flagships of few-body physics today.
Signatures of the three-body Efimov effect and associated Efimov resonances have been 
searched for in the helium trimer~\cite{brue05}  
and in nuclear systems~\cite{mazu06},
among others. However, it was
not till recently that the three-body Efimov
effect has been confirmed 
conclusively 
experimentally
via loss measurements of cold
trapped atomic 
gases~\cite{krae06,baro09,poll09,gros09,knoo09,zacc09,will09,lomp10}.
This review summarizes the key features of the Efimov
effect, or more generally of Efimov
physics, and its modifications when the
three-body system is placed in an external trap.
Extensions of the Efimov scenario to larger systems
are also considered.

The present review article
aims at providing an introduction
to the field of trapped few-atom and few-molecule systems.
Readers who would like to
place this topic
into a broader context are directed to a number of
other review articles.
Experimental and theoretical
aspects of cold and ultracold collisions
have been reviewed by Weiner {\em{et al.}} in 1999~\cite{wein99}.
The topic of Fano-Feshbach resonances has been reviewed
by Chin {\em{et al.}} in 2010~\cite{chin10}
(see also Refs.~\cite{koeh06,jone06}).
The Efimov effect in free-space
has been reviewed by Braaten and Hammer 
in 2006 and 2007~\cite{braa06,braa07}. 
A related ``Trend'' was authored by Ferlaino and Grimm~\cite{ferl10},
and a Physics Today article by Greene~\cite{gree10} in 2010.
The broad subjects of Bose and Fermi gases are covered in
reviews by Dalfovo {\em{et al.}} from 1998~\cite{dalf98} and
Giorgini {\em{et al.}} from 2008~\cite{gior08}, respectively.
Many-body aspects of cold atom gases have been
reviewed by Bloch {\em{et al.}} in 2008~\cite{bloc08}.
Various reviews have focused on specific
few-body systems:
Jensen {\em{et al.}}
discuss the physics of halo systems~\cite{jens04},
Carlson and Schiavilla that of few-nucleon systems~\cite{carl98},
Reiman {\em{et al.}} that of vortices in few-particle droplets~\cite{saar10},
and
Nielsen {\em{et al.}} that of three-body systems with short-range
interactions~\cite{niel01}.
Lastly, the treatment of few-particle systems within
the hyperspherical framework, which is 
utilized at various places
in the present work, has been reviewed
by Lin in 1995~\cite{lin95} 
and 
Rittenhouse {\em{et al.}} in 2011~\cite{ritt11}.

The remainder of this article 
is structured as follows. Section~\ref{sec_twobody}
discusses the underlying two-body physics,
both without and with confinement, while
Sec.~\ref{sec_threebody}
considers the physics of three
interacting particles under external spherically symmetric confinement.  
Section~\ref{sec_fermigas} discusses various aspects
of 
trapped Fermi gases, focusing primarily on
$s$-wave interacting 
two-component 
Fermi gases.
Section~\ref{sec_bosegas} summarizes the key characteristics
of $s$-wave interacting Bose gases under external confinement.
Lastly, Sec.~\ref{sec_summary}
comments on a number of present and future frontiers.

\section{Underlying two-body physics}
\label{sec_twobody}

\subsection{Free-space two-particle scattering in three-dimensional space}
\label{sec_twobodyfree}
This subsection considers the two-particle system in free space
while the next subsection treats the trapped two-particle system.
For two neutral atoms, 
the two-body potential at large interparticle distances $r$
is dominated by the van der Waals tail that falls of as $-C_6/r^6$
($C_6$ denotes the van der Waals coefficient).
The
characteristic length or range $l_{\rm{char}}$ of
the two-body interaction potential 
is defined as the distance at which the
kinetic and potential energies
are approximately equal,
$l_{\rm{char}} = (2 \mu C_6/ \hbar^2)^{1/4}$~\cite{wein99}, where
$\mu$ denotes the reduced mass of the two interacting
particles. 
For alkali
atoms, $l_{\rm{char}}$ 
is 
of the order of $100a_{\rm{bohr}}$, where
$a_{\rm{bohr}}$
denotes the Bohr radius, $a_{\rm{bohr}}=5.292 \times 10^{-11}m$.

The potentials that describe diatomic 
alkali systems support many two-body bound states
in free space. For example, the number of vibrational bound states
in the  rotational ground state 
is respectively 66 and 16
for the electronic
ground and excited state $X \, ^1\Sigma_g^+$ and  
$a \, ^3\Sigma_u^+$  potential
curves of $^{23}$Na$_2$~\cite{mies00}.
Throughout this review, we are concerned with the 
low-temperature regime, i.e., we are interested in
physics that occurs at energy scales that are much smaller 
than $E_{\rm{char}}$, where $E_{\rm{char}} = \hbar^2/(2\mu l_{\rm{char}}^2)$,
or 
equivalently, 
length scales that are much larger than $l_{\rm{char}}$. 
This includes at most a few weakly-bound two-body states as well
as the energetically low-lying scattering continuum (or ``quantized
scattering continuum'')
but excludes deeply bound 
two-body states and the high-energy scattering continuum.
Generally speaking, the low-energy physics is
independent of the
details of the underlying two-body potential and 
is determined by a 
few parameters such as the two-body scattering length
and the effective range.
Thus, low-energy phenomena can be described by
replacing the true or full interaction potential
by a model potential that 
is characterized by the same 
parameters as the true potential (e.g., the 
same scattering length and effective range), or by 
effective theories such as an effective range expansion approach
or an
effective
field theory approach.
Effective range and effective
field theory approaches have been
applied to a variety of systems, ranging 
from the scattering of atoms at ultracold 
temperatures~\cite{braa06}
to the low-energy scattering of 
two nucleons in free-space~\cite{kapl98} and on a 
lattice~\cite{lues86,lues91,bean04} 
to 
the description of particle physics
phenomena such as narrow charmonium resonances~\cite{braa04}.
The effective pseudopotential description,
frequently employed in 
this review, goes back 
to Fermi,
who treated the scattering between slow neutrons
and hydrogen atoms~\cite{ferm34}.

In the following,
we assume that the two particles of mass $m_1$ and $m_2$
(with position vectors
$\vec{r}_1$ and $\vec{r}_2$) 
interact through a spherically symmetric two-body potential
$V_{\rm{tb}}(r)$, where $r$ denotes the distance coordinate between the two
particles, $r=|\vec{r}_1-\vec{r}_2|$.
Separating off the center-mass degrees of freedom,
the relative scattering wave function $\psi_{k}(\vec{r})$
can be decomposed into partial waves,
$\psi_{k}(\vec{r})=\sum_{l} R_{lk}(r) P_l(\vartheta)$,
where $\vartheta$ denotes the polar angle and $P_l$ the Legendre
polynomial of degree $l$.

Asymptotically, i.e., in the regime where the interaction potential
can be neglected,
the radial wave functions are given by~\cite{taylor}
\begin{eqnarray}
\label{eq_radialasym}
R_{lk}^{\rm{out}}(r)= N_{l}(k) 
\left\{ j_{l}(kr) - \tan[\delta_l(k)] n_l(kr) \right\},
\end{eqnarray}
where $k$ is related to the scattering energy $E$ 
through $k = \sqrt{2 \mu E}/\hbar$ and where $j_l$ and $n_l$ denote
the spherical Bessel function and the Neumann function, respectively.
In the small $kr$ limit, $j_l(kr)$ and $n_l(kr)$
behave as $(kr)^l$ and $(kr)^{-(l+1)}$, respectively~\cite{taylor}.
Correspondingly, $j_l$ and $n_l$ are referred to as regular and irregular
solutions.
In Eq.~(\ref{eq_radialasym}), $N_{l}(k)$ denotes a normalization
constant.
The scattering phase shifts $\delta_l(k)$ are determined
by matching the asymptotic solution $R_{lk}^{\rm{out}}(r)$,
Eq.~(\ref{eq_radialasym}),
to the inner solution $R_{lk}^{\rm{in}}(r)$, which
depends on the details of the potential $V_{\rm{tb}}(r)$.
The phase shifts accumulate in the inner region
and quantify the admixture
of the irregular solutions $n_l(kr)$. A vanishing phase shift 
$\delta_l$ indicates
an effectively non-interacting system, i.e., the asymptotic
solution of the $l$$^{th}$ partial wave channel
is identical to that of the corresponding non-interacting system 
(differences can exist, though, in the inner region).
A small positive phase shift indicates that the asymptotic solution of the
$l^{th}$ partial wave channel is shifted to smaller $kr$ (``pulled inward'')
compared to that of the non-interacting system, thus indicating an 
effectively attractive interaction.
A small negative phase shift,
in contrast, indicates that the asymptotic solution of the
$l^{th}$ partial wave channel is shifted to larger $kr$ (``pushed outward'')
compared to that of the non-interacting system, thus indicating an 
effectively repulsive interaction.
The phase shifts $\delta_l(k)$ can be used to define
the energy-dependent generalized scattering lengths 
$a_{l}^{2l+1}(k)$~\cite{taylor} 
(note that the opposite sign convention is sometimes
employed in nuclear and particle physics),
\begin{eqnarray}
\label{eq_edepscatt}
a_l^{2l+1}(k) = - \frac{\tan[\delta_l(k)]}{k^{2l+1}}.
\end{eqnarray}
For short-range potentials, i.e., for potentials that fall off faster than
$1/r^{2 l +3}$ asymptotically, 
the energy-dependent scattering lengths
approach a constant in the zero-range limit~\cite{wein99},
$a_l^{2l+1}(0) = \lim_{k \rightarrow 0} a_l^{2l+1}(k)$
(for long-range potentials, see Subsec.~\ref{sec_twobodyother}).
For the lowest partial wave, i.e., for $l=0$, $a_0(0)$
is the usual $s$-wave scattering length, which dominates, away
from higher partial wave resonances, the
low-energy cross sections of bosonic gases, two-component 
Fermi gases and Bose-Fermi mixtures. 
In the zero-energy limit,
the scaled outside wave function $r R_{00}^{\rm{out}}(r)$
becomes $N_{0}(0) \left[ r-a_0(0) \right]$.
The leading-order
energy-dependence of the $s$-wave scattering length
can be written in terms of the effective range $r_{\rm{eff}}$~\cite{taylor},
\begin{eqnarray}
\label{eq_effrange}
\frac{1}{a_0(k)} \approx \frac{1}{a_0(0)} - \frac{1}{2} r_{\rm{eff}} k^2;
\end{eqnarray}
the validity of Eq.~(\ref{eq_effrange})
requires
that the potential
falls 
off
faster than $1/r^{2l+5}$ asymptotically.
For identical fermions, the Pauli exclusion
principle forbids $s$-wave scattering, making the $p$-wave
channel with generalized scattering length or scattering volume $a_1^3(k)$
dominant.

\begin{figure}
\vspace*{+1.5cm}
\includegraphics[angle=0,width=65mm]{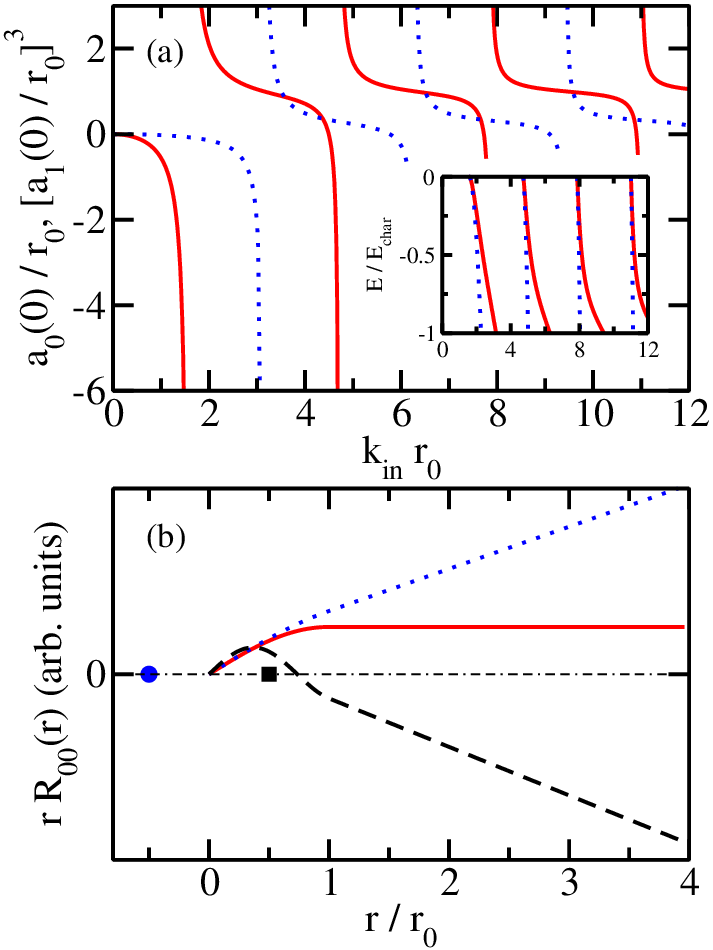}
\vspace*{-0.2cm}
\caption{(Color online)
Illustration of two-body scattering for
an attractive square well model potential with 
depth $V_0$ and range $r_0$.
(a) Solid and dotted lines show
the zero-energy scattering lengths
$a_0(0)$ 
and $a_1^3(0)$  
as a function of $k_{\rm{in}} r_0$, where 
$k_{\rm{in}}^2 = 2 \mu V_0/\hbar^2$.
Inset:
Solid lines show the energy of the most weakly-bound $s$-wave state
for the square well potential
as a function of $k_{\rm{in}}r_0$.
For comparison, dotted lines show Eq.~(\ref{eq_boundpp})
with $a_0(k)$ approximated by $a_0(0)$
of the square well potential.
(b) Solid, dotted and dashed lines
show the scaled $s$-wave scattering wave function $r R_{00}(r)$
as a function of $r/r_0$ for 
$[a_0(0)]^{-1}=0$ (one zero-energy $s$-wave bound state), 
$a_0(0)=-r_0/2$ [no $s$-wave bound state;
$a_0(0)$ is marked by a 
circle]
and $a_0(0)=r_0/2$ [one $s$-wave bound state;
$a_0(0)$ is marked by a 
square], respectively.
}\label{fig_ascattwave}
\end{figure}

Figure~\ref{fig_ascattwave}
illustrates the behavior of the scattering lengths and wave functions
for an attractive square well potential with depth $V_0$ 
and range $r_0$. This or similarly simplistic model potentials
suffice to describe generic
low-energy features of cold atom systems.
Solid and dotted lines in Fig.~\ref{fig_ascattwave}(a)
show the scattering lengths $a_0(0)$ and $a_1^3(0)$
as a function of the dimensionless ``strength parameter'' $k_{\rm{in}}r_0$,
where $k_{\rm{in}}=\sqrt{2 \mu V_0}/\hbar$.
The scattering lengths $a_0(0)$ and $a_1^3(0)$
vanish for vanishing $V_0$, and decrease with increasing 
$k_{\rm{in}}r_0$.
The $s$- and $p$-wave scattering lengths 
diverge
when $k_{\rm{in}}r_0= \pi/2$ and
$k_{\rm{in}}r_0= \pi$, respectively.
As $k_{\rm{in}} r_0$ increases further, the scattering lengths decrease
from large positive to large negative values. 
When the scattering lengths $a_0(0)$ and $a_1^3(0)$ diverge, the
square well potential supports a new $s$-wave or
a new $p$-wave bound state, respectively.
Dashed, dotted and solid lines in Fig.~\ref{fig_ascattwave}(b) 
show
the scaled $s$-wave scattering wave function $r R_{00}(r)$
for finite positive, finite negative and 
infinitely large $a_0(0)$, respectively.
The linear behavior of $r R_{00}(r)$ for $r > r_0$
is clearly visible. For finite $a_0(0)$, the scattering length
can be ``read off'' by extrapolating the linear behavior to $r < r_0$
[see symbols in Fig.~\ref{fig_ascattwave}(b)].
By construction, the 
wave functions for the same zero-energy scattering length
but different depth or number of bound states (not shown)
agree in the large $r$ limit. Differences, however, exist in the small $r$
region. These short-range features are typically not probed at 
low temperatures, making the scattering length
the only relevant quantity. This can be intuitively understood by 
realizing that the de Broglie wave length $\lambda_{\rm{dB}}$
increases with decreasing temperature,
thereby setting a resolution limit, i.e., the collision energy
(which is in typical experiments set by a combination of 
the temperature and the particle statistics) is too small
to probe features
at small length scales. Correspondingly, the
low temperature regime is known as the long wave length limit.

Mathematically, it can be advantageous to replace the
true interaction potential or shape-dependent model
potential by  a zero-range 
pseudopotential that acts only when the two
particles sit on top of each other~\cite{ferm34,brei47,huan57}. 
Using the quantum defect language 
(see, e.g., Refs.~\cite{seat83,borc03,peac04}), 
the pseudopotential can be thought
of as enforcing a particular value of the logarithmic
derivative at $r=0$. 
In three spatial dimensions, the regularized $s$-wave Fermi-Huang
pseudopotential $V_0^{\rm{pp}}$ reads~\cite{huan57}
\begin{eqnarray}
\label{eq_pseudo3d}
V_{0}^{\rm{pp}}(r)= 
g_{0}(k) \delta(\vec{r}) \frac{\partial}{\partial r}r,
\end{eqnarray}
where
the interaction strength $g_{0}(k)$ is chosen such that
$V_0^{\rm{pp}}$ reproduces the energy-dependent
$s$-wave scattering length $a_0(k)$ of the true interaction 
potential,
\begin{eqnarray}
\label{eq_pseudo3dstrength}
g_{0}(k)=
\frac{2 \pi \hbar^2 a_0(k)}{\mu}. 
\end{eqnarray}
The regularization operator $(\partial/\partial r)r$
in Eq.~(\ref{eq_pseudo3d})
ensures that the
pseudopotential is well behaved as $r$ goes to $0$~\cite{huan57}. 
The pseudopotential imposes a boundary condition on the relative 
wave function
$\psi_{k}(\vec{r})$ at $r = 0$~\cite{wign33,beth35},
\begin{eqnarray}
\label{eq_bethepeierl}
\left[ \frac{\frac{\partial}{\partial r}\left(r \psi_k(\vec{r}) \right) }
{r\psi_k(\vec{r})} \right]_{r \rightarrow 0} 
= \frac{-1}{a_0(k)}.
\end{eqnarray}
This so-called Bethe-Peierls boundary condition serves as an
alternative parametrization of $V_0^{\rm{pp}}(r)$.
The pseudopotential $V_0^{\rm{pp}}$ supports no two-body bound state
for negative $s$-wave scattering length $a_0(k)$ and a single two-body
$s$-wave bound state with binding energy
$E_{\rm{bound}}$,
\begin{eqnarray}
\label{eq_boundpp}
E_{\rm{bound}} = \frac{-\hbar^2}{2 \mu [a_0(k)]^2},
\end{eqnarray} 
for positive 
$a_0(k)$.
Equation~(\ref{eq_boundpp}) is an implicit equation for the
bound state energy, which requires knowing $a_0(k)$ for
imaginary $k$.
The size of the bound state of $V_0^{\rm{pp}}$
is determined by $a_0(k)$,
i.e., $\sqrt{\langle r^2 \rangle}=a_0(k)/\sqrt{2}$.
We note that
Eq.~(\ref{eq_boundpp}) also describes the $s$-wave bound states
of
finite-range potentials.
This can be seen by
analytically continuing
Eq.~(\ref{eq_radialasym}) 
to negative energies or by analyzing the poles of
the energy-dependent S-matrix~\cite{taylor}.
In the low-energy limit, Eq.~(\ref{eq_boundpp})
provides a good description of the bound state energies if $a_0(k)$ is
approximated by
$a_0(0)$ (see also below).
In fact, it should be noted that the original formulation of the
pseudopotential (or corresponding boundary condition)
employed the zero-energy scattering length $a_0(0)$
instead of $a_0(k)$.
In the cold atom context, the energy-dependent scattering length
$a_0(k)$ was first introduced in Refs.~\cite{blum02,bold02}.

It is important to note that $V_0^{\rm{pp}}$ only acts on the lowest
partial wave.
The partial waves $R_{lk}^{\rm{out}}$ with $l \ge 1$ have vanishing
amplitude at $r=0$, and are thus not affected by $V_0^{\rm{pp}}$.
This implies that Eq.~(\ref{eq_bethepeierl}) remains valid
if $\psi_k$ is replaced by $R_{0k}^{\rm{out}}$
and that $R_{lk}^{\rm{out}}(r)=N_l(k)j_l(kr)$ for $l \ge 1$.
Pseudopotentials $V_l^{\rm{pp}}$
for higher partial wave scattering 
have been developed in the 
literature~\cite{huan57,omon77,demk88,kanj04,stoc04,dere05a,idzi06,pric06}.
Similar to the $s$-wave pseudopotential, the higher partial
wave pseudopotentials,
applicable to spherically symmetric systems, can be parameterized
in terms of a boundary condition at $r=0$,
\begin{eqnarray}
\label{eq_bethepeierlfinitel}
\frac{\Gamma(l+1/2)}{\sqrt{\pi} \Gamma(l+1)}
\left[ \frac{
\frac{\partial^{2l+1}}{\partial r^{2l+1}}
\left( r^{l+1} \int \psi_k(\vec{r}) P_l(\cos \vartheta) d \cos \vartheta \right)}
{r^{l+1} \int \psi_k(\vec{r}) P_l(\cos \vartheta) d \cos \vartheta} 
\right]_{r \rightarrow 0} 
= \frac{-1}{a_l^{2l+1}(k)}.
\end{eqnarray}

Solid lines in the inset of 
Fig.~\ref{fig_ascattwave}(a) 
show the energy of the most weakly-bound
$s$-wave state of the square well potential as a function
of $k_{\rm{in}} r_0$. As discussed in the context of 
the main figure,
a new $s$-wave bound state is being supported when $k_{\rm{in}} r_0=n \pi/2$,
where $n=1,3,\cdots$.
For comparison, dotted lines show the  
bound state energy $E_{\rm{bound}}$, Eq.~(\ref{eq_boundpp}),
with $a_0(k)$ approximated by the zero-energy 
scattering length $a_0(0)$ of the square well potential.
For $|E|/E_{\rm{char}} \ll 1$, where $E_{\rm{char}}=\hbar^2/(2 \mu r_0^2)$
(i.e., in the low-energy regime),
the agreement between the solid and dotted lines is good.
However, as $|E|/E_{\rm{char}}$ increases,
deviations between the two data sets are visible.
This can be attributed to the fact
that the energy-dependence of $a_0(E)$
becomes stronger as $|E|$ increases.

Typical interactions between alkali atoms involve two electronic energy 
curves, the singlet and the triplet 
Born-Oppenheimer (BO) potential curves
labeled
$X \; ^1\Sigma_g^+$ and $a \; ^3 \Sigma_u^+$,
respectively, which are coupled through a 
hyperfine Hamiltonian. The coupling is responsible
for
the multi-channel nature of the scattering process. 
Multi-channel scattering is
reviewed in detail in Refs.~\cite{wein99,taylor}. 
For our purposes, it is important that the coupling between
the channels can give rise to Fano-Feshbach resonances~\cite{chin10}.
A Fano-Feshbach resonance in the $l$th partial wave
arises when the 
scattering energy coincides with the energy of a
bound state in the $l$th partial wave.
Since the ``detuning'' between the open and closed
channels in the large $r$ regime is determined
by the difference in magnetic moments,
the thresholds
of the two channels can be varied through the application of an external
magnetic field.
If the topology of the potential energy surfaces is such that
the variation of the external field
allows for
the bound state energy and
the scattering energy to coincide,
then the system exhibits a resonance.
Varying the strength $B$
of the external magnetic field
in the vicinity of a Fano-Feshbach resonance allows the coupled-channel 
$l$-wave scattering length $a_l^{2l+1}(k)$
to be tuned to essentially any value.

\begin{figure}
\vspace*{+1.5cm}
\includegraphics[angle=0,width=65mm]{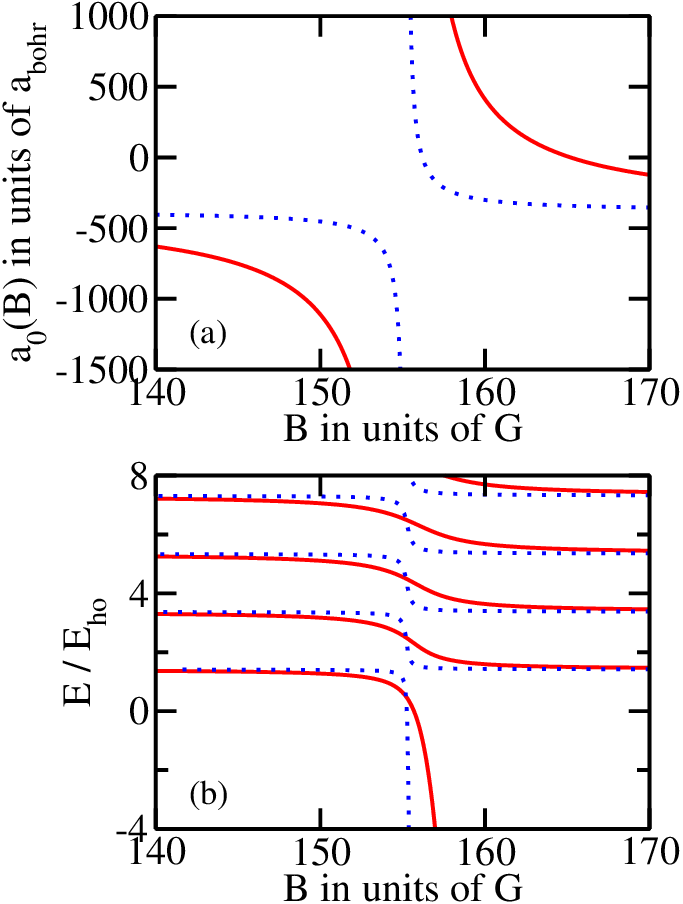}
\vspace*{-0.2cm}
\caption{(Color online)
(a) The solid line shows the field-dependent scattering length
$a_0(B)$, Eq.~(\ref{eq_scattcceff}), 
for the Fano-Feshbach resonance around 155G for
two $^{85}$Rb atoms in the $(F,M_F)=(2,-2)$ state as a function of $B$;
the parameters adopted are 
$\Delta B=10 G$, $B_{\rm{res}}=155.2G$ and 
$a_{\rm{bg}}=-380a_{\rm{bohr}}$~\cite{dunn05}.
For comparison, the dotted line shows the effective scattering length
for a ``mock resonance'' 
with $\Delta B=1 G$ but the same $a_{\rm{bg}}$ and $B_{\rm{res}}$
as for the $^{85}$Rb resonance.
(b) Solid and dotted lines show the $s$-wave energy spectra, calculated 
using Eqs.~(\ref{eq_trans}) and 
(\ref{eq_scattcceff}), for two atoms under harmonic confinement with
$a_{{\rm{ho}},\mu}=5000a_{\rm{bohr}}$.
}\label{fig_ccascatt}
\end{figure}

Assuming no overlapping Fano-Feshbach resonances exist and higher
partial wave contributions are negligible, the 
magnetic field-dependent
coupled-channel $s$-wave scattering length $a_0(B)$
can be parameterized by three effective parameters,
the background scattering length $a_{\rm{bg}}$, the resonance width 
$\Delta B$ and the resonance position $B_{\rm{res}}$~\cite{mies00},
\begin{eqnarray}
\label{eq_scattcceff}
a_0(B) = a_{\rm{bg}} \left( 1 - \frac{\Delta B}{B-B_{\rm{res}}} \right).
\end{eqnarray}
This effective description has been confirmed by 
coupled-channel calculations and experimental 
data; it can, for example, be derived using a simple
coupled-channel zero-range or square well model~\cite{holl01a,dunn05}.
The parameters $\Delta B$, $B_{\rm{res}}$ and 
$a_{\rm{bg}}$ are extracted from 
experimental data or full coupled-channel calculations.
The solid line
in Fig.~\ref{fig_ccascatt}(a) shows the 
magnetic field dependence of the
scattering length for the 
magnetic Fano-Feshbach resonance for two
$^{85}$Rb atoms in the $(F,M_F)=(2,-2)$ state near 155G
(here, $F$ denotes the hyperfine quantum number and
$M_F$ the corresponding projection quantum number).
It is often useful
to 
write
the
field-dependent scattering length in terms
of the energy width $\Gamma_E$~\cite{mies00},
\begin{eqnarray}
\label{eq_scattcceff2}
a_0(B) = a_{\rm{bg}} \left( 1 + 
\frac{    \frac{\Gamma_E}{2 k a_{\rm{bg}}}}
{E+(B-B_{\rm{res}}) \frac{\partial E_{\rm{res}}(B)}{\partial B}}
\right),
\end{eqnarray}
where $\Gamma_E= 2 k a_{\rm{bg}} \Delta B [\partial E_{\rm{res}}(B)/\partial B]$
and
where $E_{\rm{res}}(B)$ denotes the field-dependent bound state energy of
the coupled-channel 
system. 
Here, $E$ and $k$ denote, as above, the two-body scattering energy and 
corresponding wave vector, respectively.
We note that the 
$E$-term in the denominator of Eq.~(\ref{eq_scattcceff2}) 
can often be neglected.
In this case, Eq.~(\ref{eq_scattcceff2}) reduces to
Eq.~(\ref{eq_scattcceff}).
The quantity $\partial E_{\rm{res}}(B)/\partial B$ is 
known from experiment and/or theory.
The energy width $\Gamma_E$ can be used to define
broad and narrow 
resonances~\cite{gior08,bruu04,petr05,dien04,petr04,gura07}. 
If $\Gamma_E$ is larger (smaller) 
than the characteristic energy scale of the
system, the resonance is called broad (narrow).
The characteristic energy scale depends on the system
under investigation.
For the BCS-BEC crossover problem, for example,
the characteristic energy scale in the low
temperature regime is set by the Fermi energy
when $|a_0(B)|$ is small and $a_0(B)<0$,
and by the two-body binding energy when $a_0(B)$ is small and
$a_0(B)>0$~\cite{petr05}.
As a rule of thumb, broad resonances can be modeled 
quite accurately within a single-channel model;
narrow resonances, in contrast, require a coupled-channel treatment.

\subsection{Two particles in a spherically symmetric harmonic trap}
\label{sec_twobodytrap}
This subsection considers two particles
interacting through a spherically symmetric potential $V_{\rm{tb}}$
under external confinement,
i.e., we 
consider what happens to the scattering continuum and
the weakly-bound states 
when the asymptotic boundary conditions are modified by
the external confinement. 
Throughout, we consider
the regime where the confinement length is much larger
than the range of the underlying particle-particle interaction potential.
Experimentally, trapped systems
can be realized by loading atoms into an optical
lattice created by counterpropagating standing light waves.
The atom-light interaction creates a periodic potential
felt by the atoms. Near the bottom of each well, the confining potential
is approximately harmonic. Furthermore, for
atoms with short-range interactions, the spacing between the
lattice sites is so large that ``off-site'' interactions can be neglected.
Assuming that tunneling between neighboring lattice sites
can also be neglected, the lattice sites are independent
from each other
and each lattice site hosts an independent few-body system.
Experimentally, the number of atoms per lattice site can
be controlled. 
Here, we treat the two-body system; larger systems are discussed in 
Secs.~\ref{sec_threebody}-\ref{sec_bosegas}.

We assume that the external confinement is harmonic,
i.e., that the $i$$^{th}$ particle feels the external potential
$m_i \omega^2 \vec{r}_i^2/2$ ($i=1$ or 2),
where $\omega$ denotes the angular trapping frequency and
$\vec{r}_i$ the position vector of the $i$$^{th}$
particle measured with respect to the trap center. 
For this system, the center of
mass motion separates, i.e., $E_{\rm{tot}}=E_{\rm{cm}}+E$,
where the center of mass energy is given by
$E_{\rm{cm}}=(2Q_{\rm{cm}}+L_{\rm{cm}}+3/2)\hbar \omega$ with $Q_{\rm{cm}}=0,1,\cdots$
and $L_{\rm{cm}}=0,1,\cdots$, and $E$ denotes the relative energy.
Assuming a spherically symmetric interaction potential
$V_{\rm{tb}}(r)$,
the relative Hamiltonian for the two-body system reads
\begin{eqnarray} 
H = - \frac{\hbar^2}{2 \mu} \nabla^2_{\vec{r}} + 
\frac{1}{2} \mu \omega^2 r^2
+V_{\rm{tb}}(r),
\end{eqnarray}
where $\mu$ denotes, as above, the reduced two-body mass.
The relative eigen functions $\psi_{l q}(\vec{r})$ separate,
$\psi_{l q}(\vec{r})= r^{-1} u_{l q}(r) Y_{lm}(\vartheta, \varphi)$.
$u_{l q}(r)$ solves the
radial Schr\"odinger equation
\begin{eqnarray}
\label{eq_radialtrap}
\left[ \frac{-\hbar^2}{2 \mu} \frac{\partial^2}{\partial r^2} +
\frac{\hbar^2 l(l+1)}{2 \mu r^2} + 
\frac{1}{2} \mu \omega^2 r^2 + V_{\rm{tb}}(r) \right] u_{l q}(r) = 
E_{lq} u_{l q}(r),
\end{eqnarray}
where $E_{lq}=E=(2q+l+3/2)\hbar \omega$.
The non-integer quantum number $q$ labels the radial solutions for a given
relative orbital angular momentum $l$.
Equation~(\ref{eq_radialtrap}) can be solved numerically for 
essentially any $V_{\rm{tb}}(r)$. Here, we seek analytical solutions for
the $l$-wave zero-range pseudopotentials $V_l^{\rm{pp}}$.

The linearly independent solutions $f_{l q}$
and $g_{l q}$ 
of the radial Schr\"odinger equation with $V_{\rm{tb}}=V_{l}^{\rm{pp}}$
for $r>0$ are identical
to those for an isotropic harmonic oscillator with
mass $\mu$.
Introducing the harmonic oscillator length $a_{{\rm{ho}},\mu}$,
$a_{{\rm{ho}},\mu}= \sqrt{\hbar/(\mu \omega)}$, and defining the 
dimensionless variable $x$, $x = r / a_{{\rm{ho}},\mu}$,
$f_{l q}$ and $g_{l q}$ can be written as
\begin{eqnarray}
\label{eq_fregular}
f_{l q}(x) = 
x^{l+1} \exp \left( \frac{-x^2}{2} \right) \,_1F_1(-q,l+3/2,x^2)
\end{eqnarray}
and
\begin{eqnarray}
\label{eq_girregular}
g_{l q}(x) = 
x^{-l} \exp \left( \frac{-x^2}{2} \right)\, _1F_1(-q-l-1/2,-l+1/2,x^2),
\end{eqnarray}
where $_1F_1$ 
denotes the confluent hypergeometric function of the first  kind.
We write 
$u_{l q}(x)=c_{l}(q) f_{l q}(x) + d_{l}(q) g_{l q}(x)$
and  choose the ratio $c_{l}(q)/d_{l}(q)$ such that $u_{l q}(x)$ vanishes
in the large $x$ limit, i.e., so that the exponentially increasing 
pieces of $f_{l q}(x)$ and 
$g_{l q}(x)$ cancel as $x \rightarrow \infty$.
Using the $x \rightarrow \infty$ behavior of $_1F_1$~\cite{abramowitz},
we find
\begin{eqnarray}
\label{eq_radialratio}
\frac{\Gamma(l+3/2)}{\Gamma(-q)} c_{l}(q) 
=
-\frac{\Gamma(-l+1/2)}{\Gamma(-q-1/2-l)} d_l(q) .
\end{eqnarray}
Using Eq.~(\ref{eq_radialratio}) in $u_{l q}(x)$,
we find
\begin{eqnarray}
\label{eq_radialdecaying}
u_{l q}(x)= N_{l}(q) \exp \left( \frac{-x^2}{2} \right) x^{l+1}
U( -q, l+3/2, x^2),
\end{eqnarray}
where $U$ denotes the confluent hypergeometric function of the second
kind
and $N_{l}(q)$ a normalization constant.
Next, we ensure that $r^{-1} u_{l q}$ obeys the boundary 
condition in the small $x$ limit,
i.e., we enforce
Eq.~(\ref{eq_bethepeierlfinitel}). The 
resulting quantization condition reads
\begin{eqnarray}
\label{eq_trans}
\frac{(-1)^l 2^{2l+1} 
\Gamma \left( \frac{-E_{l q}}{2 \hbar \omega} + \frac{l}{2}+\frac{3}{4} 
\right) }
{\Gamma \left( \frac{-E_{l q}}{2 \hbar \omega} -\frac{l}{2} + \frac{1}{4}
\right) } = 
\frac{a_{{\rm{ho}},\mu}^{2l+1}}{a_l^{2l+1}(k)}.
\end{eqnarray}
The transcendental equation, Eq.~(\ref{eq_trans}), 
determines the eigenspectrum for
any $l$ and $a_l^{2l+1}(k)$.
The eigenenergies $E_{l q}$ can be obtained straightforwardly using
a simple rootfinding routine. 
For $l=0$, this eigenequation was first derived 
by Busch {\em{et al.}}~\cite{busc98}. For $l>0$, the solutions can be found in
Refs.~\cite{peac04,kanj04,stoc04,idzi06,chen07}.
The wave function $u_{lq}$ is normalizable
for $l=0$ but not for $l>0$.
This issue has been discussed in detail in 
Refs.~\cite{pric06,reic06,pric06a} and will
not be elaborated on here.

\begin{figure}
\vspace*{+1.5cm}
\includegraphics[angle=0,width=65mm]{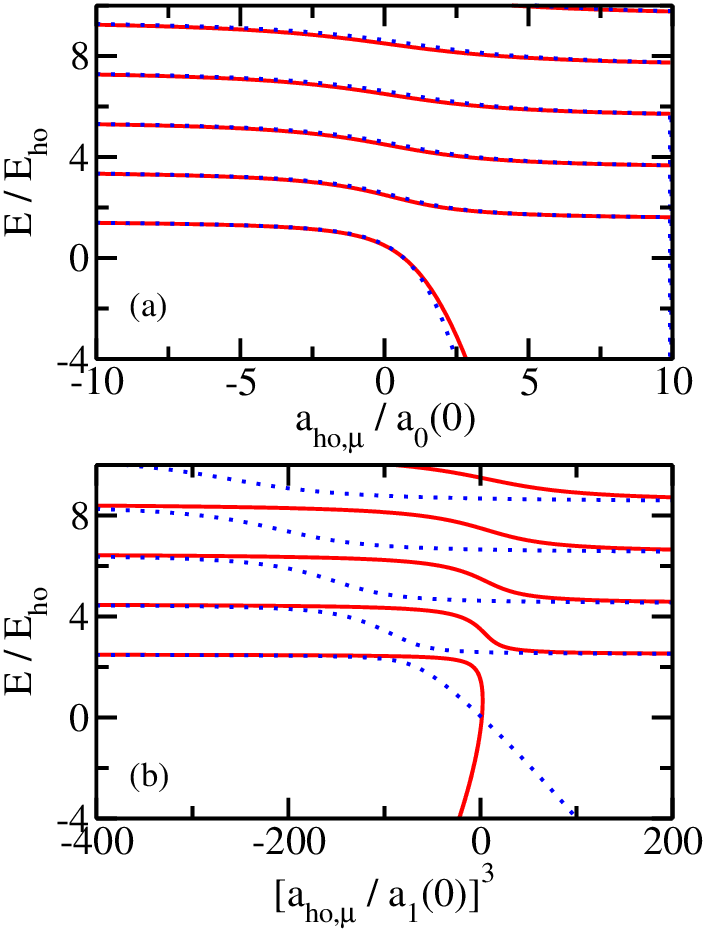}
\vspace*{-0.2cm}
\caption{(Color online)
Illustration of the energy spectrum of the trapped two-particle system
for (a) $l=0$ and (b) $l=1$.
Solid lines show the eigenenergies $E_{lq}$ obtained by solving
Eq.~(\ref{eq_trans}) with $a_l^{2l+1}(k)$ approximated
by $a_l^{2l+1}(0)$.
Dotted lines show the eigenenergies obtained for a square well potential
with $r_0=0.1 a_{{\rm{ho}},\mu}$ and $V_0$ adjusted 
to obtain the desired $a_l^{2l+1}(0)$
[and such that
the free-space system supports no (one) $l$-wave bound state
for $a_l^{2l+1}(0)<0$ 
($a_l^{2l+1}(0)>0$)].
}\label{fig_energytrap}
\end{figure}

Figures~\ref{fig_energytrap}(a) and \ref{fig_energytrap}(b) 
show the energy spectra for $l=0$ and $l=1$, respectively, as
a function of the inverse scattering lengths. 
Solid lines show the energy spectra obtained from
Eq.~(\ref{eq_trans}) using
$a_l^{2l+1}(k)=a_l^{2l+1}(0)$.
For comparison, dotted lines show the eigenspectra obtained for a 
square well potential with a fairly large range $r_0$,
$r_0 = 0.1 a_{{\rm{ho}},\mu}$.
The $s$-wave spectrum 
for the square well potential is fairly well reproduced by the 
zero-range model
for all $a_0(0)$; the deviations are largest for
negative energies where the short-range physics plays a 
non-negligible role. For $l=1$, the 
square well spectrum is qualitatively reproduced for
$ E \gtrsim 5 \hbar \omega /2$ and 
large $|a_{{\rm{ho}},\mu}/a_1(0)|^3 $. 
In other parameter regimes, significant deviations are
visible.
A significantly improved
description is obtained by treating the energy-dependence of the
generalized scattering length explicitly. To this end, we solve 
Eq.~(\ref{eq_trans}) self-consistently~\cite{blum02,bold02}, 
i.e., we seek solutions
for which
the energy on the left hand side of Eq.~(\ref{eq_trans}) 
is the same as that at which the
generalized free-space scattering length $a_l^{2l+1}(k)$ 
for the square well potential
with $r_0=0.1a_{{\rm{ho}},\mu}$ is being evaluated.
The resulting eigenenergies [not shown in 
Figs.~\ref{fig_energytrap}(a)
and \ref{fig_energytrap}(b)] 
are in excellent agreement with the eigenenergies
for the square-well potential for both $l=0$ and $l=1$.
In general, it is
found that the use of the energy-dependent scattering length is crucial 
for the description of higher partial wave
physics within the zero-range 
pseudopotential framework~\cite{kanj04,stoc04,idzi06}. 

\begin{figure}
\vspace*{+1.5cm}
\includegraphics[angle=0,width=65mm]{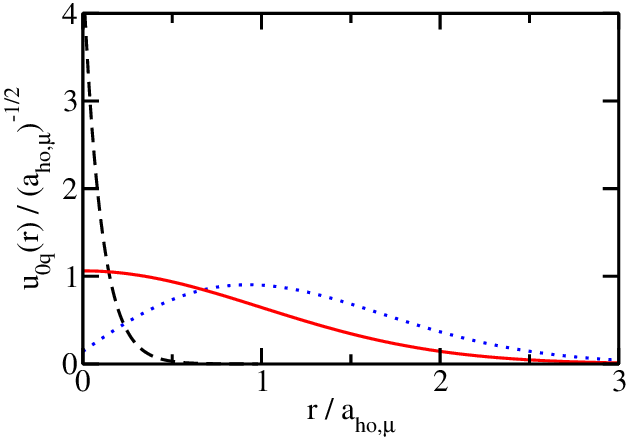}
\vspace*{-0.2cm}
\caption{(Color online)
Dotted, solid and dashed lines
show the relative $l=0$ wave function $u_{0q}(r)$, 
Eq.~(\ref{eq_radialdecaying}),
corresponding to the lowest
energy branch of the trapped two-body system
with $a_0(0)/a_{{\rm{ho}},\mu}=-0.1$,
$\infty$
and $0.1$, respectively.
}\label{fig_dimerwavefct}
\end{figure}

Figure~\ref{fig_dimerwavefct} 
shows the $l=0$ wave function 
$u_{0q}(r)$ for several $a_0(0)$ 
corresponding to the lowest energy branch. 
In the small $|a_0(0)|$ limit, $a_0(0) < 0$, $u_{0q}(r)$
approaches the non-interacting wave function.
In the small $a_0(0)$ limit, $a_0(0) > 0$, in contrast, $u_{0q}(r)$
approaches the wave function of a tightly bound 
molecule in free-space.
At unitarity, the admixture of the irregular solution
is maximal, leading to a wave function that has vanishing 
derivative at $r=0$. Formally, this wave function is identical 
to the relative even-parity wave function
of two non-interacting harmonically trapped 
particles in one dimension.

We now discuss the dependence of the
$s$-wave energy spectrum on
the external magnetic field strength $B$.
Solid and dotted lines in Fig.~\ref{fig_ccascatt}(a) show the 
magnetic field dependent scattering length $a_0(B)$,
Eq.~(\ref{eq_scattcceff}),
for two resonances
with the same $a_{\rm{bg}}$ and
$B_{\rm{res}}$ but different widths $\Delta B$,
i.e., $\Delta B = 10G$ and $1G$, respectively.
Solid and dotted lines in 
Fig.~\ref{fig_ccascatt}(b) show
the $s$-wave energy spectra 
obtained from Eq.~(\ref{eq_trans})
corresponding to the
resonances
shown in Fig.~\ref{fig_ccascatt}(a).
To this end, the
scattering length $a_0(0)$
in Eq.~(\ref{eq_trans}) is converted to $B$
via
Eq.~(\ref{eq_scattcceff}).
Figure~\ref{fig_ccascatt}(b)
shows that the energy
spectrum for  $\Delta B=10G$ is characterized by
broader avoided crossings than that for $\Delta B=1G$.
The coupling between the resonance state (or the closed
molecular channel) and the trap states increases with increasing $\Delta B$.
A Landau-Zener analysis~\cite{mies00}
shows that the energy width $\Gamma_E$,
Eq.~(\ref{eq_scattcceff2}),
determines the probability to make a transition from the
trapped state to
the molecular state. In particular, it is found that 
the transition probability decreases 
exponentially with increasing $\Gamma_E$
for the same ramp speed.
Correspondingly, the closed molecular 
channel plays a negligible role for 
resonances with relatively large $\Delta  B$
and a non-negligible role for those with
relatively small $\Delta B$. 
Thus, broad resonances 
can, generally, be described within a single channel 
framework while narrow resonances require a two-channel description
(see also Subsec.~\ref{sec_twobodyfree}).

\subsection{Beyond spherical symmetry}
\label{sec_twobodyother}
The previous subsection 
discussed the physics of two atoms interacting
through a spherically symmetric potential
under spherically symmetric confinement.
This confining geometry is fairly easy to treat both numerically
and analytically, and is directly relevant to 
on-going cold atom experiments.
This subsection considers systems for which the spherical
symmetry is broken either due to a non-spherically
symmetric confinement,
a non-spherically symmetric interaction potential or
both.
In all three cases, new two-body resonances arise that 
can be tuned by varying the confinement 
(thus motivating the term confinement-induced 
resonance~\cite{olsh98,berg03}) or 
by changing the admixture
of the non-spherical contribution to the interaction potential.
These tunable two-body resonances pave the way for investigations of
intriguing few- and many-body phenomena.

Table~\ref{tab_twobodygeometry} lists 
confinement
geometries that have been considered in the literature.
As an example,
we consider two particles interacting through
a spherically symmetric potential
$V_{\rm{tb}}$ under cylindrically-symmetric confinement in more 
detail~\cite{idzi06,olsh98,berg03,petr01,bold03,gran04,idzi06a,idzi09}.
As in Subsec.~\ref{sec_twobodytrap}, the center-of-mass degrees of freedom
separate off and the confinement in the
relative coordinate becomes 
$V_{\rm{trap}}= \mu (\lambda^2 \omega_{z}^2 \rho^2+  \omega_z^2 z^2)/2$,
where $z=z_1-z_2$ and $\rho=\sqrt{(x_1-x_2)^2+(y_1-y_2)^2}$.
The 
aspect ratio $\lambda$ is defined in terms of the
angular trapping frequencies $\omega_{\rho}$ and $\omega_z$
along the $\rho$- and $z$-directions,
$\lambda = \omega_{\rho}/\omega_z$.
Intuitively, one expects that a tight confinement in one spatial
dimension
($\lambda \ll 1$) leads to an effectively two-dimensional system in  which 
the motion along 
the tight confinement direction, or high energy coordinate, is frozen out.
Correspondingly, a tight confinement in two spatial dimensions
($\lambda \gg 1$) is expected to lead to an effectively one-dimensional system.
\begin{table}
\caption{\label{tab_twobodygeometry}
Selected trapping geometries considered
in the literature
for two-particle systems. Other geometries 
such as double well geometries and optical lattice type potentials have
also been considered.}
\begin{indented}
\item[]\begin{tabular}{l|l}
\br
Trapping geometry & References \\ \hline
hardwall at $r=R$ (three-dimensional) & \protect\cite{huan57} \\
$\frac{1}{2} \mu \omega^2 r^2$ (three-dimensional) & \protect\cite{peac04,kanj04,stoc04,busc98,chen07} \\
$\frac{1}{2} \mu \omega^2 \rho^2$ (two-dimensional) & \protect\cite{kanj07}\\
$\frac{1}{2} \mu \omega^2 z^2$ (one-dimensional) & \protect\cite{kanj04,busc98}\\
$\frac{1}{2} \mu (\omega_{\rho}^2 \rho^2 + \omega_z^2 z^2)$  & \protect\cite{idzi06,olsh98,berg03,petr01,bold03,gran04,idzi06a,idzi09} \\
$\frac{1}{2} m \sum_{i=1,2} ( \omega_x^2 x_i^2+ \omega_y^2 y_i^2+ \omega_z^2 z_i^2)$  & \protect\cite{dien06,lian08,peng10} \\
$\frac{1}{2}(m_1 \omega_1^2 \vec{r}_1^2+m_2 \omega_2^2 \vec{r}_2^2)$  & \protect\cite{deur08,gris09,blum08e} \\
cubic box; periodic boundary condition & \protect\cite{lues91}\\
box with lengths $L_x,L_y,L_z$; periodic boundary condition & \protect\cite{feng04} \\
\br
\end{tabular}
\end{indented}
\end{table}
This argument can be formalized by defining effective
one- and two-dimensional interaction potentials.

We exemplarily consider the $\lambda \gg 1$ case
with $s$-wave interactions~\cite{olsh98,berg03}.
To start with, we assume that the motion in the
$\rho$- and $z$-directions separates and, 
further, that the motion along the 
$\rho$-coordinate is frozen in the harmonic oscillator ground state
$\phi_{\rm{ho},0}(\rho)$.
Modeling the two-body interaction potential through
$V_0^{\rm{pp}}$, we define an effective
one-dimensional potential $V_{\rm{eff}}^{\rm{1d}}(z)$ through
\begin{eqnarray}
V_{\rm{eff}}^{\rm{1d}}(z)= 2 \pi \int_0^{\infty}
\phi^*_{\rm{ho},0}(\rho) V_{0}^{\rm{pp}}(\vec{r}) \phi_{\rm{ho},0}(\rho) \rho d\rho.
\end{eqnarray}
The integral evaluates to
$V_{\rm{eff}}^{\rm{1d}}(z)= g_{\rm{eff}}^{\rm{1d}} \delta(z)$ with
$g_{\rm{eff}}^{\rm{1d}}=2 \hbar^2 a_0(0) / (\mu a_{{\rm{ho}},\rho}^2)$, where 
$a_{{\rm{ho}},\rho}$ denotes  the transverse confinement length,
$a_{{\rm{ho}},\rho} = \sqrt{\hbar/(\mu \omega_{\rho})}$.
So far, we have assumed 
that the motion along the $\rho$-coordinate is frozen in the 
harmonic oscillator ground state $\phi_{\rm{ho},0}(\rho)$.
This is, however, not the case since
the pseudopotential $V_0^{\rm{pp}}$
couples the $\rho$ and $z$ degrees of freedom.
Thus, the cylindrically symmetric outside
solution 
(valid for $r>0$)
needs to be projected onto the spherically symmetric
inside solution
(valid for $r=0$), 
thereby leading to an admixture of excited
state harmonic oscillator wave functions $\phi_{\rm{ho},j}(\rho)$;
here, $j$ collectively denotes the quantum numbers
associated with the motion in the $xy$-plane.
For finite-range potentials, the outside and inside solutions can, 
using a frame transformation approach, 
be connected at a finite $r$ value~\cite{gran04}.
A full microscopic calculation 
for the zero-range potential
gives~\cite{olsh98},
\begin{eqnarray}
\label{eq_effective1d}
g_{\rm{eff}}^{\rm{1d}}= \frac{2 \hbar^2 a_0(0)}{\mu a_{{\rm{ho}},\rho}^2} 
\left[ 1 - |\zeta(1/2)| \frac{a_0(0)}{a_{{\rm{ho}},\rho}} \right]^{-1} ,
\end{eqnarray}
where $\zeta(1/2)=-1.4604$.
For small $|a_0(0)|/a_{{\rm{ho}},\rho}$, Eq.~(\ref{eq_effective1d})
reduces to the
naive estimate discussed above. 
For large
$|a_0(0)|/a_{{\rm{ho}},\rho}$, however, the strong transverse confinement 
modifies the effective one-dimensional
coupling constant. In particular, a
diverging one-dimensional coupling constant, and thus a one-dimensional
resonance, 
is found for $a_0(0)/a_{{\rm{ho}},\rho}=|\zeta(1/2)|^{-1}$.
Effectively one-dimensional many-body systems are, in many cases,
well described by strictly one-dimensional Hamiltonian
such as the Lieb-Liniger or
Gaudin-Yang Hamiltonian~\cite{lieb63,lieb63a,gaud67,yang67}.
For cold atom realizations of these 
Hamiltonian, the coupling constant 
is given by $g_{\rm{eff}}^{\rm{1d}}$, Eq.~(\ref{eq_effective1d}).
Since $g_{\rm{eff}}^{\rm{1d}}$
can be 
tuned by varying $a_{{\rm{ho}},\rho}$ or $a_0(0)$,
Eq.~(\ref{eq_effective1d}) forms the basis for studying
strongly correlated one-dimensional systems.
One-dimensional systems have attracted 
a great deal of attention from mathematicians and
physicists since they are more amenable to analytical
treatments than two- or three-dimensional systems. Strictly
one-dimensional systems with contact interactions, e.g., are integrable.
Deviations from a strictly one-dimensional  geometry may
lead to deviations from 
integrability.
Thus, important 
questions regarding integrability, near-integrability
and thermalization can be investigated by studying highly-elongated
quasi-one-dimensional Bose and Fermi 
gases~\cite{kino06,rigo07,rigo08,yuro06,maze08}.

Similar analyses have been conducted for higher partial 
waves
as well as for geometries with 
$\lambda \ll 1$~\cite{idzi06,petr01,gran04}.
In all cases, confinement induced resonances, i.e.,
diverging 
effective low-dimensional coupling constants, are
found 
for finite three-dimensional scattering lengths.
The study of small quasi-two-dimensional trapped 
Fermi gases is partially motivated by analogous studies
of two-dimensional quantum dot systems~\cite{asho96,reim02}. 
In the context of cold
Fermi gases, the filling of energy levels
has been analyzed using Hund's rules~\cite{ront09}.
An alternative trapping geometry consists of a periodic optical
lattice, which also gives rise to confinement induced resonances
that can be tuned by varying the lattice height or other system
parameters~\cite{fedi04,wout06,grup07,cui10}.

The energy spectrum of two-particle systems in a cubic box
with periodic boundary conditions has been analyzed extensively
in the context of lattice QCD calculations~\cite{lues86,lues91,hamb83}.
For this confining geometry, the center of mass of the system does not
decouple and two-body interactions lead to a modification of the plane
wave solutions applicable to the non-interacting system. In analogy to
Eq.~(\ref{eq_trans}), the
two-body  energy spectrum for the box-confinement has been
related to the phase shifts. The corresponding equation
is termed L\"uscher's formula~\cite{lues91} and has found 
applications
in QCD and other lattice simulation studies. 
An analysis of the energy spectrum determined from L\"uscher's formula
shows that the relationship between the phase shifts and 
eigenenergies is notably different from that 
found for the spherically symmetric harmonic confining
potential treated in Subsec.~\ref{sec_twobodytrap}.
This shows exemplarily that the energy spectrum
depends
in non-trivial ways
on the geometry of the external confinement.

Lastly, we comment briefly on two particles 
that interact
through a non-spherically symmetric dipole-dipole potential $V_{\rm{dd}}$
under
spherically symmetric confinement. 
For simplicity,
we assume that the electric or magnetic point dipoles
are aligned
along the $z$-axis,
$V_{\rm{dd}}(r)=d^2 (1-3 \cos^2 \vartheta)/r^3$,
where $d$ denotes the strength of the dipole moment
and $\vartheta$ the angle between the $z$-axis and $\vec{r}$.
For $r<r_0$, we assume
a spherically symmetric potential $V_{\rm{sr}}(r)$.
The dipole-dipole potential 
introduces a new
length scale, 
the dipole length $d_*$,
$d_* = 2 \mu d^2 / \hbar^2$.
Since the angle-dependence of the dipole-dipole
potential leads to the coupling of different partial waves,
the 
scattering framework discussed in Subsec.~\ref{sec_twobodyfree}
requires modifications.
One of the consequences is that one now 
has to consider a scattering matrix 
with elements $a_{l'l}$,
where 
$a_{l'l}(k) = -\tan [ \delta_{l'l}(k)]/k$~\cite{wein99,mari98,yi00,deb01}.
Compared to short-range potentials
[see Eq.~(\ref{eq_edepscatt})], 
the dipole-dipole interaction potential leads to
a modified threshold behavior, i.e.,
$a_{l'l}(0)=\lim _{k \rightarrow 0} a_{l'l}(k)$ approaches a 
constant~\cite{wein99}.
$V_{\rm{dd}}$ connects the incoming partial wave labeled 
by $l$ with the outgoing partial wave labeled by $l'$.
The multi-channel nature of the dipole-dipole
potential gives rise to novel two-body resonances,
which can be tuned by varying the dipole moment 
$d$~\cite{mari98,deb01,tick05,kanj08,roud09,roud09a}.
For electric dipoles, this can be achieved through the application
of an external electric field.
Similarly to the case of purely spherically-symmetric
two-body interaction potentials,
the energy spectrum of two dipoles under spherically
symmetric harmonic confinement can be determined
analytically if the dipole length $d_*$ is smaller than the
harmonic oscillator length $a_{{\rm{ho}},\mu}$~\cite{kanj07}. 
This review will not discuss
the few-body physics of trapped dipolar systems further.

\section{Hyperspherical coordinate
treatment of trapped three-particle system}
\label{sec_threebody}
This section considers three-particle systems with short-range
interactions under external spherically
symmetric harmonic confinement. 
For this confinement geometry,
the center-of-mass degrees of freedom $\vec{R}_{\rm{cm}}$
separate off, leaving six relative coordinates.
The trapped three-body system can be analyzed by a variety
of techniques including those developed
by chemists and nuclear physicists.
This section focuses on the hyperspherical coordinate
approach,
which provides an intuitive
picture for understanding three-body systems with bosonic,
fermionic or mixed symmetry. 
Subsection~\ref{sec_hyperspherical} introduces the formalism,
while Subsecs.~\ref{sec_threebodylimits} and
\ref{sec_efimov}
discuss 
selected aspects of
the non-interacting and 
large scattering length regimes
and the Efimov regime, respectively.
Implications of three-body 
physics
for larger systems
are discussed in Secs.~\ref{sec_fermigas} and \ref{sec_bosegas}.

\subsection{General framework}
\label{sec_hyperspherical}
The hyperspherical coordinate approach transforms
the coordinates of the system into a single length, the hyperradius $R$,
and a set of hyperangles, collectively denoted by 
$\vec{\Omega}$~\cite{wern06,lin95,ritt11,mace68,aver89,bohn98,fano99,ritt06}.
For the three-body system in the center of mass frame,
$\vec{\Omega}$ consists of five hyperangles~\cite{esry99}.
The hyperspherical coordinate approach has been employed in 
many contexts such as in the description of autoionizing states of the 
He atom~\cite{mace68} and the Efimov effect~\cite{fedo93}.
It has also been applied successfully to systems consisting of
more than three particles~\cite{wern06,bohn98,ritt06} 
(see, e.g., Subsec.~\ref{sec_fermigasunit}).
The hyperradial dynamics can
be visualized and interpreted in much the same way as 
that of a diatomic molecule described
by a set of coupled BO potential curves.

To define the hyperspherical coordinates for
three-body systems with masses $m_i$ 
and position vectors $\vec{r}_i$ ($i=1,2,3$),
we introduce Jacobi vectors $\vec{\rho}_i$,
\begin{eqnarray}
\vec{\rho}_{1} =
\vec{r}_1-\vec{r}_2
\end{eqnarray}
and
\begin{eqnarray}
\vec{\rho}_2 = 
\frac{m_1 \vec{r}_1+ m_1 \vec{r}_2}{m_1+m_2} - \vec{r}_3 .
\end{eqnarray}
The Jacobi vector $\vec{\rho}_3$ coincides with
$\vec{R}_{\rm{cm}}$.
The hyperradius $R$, which can be interpreted
as measuring the overall size of the system, is defined through
\begin{eqnarray}
\label{eq_hyperradiusthreebody}
\mu_R R^2 = \mu_{1} \vec{\rho}_1^2 + \mu_{2} \vec{\rho}_2^2,
\end{eqnarray}
where $\mu_R$ denotes the hyperradial mass 
(a scaling factor that can be chosen
conveniently), and where
$\mu_{1}$ and $\mu_{2}$ denote the 
reduced masses associated with $\vec{\rho}_{1}$ and $\vec{\rho}_{2}$,
respectively,
$\mu_{1}=m_1m_2/(m_1+m_2)$ and $\mu_{2}= (m_1+m_2)m_3/(m_1+m_2+m_3)$.
The hyperangle $\alpha$, $0 \le \alpha \le \pi/2$,
is defined as 
\begin{eqnarray}
\alpha=
\arctan \left( 
 \frac{\sqrt{\mu_{1}}\rho_{1}}{\sqrt{\mu_{2}}\rho_{2}} \right).
\end{eqnarray}
The other four hyperangles
are given by the angles $(\vartheta_1,\varphi_1)$ and
$(\vartheta_2,\varphi_2)$, which 
describe the orientation of the
unit vectors $\hat{\rho}_{1}$ and 
$\hat{\rho}_{2}$, respectively.
In these coordinates, the volume element becomes
$R^{5} dR d \vec{\Omega}$, where
$d \vec{\Omega}= 
\sin(2 \alpha) d\alpha 
d \vec{\Omega}_1 d \vec{\Omega}_2$
with
$d \vec{\Omega}_j= \sin \vartheta_j d \vartheta_j
d \varphi_j$.
We note that the definition of the 
hyperangular
coordinates 
is not unique. The ``optimal set of coordinates''
depends on the choice of the analytical or numerical  
technique employed.

In the hyperspherical framework,
the relative wave
function $\psi(R,\vec{\Omega})$
is expanded
in terms of a 
set of 
channel
functions $\Phi_{\nu}(R; \vec{\Omega})$
that depend parametrically on the hyperradius $R$ and a set
of weight functions $F_{\nu q}(R)$~\cite{niel01,mace68},
\begin{eqnarray}
\psi(R,\vec{\Omega})=
\frac{1}{R^{5/2}} {\cal{S}} 
\left(
\sum_{\nu q} F_{\nu q}(R) 
\frac{\Phi_{\nu}(R;\vec{\Omega})}{\sin (2 \alpha)}
\right),
\end{eqnarray}
where ${\cal{S}}$ denotes the symmetrization operator, which
can be written in terms of permutation operators
$P_{jk}$
that exchange
the $j$$^{th}$ and the $k$$^{th}$ particle.
For three identical bosons, e.g., ${\cal{S}}$ can be written as
$3^{-1/2}(1+P_{23}+P_{31})$.
The channel functions $\Phi_{\nu}(R;\vec{\Omega})$
are eigenfunctions of the fixed-$R$
hyperangular Hamiltonian $H_{\rm{adia}}$,
\begin{eqnarray}
\label{eq_hyperangular}
H_{\rm{adia}}
\Phi_{\nu}(R; \vec{\Omega})=
\frac{\hbar^2 s_{\nu}^2(R)}{2 \mu_R R^2} \Phi_{\nu}(R; \vec{\Omega}),
\end{eqnarray}
where 
\begin{eqnarray}
\label{eq_hamhyperangular}
H_{\rm{adia}}=
\frac{\hbar^2 \Lambda^2}{2 \mu_R R^2} +V_{\rm{int}}(R,\vec{\Omega}) .
\end{eqnarray}
For each fixed hyperradius $R$,
the channel functions
form an orthonormal set, i.e.,
$\int \Phi_{\nu}^*(R; \vec{\Omega}) \Phi_{\nu'}(R; \vec{\Omega}) d \vec{\Omega}
= \delta_{\nu \nu'}$.
In Eq.~(\ref{eq_hamhyperangular}),
$\Lambda^2$
denotes the square of the grand angular momentum operator,
which contains the contributions of the hyperangular motion
to the kinetic energy operator~\cite{aver89}. 
For the purpose of this review, the functional form 
of $\Lambda^2$ is not needed.
The potential energy surface
$V_{\rm{int}}$ accounts for the particle-particle interactions,
$V_{\rm{int}}=\sum_{j<k}V_{\rm{tb}}(r_{jk})$.

The weight functions $F_{\nu q}(R)$ are solutions to the
coupled equations
\begin{eqnarray}
\label{eq_hyperradial}
\left[ 
H_R
-\frac{\hbar^2 }{2 \mu_R}W_{\nu \nu}(R)
-E_{\nu q}
\right] F_{\nu q}(R) = \nonumber \\
\frac{\hbar^2}{2 \mu_R}\sum_{\nu' \ne \nu} 
\left[
W_{\nu \nu'}(R) + 2 P_{\nu \nu'}(R) \frac{\partial}{\partial R}
\right] F_{\nu' q}(R),
\end{eqnarray}
where
\begin{eqnarray}
\label{eq_hamhyperradial}
H_R = \frac{-\hbar^2}{2 \mu_R}\frac{\partial^2}{\partial R^2} +
\frac{\hbar^2 [ s_{\nu}^2(R)-1/4 ]}{2 \mu_R R^2} +
\frac{1}{2} \mu_R \omega^2 R^2 .
\end{eqnarray}
The last term on the right hand side of Eq.~(\ref{eq_hamhyperradial})
represents the external spherically symmetric
confining potential $V_{\rm{trap}}$
in the relative coordinates,
$V_{\rm{trap}}=\sum_i m_i \omega^2 \vec{r}_i^2/2-M_{\rm{cm}} \omega^2 \vec{R}_{\rm{cm}}^2/2$
with $M_{\rm{cm}}=\sum_i m_i$.
The coupling elements
$P_{\nu \nu'}(R)$ and $W_{\nu \nu'}(R)$
are defined through
\begin{eqnarray}
\label{eq_couplingp}
P_{\nu \nu'}(R)= \int
\Phi_{\nu} (R;\vec{\Omega}) 
\frac{\partial \Phi_{\nu'}(R;\vec{\Omega}) }{\partial R} 
d\vec{\Omega}
\end{eqnarray}
and
\begin{eqnarray}
\label{eq_couplingw}
W_{\nu \nu'}(R)=
\int
\Phi_{\nu}(R;\vec{\Omega}) 
\frac{\partial^2 \Phi_{\nu'}(R;\vec{\Omega})}{\partial R^2} 
d\vec{\Omega} .
\end{eqnarray}
No approximations have been made, and 
Eqs.~(\ref{eq_hyperangular})
and (\ref{eq_hyperradial})
are equivalent to the relative Schr\"odinger equation.

Note that
$V_{\rm{trap}}$ enters into the hyperradial Schr\"odinger equation,
Eq.~(\ref{eq_hyperradial}),
but not into the hyperangular Schr\"odinger equation,
Eq.~(\ref{eq_hyperangular}).
This implies that the three-body system under 
spherically symmetric harmonic 
confinement shares many properties with the corresponding
free-space system. In particular, the hyperangular 
eigenvalues $\hbar^2 s_{\nu}^2(R)/(2 \mu_R R^2)$ are the same for the 
free-space and trapped systems.
On the other hand,
the 
weight functions $F_{\nu q}(R)$ of the free-space system
obey, depending on the threshold, 
either bound state or scattering boundary conditions
at large $R$, while all $F_{\nu q}(R)$  of the trapped
system obey bound state
boundary conditions 
at large $R$.
A comparison of the two-body and three-body systems
reveals that
the three-body hyperradius $R$
plays a role analogous to the two-body interparticle
distance $r$, and that the hyperangles $\vec{\Omega}$
play a role analogous to the angles that describe the 
direction of $\vec{r}$. 

In general, the coupled equations in the hyperradial 
coordinate $R$, see Eq.~(\ref{eq_hyperradial}),
have to be solved numerically.
However, certain approximations can be made that lead to
strict upper and lower variational 
bounds~\cite{niel01,star79,coel91}. 
An upper bound to the energies
can be  obtained by neglecting the coupling matrix elements
$W_{\nu \nu'}$ and $P_{\nu \nu'}$ with $\nu \ne \nu'$,
i.e., by setting the right hand side of Eq.~(\ref{eq_hyperradial})
to zero. The energies obtained by solving the radial equation
in this approximation, the so-called adiabatic
approximation, provide an upper bound to the exact
eigenenergies.
A lower bound to the exact ground state energy
can be obtained by additionally neglecting the
diagonal coupling element or so-called
adiabatic correction $W_{\nu \nu}$.

If $V_{\rm{int}}$ is given by a sum of two-body zero-range
potentials, i.e., $V_{\rm{int}}=\sum_{j<k} V_0^{\rm{pp}}(\vec{r}_{jk})$,
the eigenvalues $\hbar^2 s_{\nu}^2(R)/(2 \mu_R R^2)$ of the
hyperangular Schr\"odinger equation can be obtained by
finding the roots of a
transcendental equation~\cite{ritt10} for any combination of
masses $m_i$, any combination
of $s$-wave scattering lengths $a_0^{(jk)}(0)$, and any 
symmetry (for earlier work see, e.g., 
Refs.~\cite{niel01,fedo93,kart07,wern06a,kart02}).
For this review, the BBB system (three identical bosons)
and the FFX system (two identical fermions and a third particle,
either boson or fermion)
are the most relevant.
The FFX system with mass ratio $\kappa=1$ can be realized 
by occupying
two different hyperfine
states 
of $^{6}$Li or $^{40}$K.
FFX systems with unequal masses of current experimental interest include
$^6$Li-$^{84}$Sr ($\kappa \approx 0.071$),
$^6$Li-$^{87}$Rb ($\kappa \approx 0.069$),
$^{40}$K-$^{84}$Sr ($\kappa \approx 0.48$),
$^{173}$Yb-$^{40}$K ($\kappa \approx 4.3$),
$^{40}$K-$^6$Li ($\kappa \approx 6.7$),
and
$^{87}$Sr-$^6$Li ($\kappa \approx 14.5$).

\begin{figure}
\vspace*{+1.5cm}
\includegraphics[angle=0,width=65mm]{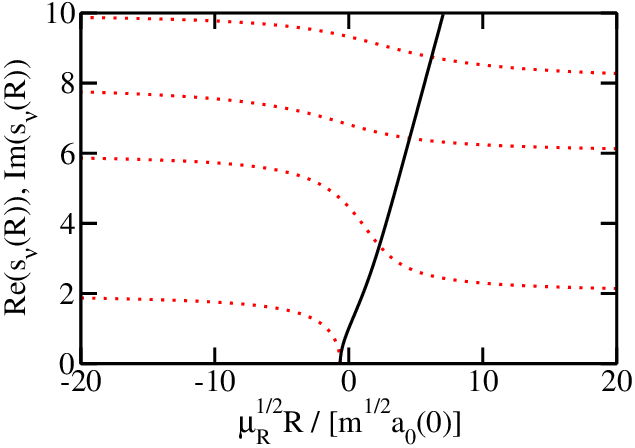}
\vspace*{-0.2cm}
\caption{(Color online)
$s_{\nu}(R)$ values for the BBB system with $L=0$.
Dotted and solid lines
show the purely real and purely imaginary
roots $s_{\nu}(R)$, obtained by solving Eq.~(\ref{eq_transbbb}), 
as a function of $\sqrt{\mu_R}R/[\sqrt{m}a_0(0)]$.
}\label{fig_snuroot}
\end{figure}

For the BBB system with relative angular momentum 
$L=0$ and $s$-wave scattering length
$a_0(0)$
[i.e., all $a_0^{(jk)}(0)$ are identical], the transcendental equation 
reads~\cite{fedo93,ritt10,jons02}
\begin{eqnarray}
\label{eq_transbbb}
s_{\nu}(R) \cos \left[ s_{\nu}(R) \frac{\pi}{2} \right]
-\frac{8}{\sqrt{3}} \sin \left[ s_{\nu}(R) \frac{\pi}{6} \right]
= \nonumber \\
\sqrt{\frac{\mu_R}{m}}\frac{R}{a_{0}(0)} \sqrt{2}
\sin \left[ s_{\nu}(R) \frac{\pi}{2} \right].
\end{eqnarray}
Equation~(\ref{eq_transbbb}) is 
written in terms of $\sqrt{\mu_R} R$ as opposed to $R$ since 
the definition of $R^2$, Eq.~(\ref{eq_hyperradiusthreebody}), 
contains the scaling factor $\mu_R$; 
in other words, $\mu_R R^2$ is an invariant.
Figure~\ref{fig_snuroot} shows the quantity
$s_{\nu}(R)$ for the BBB system with $L=0$
as a function of $\sqrt{\mu_R}R/[\sqrt{m}a_0(0)]$.
$s_{\nu}(R)$ is either purely real 
or purely imaginary.
In particular, the lowest $s_{\nu}(R)$ value becomes
imaginary at $\sqrt{\mu_R}R/[\sqrt{m}a_{0}(0)] \approx -0.6385$.
Although $s_{\nu}$ changes from  being
purely real to being purely imaginary, the quantity
$s_{\nu}^2$, which enters into the hyperradial
Schr\"odinger equation [see Eqs.~(\ref{eq_hyperradial})
and (\ref{eq_hamhyperradial})],
changes smoothly (it simply changes sign).
The purely imaginary root has profound implications
on the system properties. As discussed in more
detail in Subsec.~\ref{sec_efimov}, the imaginary root 
in the large $|a_0(0)|$ regime gives rise to 
Efimov physics. 
Expressions analogous to Eq.~(\ref{eq_transbbb})
can be obtained for other $L$ values,
i.e.,
there exists a sequence of hyperangular eigenvalues $s_{\nu}(R)$ for
each $L$.
Efimov physics
is absent 
for BBB systems with $L>0$. 

\begin{figure}
\vspace*{+1.5cm}
\includegraphics[angle=0,width=65mm]{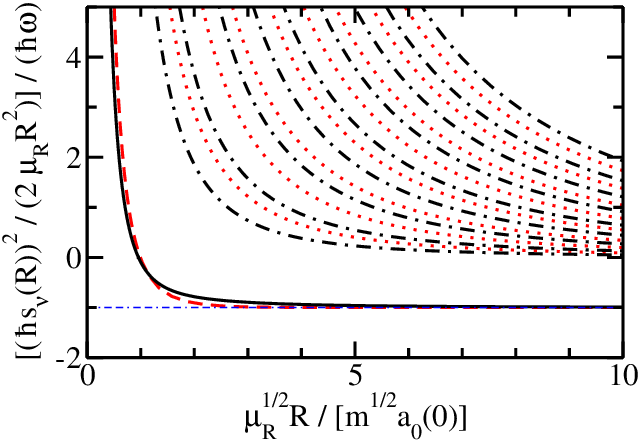}
\vspace*{-0.2cm}
\caption{(Color online)
Hyperradial potential curves 
$\hbar^2 s_{\nu}^2(R) / (2 \mu_R R^2)$ for the FFX system 
($\kappa=1$) as a function of 
$\sqrt{\mu_R}R/[\sqrt{m}a_0(0)]$ for $a_0(0)>0$.
Dashed and solid lines show the potential curves that asymptotically
approach the atom-dimer threshold
for $L=0$ and $L=1$, respectively.
Dotted and dash-dotted lines show the potential curves that asymptotically
approach the atom-atom-atom threshold
for $L=0$ and $L=1$, respectively.
As a reference, the thin dash-dash-dotted line
indicates the atom-dimer threshold.
}\label{fig_pothyper}
\end{figure}

The FFX system contains two
$s$-wave interactions, one for each of the two FX pairs.
The implicit eigenequations for $s_{\nu}(R)$ read~\cite{ritt10}
\begin{eqnarray}
\label{eq_transfffp}
\frac{s_{\nu}(R) 
\kappa^{1/2}
\cos \left[ s_{\nu}(R)\frac{\pi}{2} \right]}{(1+\kappa)^{1/2}} + \nonumber \\
\frac{(1+\kappa)^{3/2} 
\sin \left[ s_{\nu}(R) \sin^{-1} \left(\frac{\kappa}{1+\kappa} \right) \right]}
{\left[\kappa (1 + 2 \kappa)\right]^{1/2}}= \nonumber \\
\sqrt{\frac{\mu_R}{m}} \frac{R}{a_0(0)} 
\sin \left[ s_{\nu}(R)\frac{\pi}{2} \right]
\end{eqnarray}
for $L=0$ 
and~\cite{ritt10,kart07}
\begin{eqnarray}
\label{eq_transfffp2}
\frac{\kappa^{3/2} [s_{\nu}^2(R)-1] 
\,
_2F_1 \left(\frac{3-s_{\nu}(R)}{2},\frac{3+s_{\nu}(R)}{2},\frac{5}{2},\frac{\kappa^2}{(1+\kappa)^2} \right)}
{3(1+\kappa)^{3/2}} - \nonumber \\
\frac{[s_{\nu}^2(R)-1]
\kappa^{1/2}
\sin \left[ s_{\nu}(R) \frac{\pi}{2} \right]}
{s_{\nu}(R)
(1 + \kappa)^{1/2}}= \nonumber \\
\sqrt{\frac{\mu_R}{m}} \frac{R}{a_0(0)} 
\cos \left[ s_{\nu}(R)\frac{\pi}{2} \right]
\end{eqnarray}
for $L=1$,
where $\kappa$ 
denotes the ratio between the mass of the F and 
X atoms and $m$ denotes the mass of the $X$ atom.
Similar expressions can be obtained for $L>1$.
Figure~\ref{fig_pothyper} shows the
effective potentials $\hbar^2 s_{\nu}^2(R) / (2 \mu_R R^2)$
for the FFX system ($\kappa=1$) with $L=0$ and $1$
as a function of $\sqrt{\mu_R}R/[\sqrt{m}a_0(0)]$, where $a_0(0)$ is positive.
For the three-body system,
a large hyperradius $R$ corresponds
to one of two scenarios.
Either all particles are far apart from
each other or one particle is far
apart from the other two. The former is associated with the 
atom-atom-atom continuum while the latter
is associated with the atom-dimer continuum.
The hyperradial potential curves that approach the
atom-dimer threshold are shown by
dashed and solid lines for $L=0$ and $L=1$, respectively.
Figure~\ref{fig_pothyper} shows that the $L=0$ curve approaches the
atom-dimer threshold
(thin dash-dash-dotted line)
faster
than
the $L=1$ curve.
The ordering of the hyperspherical potential
curves in the small $R/a_0(0)$ regime reflects the
values of $s_{\nu}$ at unitarity
[$R/a_0(0)=0$ corresponds to the infinite scattering length
or unitary limit]. 
As elaborated on further in 
Subsec.~\ref{sec_threebodylimits} (see Fig.~\ref{fig_kappa1}), the 
lowest $s_{\nu}$ value for $L=1$ lies below that for $L=0$
at unitarity.
The fact that Fig.~\ref{fig_pothyper}
contains only a single atom-dimer threshold 
is unique to the zero-range potential model employed. More realistic 
two-body potentials support several two-body bound states, and the 
corresponding three-body
hyperradial potential curves contain as many atom-dimer thresholds
as the two-body potential supports bound states.
The fact that the lowest hyperradial potential curves
for $L=0$ and $L=1$ are
purely repulsive for $\kappa=1$
signals the absence of three-body
bound states for the FFX system with zero-range interactions.

\subsection{Analytical solutions in the non-interacting
and 
large scattering length limits}
\label{sec_threebodylimits}
This subsection applies the hyperspherical
framework to 
non-interacting 
three-body systems and three-body systems with large scattering length.
Throughout this subsection, we assume that $V_{\rm{int}}$
is given by a sum of two-body potentials $V_{\rm{tb}}(r_{jk})$,
which are characterized by zero-energy
$s$-wave scattering lengths
$a_0^{(jk)}(0)$.
If all $a_0^{(jk)}$ are either zero or infinity,
the scattering lengths
do
not define meaningful length scales.
In the zero-range limit, the hyperradial and hyperangular
degrees of freedom decouple, i.e., the coupling
matrix elements $W_{\nu \nu'}$ and $P_{\nu \nu'}$ vanish
identically~\cite{wern06,ritt10,jons02}, 
and $\Phi_{\nu}$ and $s_{\nu}$ are
independent of $R$.
For finite-range 
two-body interactions with infinite scattering lengths, 
the decoupling is approximate. In this case,
the zero-range solutions apply in the regime 
where $R$ and $a_0(0)$ are much greater than $r_0$, where
$r_0$ denotes the 
range of $V_{\rm{tb}}$ [or, if the $V_{\rm{tb}}(r_{jk})$
are not the same for the different pairs, the largest
of the ranges of the $V_{\rm{tb}}(r_{jk})$].

For now,
we assume that the $s_{\nu}$ are known
[see, e.g., Eqs.~(\ref{eq_transbbb}) through (\ref{eq_transfffp2})]
and independent of $R$.
The hyperradial Schr\"odinger equation
then reads
\begin{eqnarray}
\label{eq_hyperradiallimits}
\left[
\frac{-\hbar^2}{2\mu_R} \frac{\partial ^2}{\partial R^2}
+ V_{\rm{eff}}(R)+
\frac{1}{2} \mu_R \omega^2 R^2 \right]
F_{\nu q}(R) = 
 E_{\nu q} F_{\nu q}(R),
\end{eqnarray}
where
\begin{eqnarray}
\label{eq_hyperradialeffective}
V_{\rm{eff}}(R)=\frac{\hbar^2(s_{\nu}^2 -1/4)}{2 \mu_R R^2} .
\end{eqnarray}
Here, $q$ is the hyperradial quantum number [see
Eqs.~(\ref{eq_hyperradial}) and (\ref{eq_hamhyperradial})].
Equations~(\ref{eq_hyperradiallimits}) and (\ref{eq_hyperradialeffective})
are identical to Eq.~(\ref{eq_radialtrap}) 
with $V_{\rm{tb}}=0$ if the
identifications $R=r$, $\mu_R=\mu$ and $s_{\nu}-1/2=l$ are
made.
This subsection assumes that $s_{\nu}$ is real
and greater than zero; the next subsection discusses the case
where $s_{\nu}$ is purely imaginary. 
The two linearly independent solutions of 
Eq.~(\ref{eq_hyperradiallimits})
are thus identical to those given in
Eqs.~(\ref{eq_fregular}) and (\ref{eq_girregular})
if the appropriate substitutions are made.
The main difference between Eq.~(\ref{eq_hyperradiallimits})
and Eq.~(\ref{eq_radialtrap}) with $V_{\rm{tb}}=0$
is that the quantity
$s_{\nu}-1/2$ can take non-integer values ($s_{\nu}-1/2>-1/2$)
while the orbital angular momentum quantum number $l$
takes the 
values $0,1,\cdots$.
Enforcing the 
bound state boundary condition at
large $R$, the hyperradial wave function
is given by Eq.~(\ref{eq_radialdecaying})
with the substitutions introduced above.

To determine the quantization condition
for $q$, we investigate the small $R$
behavior of the two linearly independent
functions $f_{\nu q}(R)$ and $g_{\nu q}(R)$ (here
and in the following, the first subscript is 
written as
$\nu$ as opposed to $s_{\nu}-1/2$ 
for notational convenience).
Using 
$\lim_{x \rightarrow 0} \left( {_1}F_1(a,b,x^2) \right) =1$~\cite{abramowitz}
in Eqs.~(\ref{eq_fregular})
and (\ref{eq_girregular}), we find
$f_{\nu q}(x) \rightarrow x^{s_{\nu}+1/2}$
and
$g_{\nu q}(x) \rightarrow x^{-s_{\nu}+1/2}$ as
$x \rightarrow 0$,
where $x=R/a_{{\rm{ho}},R}$ with $a_{{\rm{ho}},R}=\sqrt{\hbar/(\mu_R \omega)}$.
$f_{\nu q}(x)$
is normalizable for all $s_{\nu}+1/2>-1/2$
while $g_{\nu q}(x)$ is only normalizable
if $s_{\nu}$ is less than one~\cite{nish08}.
The latter follows from the fact
that 
$\int_0^{\epsilon} |g_{\nu q}(x)|^2 dx = \mbox{ln} x |_0^{\epsilon} +C$
for $s_{\nu}=1$, where $C$ is an integration constant and $\epsilon$ 
denotes a small positive quantity;
thus, the integral shows divergent
(convergent) behavior in the $x \rightarrow 0$ limit for
$s_{\nu}>1$ ($s_{\nu}<1$).
Correspondingly, the irregular function 
$g_{\nu q}(x)$ must be eliminated for 
$s_{\nu}>1$~\cite{wern06,nish08,petr03}. Enforcing the boundary condition
$F_{\nu q}(x) \rightarrow 0$ on the exponentially decaying
piece of $f_{\nu q}(x)$, one finds
$q=0,1,\cdots$ for $s_{\nu} >1$.
For $0<s_{\nu}<1$, the irregular solution is normalizable
and 
the hyperradial solution
is 
given by a linear combination
of the regular and irregular solutions.
The ratio between the two solutions is determined by the
logarithmic derivative or hyperradial boundary condition at small $R$,
which can be thought of as being imposed by an effective 
hyperradial three-body
potential.
If
the irregular solution 
is suppressed naturally,
which is expected to be the case 
for ``generic 
interactions'',
the properties of the system 
with $1>s_{\nu}>0$ are fully determined by 
$a_0(0)$~\cite{nish08,petr03}.
However, as discussed in
Refs.~\cite{blum10,blum10a,gand10} 
and Subsec.~\ref{sec_fermigasunequalmasses}, intriguing
physics emerges if both the regular
and irregular solutions contribute.
Table~\ref{tab_hyperradial} summarizes the 
\begin{table}
\caption{\label{tab_hyperradial}
Dependence of the solutions to
the hyperradial Schr\"odinger equation with
$R$-independent $s_{\nu}$, Eqs.~(\ref{eq_hyperradial})
and (\ref{eq_hamhyperradial}),
on the value of $s_{\nu}$.}
\begin{indented}
\item[]\begin{tabular}{l|l}
\br
$s_{\nu}$ real, $s_{\nu} > 1$       & $q=0,1,\cdots$     \\
$s_{\nu}$ real, $1>s_{\nu} > 0$       & $q$ depends on hyperradial boundary condition    \\
$s_{\nu}$ imaginary     & $q$ depends on hyperradial boundary condition \\
& (Efimov physics)     \\
\br
\end{tabular}
\end{indented}
\end{table}
dependence of the quantization condition on the value of $s_{\nu}$.

We now apply the outlined formalism to the BBB and FFX 
systems~\cite{wern06a,jons02}.
For three identical bosons, the wave function
has to be fully symmetric under the exchange of
any pair of bosons. This symmetry constraint reduces
the density of states compared to that of 
a XYZ system,
which consists of three distinguishable particles
that interact through the same 
potential energy surface as the BBB system.
Since the hyperradius
$R$ is unchanged under the exchange of any pair of particles,
the symmetrization operator ${\cal{S}}$ only affects
$\vec{\Omega}$ and, correspondingly, only 
the hyperangular channel functions.
The ground state energy 
of the
non-interacting BBB system,
which has vanishing relative angular momentum 
and natural parity (i.e.,  $L^{\Pi}=0^+$), 
corresponds to a situation in
which all bosons occupy the lowest single
particle state.
Since each single particle state
has an energy of $3 \hbar \omega /2$,
the relative ground state energy $E$
is 
$E=3\hbar \omega$ (or $s_0=2$) and the center of
mass
energy is $E_{\rm{cm}}=3 \hbar \omega /2$.
The allowed $s_{\nu}$ values of the excited non-interacting states
can be found by
solving Eq.~(\ref{eq_transbbb}) for $a_0(0)=0$
(see also asymptotic values in Fig.~\ref{fig_snuroot}).
For $L=0$, 
we find
$s_{\nu}=2,6,8,10,12,14,16,18,\cdots$ ($s_{\nu}=4$ 
is forbidden due to symmetry constraints),
with degeneracies $1,1,1,1,1,2,1,2,\cdots$; 
the $s_{\nu}$ values
for 18 and higher are doubly-degenerate~\cite{blum02b}.
This non-interacting $L=0$ example shows that the $s_{\nu}$ values do not
necessarily follow a simple, easy to guess pattern.
For each $s_{\nu}$ (which depends on $L$), 
$q$ takes the values $q=0,1,\cdots$.

Next, we consider the BBB system with infinite 
scattering lengths~\cite{wern06a,jons02}.
The lowest few $s_{\nu}$ values for $L=0$ are
$s_{\nu} \approx 1.00624 i$, $4.46529$, $6.81836$ and $9.32469$
(see Fig.~\ref{fig_snuroot}).
To connect the allowed $s_{\nu}$ values for vanishing
and infinitely large
scattering lengths,
we follow the dotted lines in Fig.~\ref{fig_snuroot}
from the left to the right.
As $\sqrt{\mu_R}R/[\sqrt{m}a_0(0)]$ increases from 
$-\infty$ to $-0.6385$,
the lowest $s_{\nu}$ value
decreases from $2$ to $0$.
At $\sqrt{\mu_R}R/[\sqrt{m}a_{0}(0)] \approx -0.6385$, 
the lowest $s_{\nu}$ 
value becomes purely imaginary
and the dotted line ``goes over'' to the solid line.
The $s_{\nu}$ values corresponding to excited hyperangular states decrease 
as $\sqrt{\mu_R}R/[\sqrt{m}a_0(0)]$
increases from $-\infty$ to 0 to $\infty$.
The overall behavior of the $s_{\nu}$ values ($s_{\nu}$ real) as a 
function of $\sqrt{\mu_R}R/[\sqrt{m}a_0(0)]$
is similar to that of the eigenenergies
of the two-body system as a function of $a_{{\rm{ho}},\mu}/a_0(0)$
[see Fig.~\ref{fig_energytrap}(a)].
However,
the amount by which $s_{\nu}$ drops
when going from the non-interacting to the 
large scattering length limits depends, unlike in
the two-body case, on the state under consideration.
Furthermore, it must be kept in mind that the $s_{\nu}$ values
are
related to the hyperangular eigenvalues and that coupling between the different 
hyperangular channels modifies the energy spectrum for finite $a_0(0)$.

\begin{figure}
\vspace*{+1.5cm}
\includegraphics[angle=0,width=65mm]{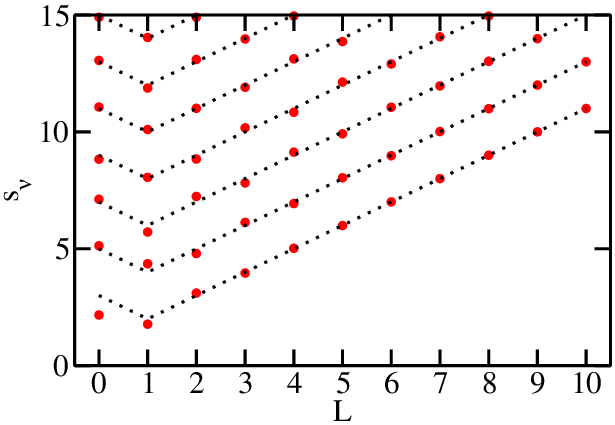}
\vspace*{-0.2cm}
\caption{(Color online)
Symbols show the $s_{\nu}$ coefficients
for the FFX system ($\kappa=1$) as a function of $L$ for
$a_0(0)=\infty$.
Note that the
$s_{\nu}$ values are
purely real for all $L$.
The $s_{\nu}$ values have been obtained by solving Eq.~(17)
of Ref.~\cite{wern06a}
[for $L=0$ and 1, see Eqs.~(\ref{eq_transfffp})
and (\ref{eq_transfffp2})].
Dotted lines show approximate semi-classically derived expressions
(see text).
}\label{fig_kappa1}
\end{figure}

Next, we consider FFX systems with $\kappa=1$ and $\kappa>1$.
Figure~\ref{fig_kappa1}
shows the $s_{\nu}$ coefficients
for the FFX system ($\kappa=1$) as a function of $L$
for $a_0(0)=\infty$~\cite{wern06a}.
The $s_{\nu}$ coefficients are purely
real and greater than $1$ for all
$L$ and correspondingly $q$ takes the
values $0,1,\cdots$ for each $s_{\nu}$ 
(see Table~\ref{tab_hyperradial}).
In Fig.~\ref{fig_kappa1},
dotted lines show the approximate semi-classically
derived $s_{\nu}$ values for $a_0(0) = \infty$~\cite{wern06a}, 
$s_{\nu}=2 \nu + 3$ for $L=0$
and $s_{\nu}=2 \nu + L + 1$ for
$L \ge 1$ ($\nu=0,1,\cdots$).
These expressions, which have also been
deduced using an 
``atom-dimer type picture''~\cite{dail10,dail10explain},
become more accurate with
increasing $\nu$ and/or $L$.
The $s_{\nu}$ coefficients at unitarity constitute a key ingredient
for determining the finite temperature equation of
state
of strongly-interacting
two-component Fermi gases
in the bulk limit via a high temperature
virial expansion approach (see 
Subsec.~\ref{sec_fermigasvirial} and Refs.~\cite{liu09,liu10}).

\begin{figure}
\vspace*{+1.5cm}
\includegraphics[angle=0,width=65mm]{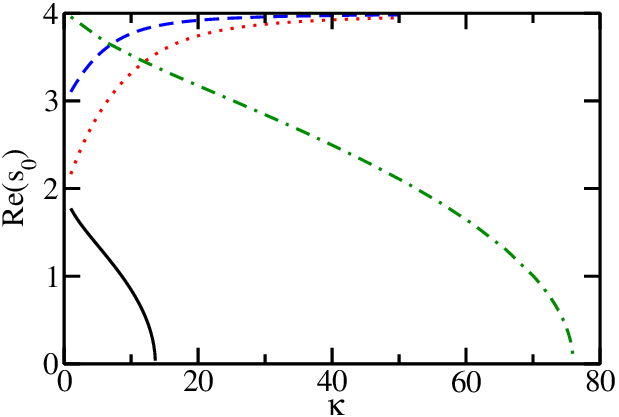}
\vspace*{-0.2cm}
\caption{(Color online)
Dotted, solid, dashed and dash-dotted lines
show the lowest $s_{0}$ value 
as a function of $\kappa$ for the FFX system at unitarity
for $L=0$, 1, 2 and 3, respectively, and $\Pi=(-1)^L$.
The $s_{0}$ values are 
obtained by solving the equations
given in Sec.~III of Ref.~\cite{ritt10}
[for $L=0$ and 1, see Eqs.~(\ref{eq_transfffp})
and (\ref{eq_transfffp2})].
}\label{fig_snufffp}
\end{figure}

Figure~\ref{fig_snufffp} shows 
the real part of 
the lowest $s_{\nu}$ 
coefficients of the FFX system
for $a_0(0)=\infty$ as a function of 
the mass ratio $\kappa$ for $L=0-3$.
The $s_{\nu}$ coefficient for $L=1$, which is lower than that for $L=0$,
2 and 3,
decreases
from $1.773$ to $1$ to $0$ as $\kappa$ increases from $1$ 
to $8.619$ to $13.607$.
For $\kappa>13.607$, the lowest $s_{\nu}$ 
coefficient for $L^{\Pi}=1^-$ becomes purely imaginary,
indicating that Efimov
physics plays a role for two-component Fermi systems with 
$\kappa>13.607$.
Applying the results from Table~\ref{tab_hyperradial}
to the FFX system with $L^{\Pi}=1^-$, 
$q$ takes integer values for $\kappa < 8.619$,
depends on the short-range hyperradial boundary condition
for $8.619 < \kappa < 13.607$,
and depends on the three-body Efimov parameter for $\kappa>13.607$.
The fact that the lowest $s_{\nu}$ coefficient of the
FFX system for $L=1$ decreases with increasing $\kappa$
can be attributed to an effective attraction
between the two heavy fermions, which is induced by the light
particle. This interpretation emerges 
within a BO treatment~\cite{petr03}, in which the light
particle (``electron'')
moves in an effective potential created
by the two heavy particles (``nuclei''); thus, the FFX
system with large $\kappa$ can be interpreted in much the same
way as, for example, the H$_2^+$ molecule.

\subsection{Efimov physics}
\label{sec_efimov}
As already alluded to, the Efimov
effect arises if two or more of the 
$s$-wave scattering lengths that characterize the
two-body subsystems diverge and if $s_{\nu}$ is purely
imaginary.
Examples include the BBB system with
$L^{\Pi}=0^+$ symmetry
and
the FFX system with $\kappa > 13.607$ and $L^{\Pi}=1^-$
symmetry.
In the limit that $r_0 \rightarrow 0$,
the hyperradial potential $V_{\rm{eff}}$, Eq.~(\ref{eq_hyperradialeffective}), 
takes the form 
$\hbar^2(-\bar{s}_{\nu}^2-1/4)/(2 \mu_R R^2)$, where
$\bar{s}_{\nu}$ is purely
real, i.e., $\bar{s}_{\nu}=\mbox{Im}(s_{\nu})$.
The bound state solution of the free-space
hyperradial Schr\"odinger, Eq.~(\ref{eq_hyperradiallimits}) with $\omega=0$
and $V_{\rm{eff}}$ given above,
reads 
$F_{\nu \bar{\kappa}}(R)=
(\bar{\kappa} R)^{1/2}K_{i \bar{s}_{\nu}}(\bar{\kappa} R)$,
where 
$\bar{\kappa} = \sqrt{2 \mu_R |E|/\hbar}$
and where $K_{i\bar{s}_{\nu}}$ 
denotes the modified Bessel function of the second kind with imaginary
index~\cite{braa06}.
Here, the hyperradial solution
is labeled by $\bar{\kappa}$ and not by $k$, as
done in Subsec.~\ref{sec_twobodyfree}, since we are considering
bound states and not scattering states.
To be consistent with
our previously introduced conventions,
our $\bar{\kappa}$ includes
an extra factor of $\sqrt{2}$ 
compared to Eq.~(152) of Ref.~\cite{braa06}.
An analysis of the solution
shows that 
the three-body system supports an infinite
number of three-body bound states. While
the bound state
energies $E^{(n)}$ depend on the short range boundary
condition, the ratio between two consecutive
bound state energies $E^{(n)}$ and $E^{(n+1)}$
is independent of the three-body
parameter and fully determined by $\bar{s}_{\nu}$,
\begin{eqnarray}
\label{eq_energyspacing}
\frac{E^{(n+1)}}{E^{(n)}} = \exp \left(- \frac{2 \pi}{\bar{s}_{\nu}}
\right).
\end{eqnarray}
For the BBB system in free space,
for example, the spacing between the $(n+1)$$^{\mathrm{th}}$ and 
$n$$^{\mathrm{th}}$
bound state is approximately $1/515$. 
This geometric spacing reflects a discrete scale invariance and is the 
key characteristic of the Efimov effect.
Other observables 
such as the three-body recombination rate $K_3$
to weakly bound dimers and scattering resonances also
carry signatures of the discrete
scale invariance.
For example, $K_3$ can be written as $f \hbar |a_0(0)|^4/m$,
where the ``amplitude'' $f$
depends on the two-body $s$-wave scattering
length $a_0(0)$ and the three-body phase $\theta_b$, 
which is---in turn---related
to $E^{(n)}$~\cite{esry99b,niel99,beda00,braa01}.
The short-range phase $\theta_b$ can thus be determined by 
fitting the experimentally determined 
three-body recombination rate, measured over 
a wide range of $s$-wave scattering 
lengths~\cite{krae06},
to the known functional
form of $K_3$.
Although the discussion so far is, strictly speaking,
only valid in the $r_0 \rightarrow 0$ and $a_0(0) \rightarrow \infty$
limits, the key features survive as long as $r_0 / |a_0(0)| \ll 1$.
See, e.g., Refs.~\cite{efim93,plat09,thog09} for a discussion of
finite-range effects.

We now investigate how the external
spherically symmetric confinement modifies the 
Efimov spectrum~\cite{wern06a,jons02,stol05,port11}.
In particular,
we seek solutions to Eqs.~(\ref{eq_hyperradiallimits})
and (\ref{eq_hyperradialeffective})
for 
purely imaginary and $R$-independent 
$s_{\nu}$.
As before, the hyperradial solutions
are given by Eq.~(\ref{eq_radialdecaying})
with $l$ replaced by $i\bar{s}_{\nu}-1/2$ and 
$x$ equal to $R/a_{{\rm{ho}},R}$. The 
radial solutions can be conveniently written in terms of the
Whittaker function $W$~\cite{abramowitz,wern06a},
$F_{\nu \kappa}(x) = 
N_{i \bar{s}_{\nu}}(\kappa) x^{-1/2}W _{E/2,i\bar{s}_{\nu}/2}(x^2)$, where
$\kappa$ is used instead of $q$ to label the hyperradial
solutions
and where
$N_{i \bar{s}_{\nu}}(\kappa)$ denotes a normalization constant.
In the small $x$ limit,
the radial solutions become
\begin{eqnarray}
\label{eq_smallxlimitefimov}
F_{{\nu} \kappa}(x) \rightarrow
N_{i \bar{s}_{\nu}}(\kappa) \sqrt{x}
(b x^{i \bar{s}_{\nu}} - b^* x^{-i \bar{s}_{\nu}}),
\end{eqnarray}
where~\cite{jons02}
\begin{eqnarray}
\label{eq_bvalue}
b = \frac{\Gamma
\left(\frac{1}{2}-\frac{E}{2 \hbar \omega} +\frac{i \bar{s}_{\nu}}{2}
\right)}
{\Gamma(1+ i \bar{s}_{\nu})}.
\end{eqnarray}
We emphasize that the quantity $b$ depends on the energy
of the trapped three-body system.
Thus, for a given three-body system,
$b$ can be varied by changing the angular trapping frequency.
Defining 
$b=|b| \exp(i \theta_b)$, 
Eq.~(\ref{eq_smallxlimitefimov}) becomes
\begin{eqnarray}
\label{eq_smallxlimitefimov2}
F_{\nu \kappa}(x) \rightarrow
N_{i \bar{s}_{\nu}}(\kappa) \sqrt{x} \sin( \bar{s}_{\nu} \ln x + \theta_b)
\end{eqnarray}
with
\begin{eqnarray}
\label{eq_efimovphasetrap}
\theta_b = \mbox{arg}(b).
\end{eqnarray}
For a fixed $\theta_b$, the allowed energy eigenvalues
are determined by Eq.~(\ref{eq_efimovphasetrap}).

\begin{figure}
\vspace*{+1.5cm}
\includegraphics[angle=0,width=65mm]{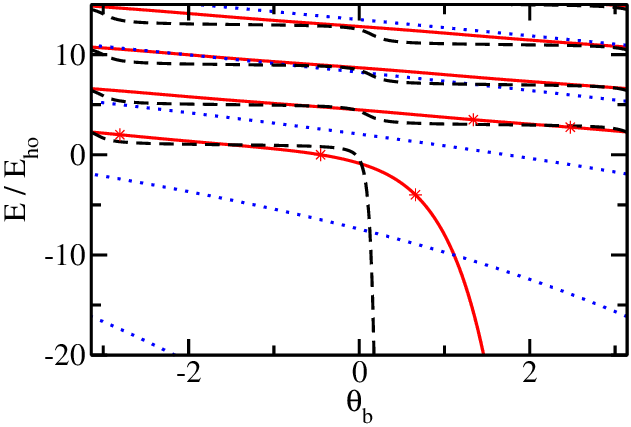}
\vspace*{-0.2cm}
\caption{(Color online)
The solid lines show
the relative eigenenergies $E$,
calculated using Eq.~(\ref{eq_efimovphasetrap}), 
as a function of $\theta_b$ 
for the BBB system with $L^{\Pi}=0^+$ at unitarity
($\bar{s}_{\nu} \approx 1.00624$). 
For comparison, dashed and dotted lines 
show the eigenenergies for a 10 times smaller and a
10 times larger $\bar{s}_{\nu}$ value,
corresponding to the FFX system with $L^{\Pi}=1^-$
and $\kappa$ just larger
than 13.607 
and very large, respectively.
Asterisks mark the $(\theta_b,E)$ values for which 
Fig.~\ref{fig_radialefimov} shows the radial wave functions.
}\label{fig_efimovphasetrap}
\end{figure}

Figure~\ref{fig_efimovphasetrap} shows the three-body eigenenergies
as a function of the three-body phase $\theta_b$
at unitarity for three
different $\bar{s}_{\nu}$ values, i.e.,
$\bar{s}_{\nu}=1.00624$, $0.100624$, and $10.0624$.
It can be seen that the energy spectrum depends 
quite strongly on $\theta_b$. In particular, the
energy spacing between consecutive eigenenergies 
varies notably with $\theta_b$ for fixed $\bar{s}_{\nu}$.
For negative energies, the spectrum
looks Efimov-like; by this we mean that the spacing between 
consecutive energy levels follows, very roughly, 
that of the free-space system.
For positive energies, in contrast,
the spectrum shows similarities with
the gas-like spectrum of three-body systems for which Efimov physics is
absent.
We emphasize that
the three-body phase $\theta_b$ is not determined by the 
zero-range model and needs to be extracted from experimental data or
{\em{ab initio}} calculations.
Spectroscopic measurements for the trapped three-body system
may, e.g., allow one to extract the three-body phase 
$\theta_b$~\cite{port11}.

\begin{figure}
\vspace*{+1.5cm}
\includegraphics[angle=0,width=65mm]{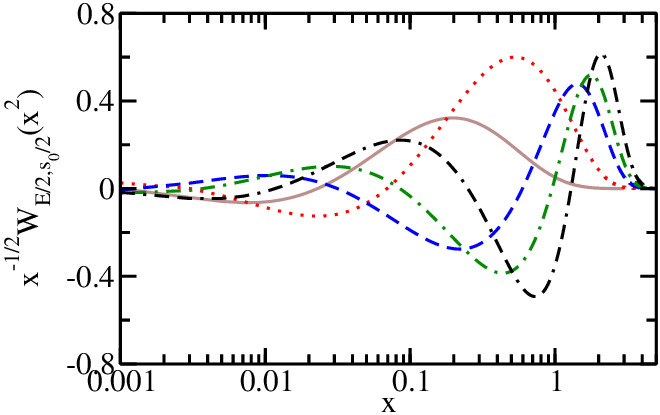}
\vspace*{-0.2cm}
\caption{(Color online)
Solid, dotted, dashed, dash-dotted and dash-dash-dotted lines show the
radial wave function $W_{E/2,s_{0}/2}(x^2)/\sqrt{x}$ 
for the BBB system at unitarity 
($\bar{s}_0=1.00624$) for 
$E=-4\hbar \omega$,
$0$,
$2\hbar \omega$,
$11 \hbar \omega/4$ and 
$7 \hbar \omega/2$, respectively.
}\label{fig_radialefimov}
\end{figure}

The short-range phase $\theta_b$
determines the small
$R$ hyperradial boundary condition.
The limiting behavior of $F_{\nu \kappa}(x)$,
Eq.~(\ref{eq_smallxlimitefimov2}), indicates that 
the radial eigenfunction shows oscillatory behavior at small $R$.
Figure~\ref{fig_radialefimov} shows the radial solutions
$F_{\nu \kappa}(x)$ for the BBB
system 
at unitarity ($\bar{s}_{\nu}=1.00624$)
for five different relative energies $E/(\hbar \omega)$ or, alternatively,
for five different three-body phases $\theta_b$.
The corresponding $(\theta_b,E)$ values are marked by 
asterisks in Fig.~\ref{fig_efimovphasetrap}.
It can be seen that 
the wave length of the oscillations decreases 
with decreasing $x$
and that the wave function extends to larger
hyperradii with increasing energy.

A more detailed discussion of Efimov physics
can be found in Refs.~\cite{braa06,braa07}.
The key
point for the discussion of larger systems is that the 
three-body energy spectrum in the Efimov regime
depends, in addition to the scattering length, on a three-body 
parameter (parameterized through $\theta_b$ above).
Although the behavior at unitarity is particularly transparent
since the channel coupling vanishes,
the three-body phase also plays a role for systems with finite
scattering lengths. As shown in Subsec.~\ref{sec_bosegasthreebodyparam},
the energy of weakly-interacting trapped Bose gases
also depends---although in a rather different
manner---on the three-body parameter.

\section{Fermi gases
under external confinement}
\label{sec_fermigas}
This section discusses the behavior of 
$s$-wave interacting 
Fermi gases under
external confinement.
Subsection~\ref{sec_fermigascrossover}
provides a qualitative and, where possible, quantitative description of the 
crossover from the weakly-attractive atomic Fermi gas to the
weakly-repulsive molecular Bose gas regime of equal-mass
systems and introduces a perturbative description
of the weakly-interacting regimes.
Subsection~\ref{sec_fermigasunit}
focuses on the unitary regime, where the $s$-wave scattering length
diverges. 
Subsection~\ref{sec_fermigasunequalmasses} extends the discussion
of equal-mass Fermi gases
to unequal-mass Fermi gases.
Lastly,
Subsec.~\ref{sec_fermigasvirial}
shows that the microscopic few-body energy spectra 
can be used to predict the thermodynamic properties of
inhomogeneous and homogeneous 
equal-mass two-component
Fermi gases via a high temperature virial expansion approach, which---somewhat
surprisingly---remains valid down to about 
half the Fermi 
temperature.

\subsection{Microscopic
description of BCS-BEC crossover of trapped 
two-component Fermi gas with equal masses}
\label{sec_fermigascrossover}
This subsection considers two-component Fermi gases with $N_1$ atoms of the
first species and $N_2$ atoms of the second species ($N=N_1+N_2$).
We assume that like fermions are non-interacting and that unlike
fermions interact through a 
spherically symmetric short-range two-body
potential $V_{\rm{tb}}(r_{jk})$, where $r_{jk} = |\vec{r}_j-\vec{r}_k|$
[$\vec{r}_j$
denotes the position vector
of the $j$$^{{th}}$
fermion ($j=1,\cdots,N$)  measured with respect to the trap center].
Restricting ourselves to spherically
symmetric harmonic confining potentials
with angular trapping frequency $\omega$,
the model
Hamiltonian $H$ reads
\begin{eqnarray}
\label{eq_hamfermigas}
H = \sum_{j=1}^N 
\left(
\frac{-\hbar^2}{2m} \nabla_{\vec{r}_j}^2
+
\frac{1}{2} m \omega^2 \vec{r}_j^2 
\right)
+\sum_{j=1}^{N_1} \sum_{k=N_1+1}^N V_{\rm{tb}}(r_{jk}),
\end{eqnarray}
where 
$m$ denotes the atom mass. The Hamiltonian
given in Eq.~(\ref{eq_hamfermigas})
describes equal-mass two-component Fermi gases such
as $^6$Li atoms in two different hyperfine states
or $^{40}$K atoms
in two different hyperfine states.
The 
Hamiltonian 
is
single-channel
in nature,
which---following the discussion of 
Subsec.~\ref{sec_twobodyfree}---implies that our treatment applies to
Fermi gases near broad but not near narrow Fano-Feshbach
resonances. 
This subsection discusses the properties
of systems governed by
$H$,
Eq.~(\ref{eq_hamfermigas}),
as functions of the number of particles $N$
(throughout, we use 
$N_1 \ge N_2$) and the $s$-wave scattering length $a_0(0)$
of the two-body potential $V_{\rm{tb}}$.

Equal-mass two-component
Fermi gases are fully universal, i.e., the system behavior is fully
determined by $a_0(0)$, provided the range $r_0$ of the
two-body potential $V_{\rm{tb}}$ is much smaller than the $s$-wave
scattering length $a_0(0)$,
the average interparticle spacing 
$\langle r_{ij} \rangle$
and the
harmonic oscillator length $a_{\rm{ho}}$, 
where $a_{\rm{ho}}=\sqrt{\hbar/(m \omega)}$.
The evidence for universality comes from 
studies that
show that the system behavior remains essentially
unchanged when the shape of the two-body
interaction potential is varied in the regime 
$r_0 \ll a_{\rm{ho}}$ and $r_0 \ll \langle r_{ij} \rangle$.
This characteristic is closely linked to the fact that the $(2,1)$ system
does not support a weakly-bound three-body state~\cite{petr03,skor57}
(see Subsecs.~\ref{sec_hyperspherical} and \ref{sec_threebodylimits}).
Furthermore, weakly-bound four-body or higher-body 
bound states are 
absent in the zero-range 
limit~\cite{carl03,astr04c,chan04a,petr04aa,blum07}, 
and dilute two-component Fermi gases are stable even
for infinitely 
large
$s$-wave scattering length
The Pauli exclusion principle can be thought of as 
producing an effective repulsive force that stabilizes the 
system with attractive interactions against collapse.
This is reminiscent of the stabilization of white
dwarfs against gravitational collapse 
by
the electron degeneracy pressure~\cite{freg06}. 

\begin{figure}
\vspace*{+1.5cm}
\includegraphics[angle=0,width=65mm]{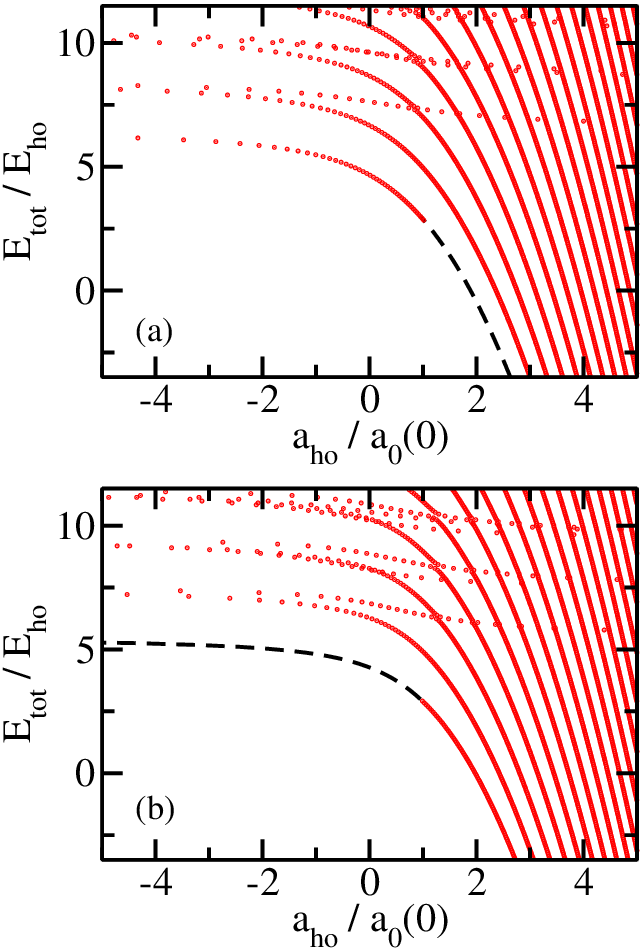}
\vspace*{-0.2cm}
\caption{(Color online)
Energy spectra for the FFX system
with $\kappa=1$ for
(a) $L^{\Pi}=0^+$ and
(b) $L^{\Pi}=1^-$ 
as a function of the inverse scattering length $[a_0(0)]^{-1}$
($E_{\rm{tot}}$ 
includes the center-of-mass energy of $3 \hbar \omega/2$).
The BCS-BEC crossover curve is shown by dashed lines. 
The energy spectra are calculated numerically
following the formulation introduced in Ref.~\protect\cite{kest07}.}
\label{fig_threefermion}
\end{figure}

We start our analysis of the Hamiltonian $H$, Eq.~(\ref{eq_hamfermigas}),
by considering the $(N_1,N_2)=(2,1)$
system. 
The energy spectrum
of the $(2,1)$ system can be obtained using the hyperspherical framework 
outlined in Subsec.~\ref{sec_hyperspherical}.
Here,
we instead employ an approach based on the
Lippmann-Schwinger equation, which
allows for the determination of the energy spectrum 
of the equal-mass three-fermion system with zero-range interactions
with comparatively little computational effort~\cite{kest07}.
The success of this technique rests in the fact that
two-body correlations are build into the
three-body wave function from the outset.
Figure~\ref{fig_threefermion} shows the energy
spectra for the two lowest relative angular momenta, i.e., 
for $L=0$ and $1$, of the three-fermion
system as a function of $a_{\rm{ho}}/a_0(0)$.
In determining the eigenenergies, we assumed that the center-of-mass 
degrees of freedom are in the ground state, i.e., 
that $E_{\rm{cm}}=3 \hbar \omega /2$.
The key characteristics of the three-fermion spectra
can be summarized as follows:

(i) In the $a_0(0) \rightarrow 0^-$ limit, the three-fermion
system behaves like a non-interacting atomic Fermi gas. The 
energy of the lowest $L=0$ and 1 states is $E_{\rm{tot}}=13 \hbar \omega/2$
and $11 \hbar \omega /2$, respectively.
The fact that the lowest $L=1$ state has a lower energy
than the lowest $L=0$ state can be understood readily
by considering the non-interacting limit,
where the wave function separates into
a component that depends on $\vec{r}_{12}$
and a component that depends on the second Jacobi vector
(when the interactions are turned on, the wave functions
no longer separate and the argument needs to be modified accordingly). The 
three-fermion wave function is antisymmetric
under the exchange of the two spin-up atoms, implying that the 
wave function component along the $\vec{r}_{12}$
vector carries one unit of angular momentum.
For the lowest state with $L=1$, the wave function
component along the second Jacobi vector carries no angular momentum
(implying that the wave function
component along the second Jacobi vector
contributes an energy of $3 \hbar \omega /2$).
For the lowest state with $L=0$, in contrast,
the wave function
component along the second Jacobi vector  carries one
unit of angular momentum, which  
couples to the angular momentum associated with the $\vec{r}_{12}$
coordinate
such that $L=0$
(implying that
the wave function component along the second Jacobi vector
contributes an energy of $5 \hbar \omega /2$).

(ii) In the $a_0(0) \rightarrow 0^+$ limit,
the three-fermion spectra consist of two different
energy families.
The first energy family ``dives down''
as $a_0(0)$ approaches $0^+$ (this is the ``molecule+atom
family'') while the second energy family, the so-called
``atom+atom+atom family'', 
approaches the energy spectrum of the non-interacting 
atomic Fermi gas [i.e., the 
atom+atom+atom part of the spectrum is the same as that in the
$a_0(0) \rightarrow 0^-$ limit (see above)].
The lowest energy level of
the $(2,1)$ system belongs to
the molecule+atom family, for which the system consists of a tightly-bound 
$s$-wave molecule 
and a spare atom
that
carries the angular momentum $L$ of the three-body system.
The energies of states belonging to
the molecule+atom family approach
$E_{\rm{tot}}=E_{\rm{bound}} + (2n_{\rm{eff}}+L+3) \hbar \omega$,
where $n_{\rm{eff}}=0,1,\cdots$ and where $E_{\rm{bound}}$
denotes the energy of the tightly-bound $s$-wave molecule
[see Eq.~(\ref{eq_boundpp})],
as $a_0(0) \rightarrow 0^+$.
Thus, the lowest $L=0$ state has an energy
that is approximately $\hbar \omega$ below that of the
lowest $L=1$ state.
The energy shift due to the effective interaction between
the molecule and the atom is discussed below
[see Eq.~(\ref{eq_pertbec})].

(iii) 
As discussed in (i) and (ii), the lowest energy of the $(2,1)$
system has angular momentum $L=1$ in the $a_0(0) \rightarrow 0^-$
limit
and $L=0$ in the 
$a_0(0) \rightarrow 0^+$
limit.
The lowest $L=1$ and $L=0$ states cross at 
$a_{\rm{ho}}/a_0(0) \approx 1$~\cite{kest07,stet07,stec08}.
The energy branch shown by
dashed lines in Figs.~\ref{fig_threefermion}(a)
and \ref{fig_threefermion}(b) corresponds to the trap analog 
of the
BCS-BEC crossover curve.
While the BCS-BEC crossover curve of the homogeneous system
depends on two dimensionless parameters, i.e., $k_F a_0(0)$ ($k_F$
denotes the Fermi vector) and the population imbalance 
$p$, $p=(N_1-N_2)/N$,
that of the harmonically trapped system depends on three 
dimensionless parameters, i.e., $a_0(0)/a_{\rm{ho}}$, $N$ and $N_1-N_2$.
The system properties of the $(4,2)$ and $(6,3)$
systems,
for example, differ somewhat, despite the fact that
both are characterized by
$p=1/3$.

(iv) The $(2,1)$ energy spectra show sequences 
of avoided crossings in the strongly-interacting regime, i.e.,
for 
$a_{\rm{ho}} /|a_0(0)| \lesssim 1$. These
avoided crossings have been analyzed within a Landau-Zener 
framework~\cite{stec07c} and play a 
crucial role in time-dependent
studies. One can, for example, prepare an initial state at 
equilibrium and analyze the system response to slow 
(adiabatic) and fast (non-adiabatic) magnetic field ramps,
modeled using a time-dependent $s$-wave scattering 
length.
At the end of the field ramp, the occupation of the 
energy families, and branches within the energy families, can be monitored.
Calculations along these lines are of direct relevance 
to ongoing experiments. On the one hand, these calculations
provide insights into the characteristic time scales
that may be extrapolated to larger systems. On the other hand,
these calculations describe the dynamics of an isolated
optical lattice site. Knowledge of the single optical lattice
site dynamics constitutes a crucial building block for understanding
the dynamics of systems in which lattice sites
are coupled.

The features pointed out in (i)-(iv) for the
$(2,1)$ equal-mass system apply, with appropriate modifications,
also to larger trapped two-component Fermi gases.
In particular, 
for small  $|a_0(0)|$, $a_0(0) <0$,
the gound state of the
trapped $(N_1,N_2)$ system behaves like a weakly-attractive atomic Fermi
gas.
For small  $|a_0(0)|$, $a_0(0) > 0$,
the ground state of the 
trapped system behaves like a weakly-repulsive molecular
Bose
gas consisting of $N_2$ diatomic
$s$-wave molecules and $N_1-N_2$ impurity fermions (if $N_1 \gg N_2$,
it is more appropriate to refer to the molecules
as impurities immersed in a Fermi sea).
In both regimes 
[negative and positive $a_0(0)$], perturbative treatments reveal the
leading order behavior of the energy levels.

\begin{figure}
\vspace*{+1.5cm}
\includegraphics[angle=0,width=65mm]{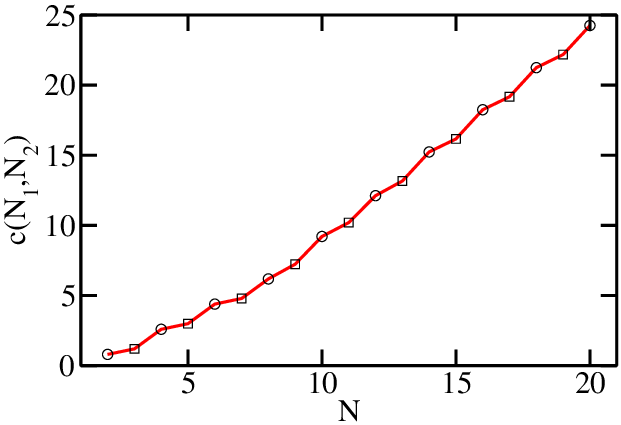}
\vspace*{-0.2cm}
\caption{(Color online)
Coefficients $c(N_1,N_2)$ that determine the energy shift
of the energetically lowest lying state
of the weakly-interacting $N$-atom family [see Eq.~(\ref{eq_pertbcs})]
as a function of $N$.
Circles and squares correspond to $N_1-N_2=0$ and 1, respectively;
the solid line serves as a guide to the eye.
The values of the coefficients are taken from 
Table~I of Ref.~\protect\cite{stec08}.}
\label{fig_pertbcs}
\end{figure}

For small $a_0(0)$, the unperturbed eigenstates of the $N$-atom
family are
given by the non-interacting gas-like atomic states and the 
sum of atom-atom potentials between unlike fermions
is treated as the perturbation.
The non-interacting wave functions can be written as a product
of two Slater determinants (one for each component), which are
constructed from the single particle harmonic oscillator
orbitals~\cite{stec08,stec07b}.
If the atom-atom interaction potential is approximated
by the Fermi pseudopotential $V_F$ with $s$-wave
scattering length $a_0(0)$~\cite{ferm34},
\begin{eqnarray}
\label{eq_fermipp}
V_F(\vec{r}_{jk}) = \frac{2 \pi \hbar^2 a_{0}(0)}{\mu} \delta(\vec{r}_{jk}),
\end{eqnarray}
where $\mu$ denotes the reduced mass of the two colliding bodies, the leading
order energy shifts can be calculated semi-analytically
within first order degenerate perturbation
theory.
Note that $V_F$ is identical to $V_0^{\rm{pp}}$,
Eq.~(\ref{eq_pseudo3d}), without the regularization operator
$(\partial/\partial r_{jk})r_{jk}$.
Diagonalization of the perturbation matrix results in the
approximate expression for $E_{\rm{tot}}(N_1,N_2)$~\cite{stec08,stec07b},
\begin{eqnarray}
\label{eq_pertbcs}
E_{\rm{tot}}(N_1,N_2) \approx E_{\rm{tot}}^{\rm{NI}}(N_1,N_2) + c(N_1,N_2) 
\frac{a_{0}(0)}{a_{\rm{ho}}} \hbar \omega,
\end{eqnarray}
where $E_{\rm{tot}}^{\rm{NI}}(N_1,N_2)$
denotes the total energy of
the unperturbed (non-interacting) atomic state
under consideration and $c(N_1,N_2)$ a  coefficient.
Figure~\ref{fig_pertbcs} shows the  coefficients $c(N_1,N_2)$ 
corresponding to
the lowest energy state for systems with $N \le 20$
and $N_1-N_2=0$ or $1$.
Interestingly, the $c(N_1,N_2)$
coefficients show distinct even-odd oscillations, with $c$ being larger
for even $N$ than for odd $N$. If $a_0(0)$ is negative, as in the
weakly-interacting BCS regime, then a larger $c$ coefficient
corresponds to a lower energy.
Thus, the perturbative energy expression predicts the existence
of a finite energy gap $\Delta E(N)$ 
in the weakly-interacting regime
[see Sec.~\ref{sec_fermigasunit}, Eq.~(\ref{eq_excitationgap}), for the 
definition of $\Delta E(N)$].

For small and positive $a_0(0)$,
the perturbative description of the lowest energy family is based
on the assumption that $N_2$ 
$s$-wave molecules, consisting of one spin-up and one spin-down
atom, form and 
that the remaining $N_1-N_2$
spin-up fermions are unpaired.
Effectively, the Fermi gas consists of two constituents 
(fermions of mass $m$ and molecules of mass $2m$) and the system
properties are governed by the molecule-molecule
(or dimer-dimer) scattering length $a_{\rm{dd}}$ 
and the atom-molecule (or atom-dimer)
scattering length $a_{\rm{ad}}$.
Following this logic, the unperturbed wave function
is written as a product of a permanent, constructed from
$N_2$ harmonic oscillator orbitals with width $\sqrt{\hbar /(2m \omega)}$
that depend
on the center of mass vectors of the molecules, and a Slater determinant,
constructed from $N_1-N_2$ single particle harmonic oscillator
orbitals with width $\sqrt{\hbar /(m \omega)}$.
The interaction between an atom and a molecule (between two molecules)
is parameterized by Eq.~(\ref{eq_fermipp}) with $a_0(0)$ replaced
by $a_{\rm{ad}}$ ($a_{\rm{dd}}$) and $\mu$ replaced by
$\mu_{\rm{ad}}=2m/3$ ($\mu_{\rm{dd}}=m$).
For the lowest energy family, the perturbative
treatment gives~\cite{stec08,stec07b}
\begin{eqnarray}
\label{eq_pertbec}
E_{\rm{tot}}(N_1,N_2) \approx N_2 E_{\rm{dimer}} +  
E_{\rm{tot}}^{\rm{NI}}(N_1-N_2,0) + \nonumber \\
\frac{N_2 (N_2-1)}{2} \sqrt{\frac{2}{\pi}} 
\frac{a_{\rm{dd}}}{a_{{\rm{ho}},\mu_{\rm{dd}}}} \hbar \omega
+ 
N_2 (N_1-N_2) c_{\rm{ad}}
\frac{a_{\rm{ad}}}{a_{{\rm{ho}},\mu_{\rm{ad}}}} \hbar \omega ,
\end{eqnarray}
where $a_{{\rm{ho}},\mu_i}=\sqrt{\hbar/(\mu_i \omega)}$ with
$i= {\rm{ad}}$ or ${\rm{dd}}$.
In Eq.~(\ref{eq_pertbec}), $E_{\rm{dimer}}$ denotes the ground 
state energy of the 
trapped
dimer interacting through $V_0^{\rm{pp}}$ with scattering length $a_0(0)$
[i.e., the lowest $l=0$ eigenenergy of Eq.~(\ref{eq_trans}),
with the center of mass energy of $3 \hbar \omega/2$ added].
The coefficient $c_{\rm{ad}}$ depends on 
how many unpaired fermions there are
and which
single-particle states the unpaired fermions occupy.
For the ground state of systems with $N_1-N_2=0$ and 1,
we have $c_{\rm{ad}}=0$ and $c_{\rm{ad}}=\sqrt{2/\pi}$, respectively.

For the 
$(2,2)$ system, for example,
Eq.~(\ref{eq_pertbec}) describes the
energies of the 
molecule+molecule family
[$a_0(0)$ small and positive] while 
Eq.~(\ref{eq_pertbcs}) 
describes those of 
the four-atom family
[$|a_0(0)|$ small and either positive or negative].
The $(2,2)$ system additionally contains a molecule+atom+atom
family, whose limiting $a_0(0) \rightarrow 0^+$
behavior can be obtained by
assuming that only one dimer forms.

Table~\ref{tab_scatteringlength} summarizes
\begin{table}
\caption{\label{tab_scatteringlength}
Summary of selected studies that
determined the three-dimensional 
atom-dimer and dimer-dimer scattering lengths
$a_{\rm{ad}}$ and $a_{\rm{dd}}$, respectively.
}
\begin{indented}
\item[]\begin{tabular}{l|l}
\br
$a_{\rm{ad}}/a_0(0)$ & method \\ \hline
$\kappa=1$: $a_{\rm{ad}}/a_0(0)=1.2$ & free-space scattering~\protect\cite{skor57} \\
any $\kappa$ & free-space scattering~\protect\cite{petr03} \\ 
$\kappa=1$: $a_{\rm{ad}}/a_0(0)=1.1790662349$ & free-space scattering~\protect\cite{tan08a} \\
$\kappa=1$: $a_{\rm{ad}}/a_0(0)=1.18(1)$ & trap spectrum~\protect\cite{stec08} \\ \hline \hline
$a_{\rm{dd}}/a_0(0)$ & method \\ \hline
$\kappa=1$: $a_{\rm{dd}}/a_0(0)=0.6^{(a)}$ & free space scattering (four-body)~\cite{petr04aa} \\ 
$\kappa=1$: $a_{\rm{dd}}/a_0(0)=0.60^{(b)}$ & free space scattering~\cite{bulg03} \\
$\kappa=1$: $a_{\rm{dd}}/a_0(0)=0.60^{(b)}$ & diagramatically~\cite{brod05,levi06} \\
$1\le\kappa\le 13.607$ & free-space scattering~\cite{petr05} \\
$\kappa=1$: $a_{\rm{dd}}/a_0(0)=0.608(2)$ & trap spectrum~\cite{stec08,stec07b} \\
$1 \le \kappa \le 20$ & trap spectrum~\cite{stec08,stec07b} \\
$\kappa=1$: $a_{\rm{dd}}/a_0(0)=0.62(1)$ & homogeneous system~\protect\cite{astr04c} \\
$\kappa=1$: $a_{\rm{dd}}/a_0(0)=0.605(5)$ & free-space scattering~\protect\cite{dinc09a} \\
\br
\end{tabular}
$^{(a)}$The accuracy is
reported to be 2~\%.
$^{(b)}$No errorbar is reported.
\end{indented}
\end{table}
the values obtained for $a_{\rm{ad}}$ and $a_{\rm{dd}}$
in the literature by a variety of means.
Here, we discuss the determination of the dimer-dimer scattering length
in more detail.
In the first Born approximation,
the dimer-dimer
scattering length $a_{\rm{dd}}$ equals $2 a_0(0)$.
Microscopic free-space scattering calculations
confirm that $a_{\rm{dd}}$ is 
directly proportional to the
atom-atom scattering length 
and that no other parameter enters
but find that the proportionality factor
is $\approx 0.6$ as opposed to $2$~\cite{petr04aa}.
The microscopically derived
proportionality factor has, for example, been confirmed 
using a diagrammatic approach~\cite{brod05,levi06} and 
by
analyzing the equation of state of the homogeneous two-component
Fermi gas, calculated using Monte Carlo techniques~\cite{astr04c},
within a perturbative framework analogous to that given 
in Eq.~(\ref{eq_pertbec}).
Here, we show that the  dimer-dimer scattering length
can alternatively be extracted from the energies of states
corresponding to the dimer-dimer family of the trapped 
$(2,2)$ system~\cite{stec08,stec07b}.

We start our discussion with the normalized
energy crossover curve $\Lambda_{N_1, N_2}$~\cite{stec08,stec07b,jaur07},
\begin{eqnarray}
\label{eq_crossover}
\Lambda_{N_1, N_2} = \frac{E_{\rm{tot}}(N_1,N_2)-N_2 E_{\rm{dimer}} - 
\frac{3 (N_1-N_2)\hbar \omega}{2}}{E_{\rm{tot}}^{\rm{NI}}(N_1,N_2)-
\frac{3}{2}N \hbar \omega},
\end{eqnarray}
which depends on $N_1$, $N_2$ and $a_0(0)$.
For the BCS-BEC branch, 
$\Lambda_{N_1, N_2}$
changes from $1$ to $0$ as $1/a_0(0)$ changes from $-\infty$ to $\infty$.
In addition to providing a convenient means to plot
the energies throughout the crossover,
$\Lambda_{N_1,N_2}$ 
significantly reduces
the
dependence on the underlying two-body 
potential~\cite{stec08,stec07b,jaur07}.

\begin{figure}
\vspace*{+1.5cm}
\includegraphics[angle=0,width=65mm]{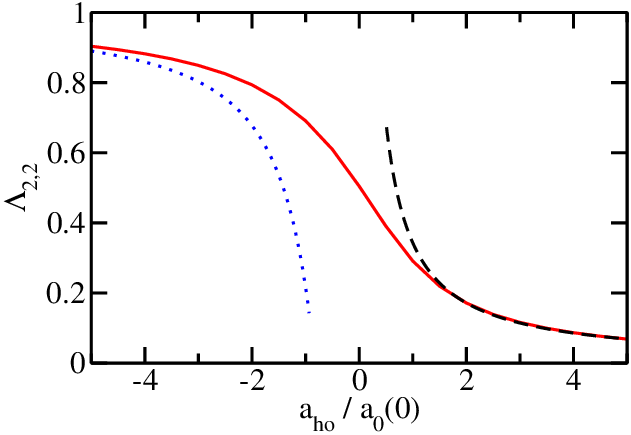}
\vspace*{-0.2cm}
\caption{(Color online)
Normalized energy crossover curve $\Lambda_{2,2}$ 
for $\kappa=1$ 
as a function of $a_{\rm{ho}}/a_0(0)$.
The solid line shows $\Lambda_{2,2}$ derived from the 
$L=0$ extrapolated zero-range ground state energies 
reported in the supplemental material
of Ref.~\cite{dail10}.
The dotted and dashed lines are obtained using the perturbative expressions
[Eqs.~(\ref{eq_pertbcs}) and (\ref{eq_pertbec})]
in Eq.~(\ref{eq_crossover}).
}
\label{fig_crossovern4}
\end{figure}

The solid line in Fig.~\ref{fig_crossovern4} shows $\Lambda_{2,2}$
as a function of $1/a_s(0)$.
For $a_s(0) \rightarrow 0^-$, $E_{\rm{tot}}(N_1,N_2)$ 
equals $E_{\rm{tot}}^{\rm{NI}}(N_1,N_2)$
and $E_{\rm{dimer}}$ equals $3 \hbar \omega$ (we have two fermions 
with energy $3\hbar \omega/2$ each), yielding
$\Lambda_{2,2}=1$. 
For $a_s(0) \rightarrow 0^+$, the total energy can be thought of
as being due to $N_2$ dimers (with relative energy
$3 \hbar \omega/2$ and center of mass energy
$3 \hbar \omega/2$), yielding $\Lambda_{2,2}=0$.
Inserting Eq.~(\ref{eq_pertbec}) into Eq.~(\ref{eq_crossover}),
we find that
$\Lambda_{2,2} \approx (2 \pi)^{-1/2} a_{\rm{dd}}/a_{{\rm{ho}},\mu_{\rm{dd}}}$
in the $a_0(0) \rightarrow 0^+$ regime.
Thus, $a_{\rm{dd}}$ can be extracted from the full
four-particle energy spectrum by fitting the crossover curve
to the limiting expression. 
One arrives at the same result by expanding the eigenenergy
of the lowest energy branch of Eq.~(\ref{eq_trans}) with $l=0$ and
$\mu$ and $a_{{\rm{ho}},\mu}$ replaced by
$\mu_{\rm{dd}}$ and $a_{{\rm{ho}},\mu_{\rm{dd}}}$, respectively,
around $a_0(0)=0^+$ (in this approach, 
the internal dimer energy and the center-of-mass
energy need to be added).
The analysis can be refined 
by using the full two-body energy spectrum
as opposed to the Taylor-expanded 
expression and by allowing $a_{\rm{dd}}$ to be energy-dependent
via Eq.~(\ref{eq_effrange}). 
The refined analysis gives $a_{\rm{dd}}=0.608(2) a_0(0)$~\cite{stec08,stec07b}
in good agreement with the microscopic
free-space results (see Table~\ref{tab_scatteringlength})
and also provides an estimate of the dimer-dimer effective
range $r_{{\rm{eff}},{\rm{dd}}}$, $r_{{\rm{eff}},{\rm{dd}}}=0.13(2)a_0(0)$~\cite{stec08,stec07b}.
The positive value of $r_{{\rm{eff}},{\rm{dd}}}$ is consistent
with the conclusion by Petrov {\em{et al.}}~\cite{petr04aa} 
that the two dimers
interact through an effective potential with soft core repulsion.
Dotted and dashed lines show $\Lambda_{2,2}$ 
obtained using the perturbative energy expressions,
i.e., Eqs.~(\ref{eq_pertbcs}) and (\ref{eq_pertbec}). 
It can be seen that
the perturbative treatment
provides a fairly accurate 
description for $a_{\rm{ho}}/a_0(0) \lesssim -4$
and $a_{\rm{ho}}/a_0(0) \gtrsim 1.5$, respectively.

The energy spectra and crossover curve shown in 
Figs.~\ref{fig_threefermion} and \ref{fig_crossovern4}
are for the $(2,1)$ and $(2,2)$ systems.
Crossover studies for systems with $N>4$ can be found in 
Refs.~\cite{stec07b,jaur07,blum11}.
The energies of
the $(2,2)$ system were obtained using the stochastic variational 
approach~\cite{cgbook},
a basis set expansion approach
that results in stable and highly
accurate results for trapped systems with
up to $N=6$ fermions throughout the entire crossover 
regime~\cite{dail10,blum07,stec08,stec07c,stec07b,blum11,blum09a}.
This numerical approach cannot, 
at least at present, be applied
to significantly 
larger systems since the computational effort increases
notably with increasing $N$.
Alternative numerical microscopic
approaches include an effective field theory
approach~\cite{stet07,rotu10,toll11},
an
effective interaction approach~\cite{alha08}, 
and Monte Carlo approaches.
The applicability of the effective field theory 
approach is presently also limited to 
relatively small 
particle numbers ($N \le 4$) 
while Monte Carlo methods have been applied to larger systems.
The Monte Carlo techniques 
can, for the purpose of this review,
be divided into two classes. The first class,
which includes the diffusion Monte Carlo (DMC, sometimes
also referred to as Green's function Monte Carlo) and variational
Monte Carlo (VMC) approaches, 
is applicable to essentially any interaction 
strength~\cite{blum07,stec08,stec07b,jaur07,chan07}.
These techniques employ the fixed node approximation
and
result in 
variational upper bounds to the exact 
eigenenergies~\cite{reyn82}.
Note
that
the accuracy may diminish for certain regions of 
interaction strengths.
The second class of Monte Carlo
approaches such as the lattice MC
approach from Refs.~\cite{endr10,nich10}
have thus far only been applied to the
unitary regime.

\subsection{Large scattering length regime: Unitarity}
\label{sec_fermigasunit}
This subsection considers
two-component equal-mass Fermi gases
with infinitely large $s$-wave scattering length under external
spherically-symmetric harmonic confinement. As discussed in
Subsec.~\ref{sec_threebodylimits},
an infinitely large $s$-wave scattering length does 
not define a meaningful length scale
for the system. Thus, two-component Fermi
gases with infinite $a_0(0)$ and zero-range
interactions are characterized by the 
same number of length
scales as the non-interacting system.
This implies, as in the three-body
case, that
the hyperradial and hyperangular 
degrees of freedom separate in the limit
of zero-range interactions with infinitely large
$a_0(0)$~\cite{wern06,cast04}.
Correspondingly, 
the hyperradial Schr\"odinger equation is 
given by Eq.~(\ref{eq_hyperradiallimits})
with $R$ and $\mu_R$ now denoting the hyperradius
and the hyperradial mass
of the $N$-body system
(i.e., $\mu_R R^2 = \sum_{i=1}^{N-1}\mu_i \vec{\rho}_i^2$,
where the $\vec{\rho}_i$ denote the Jacobi vectors---excluding the center of
mass vector---and the $\mu_i$ denote the associated Jacobi masses).
Thus, the hyperradial solutions
discussed in Subsec.~\ref{sec_threebodylimits} 
apply to two-component Fermi gases with arbitary $(N_1,N_2)$.
For equal mass systems
with zero-range interactions, the $s_{\nu}$ are all greater than $1$.
This implies that the energies $E_{\rm{tot}}^{\rm{unit}}$
at unitarity can be written as~\cite{wern06,cast04}
\begin{eqnarray}
\label{eq_energyunit}
E_{\rm{tot}}^{\rm{unit}}= (2q + s_{\nu} + 1) \hbar \omega + E_{\rm{cm}},
\end{eqnarray}
where the hyperradial quantum number
$q$ takes the values $0,1,\cdots$ and where the $s_{\nu}$,
which depend on $N_1$ and $N_2$,
are determined from the solution of the $3N-4$ dimensional hyperangular
differential equation.
The separability of the hyperradial and hyperangular degrees
of freedom implies that the dynamics of the $N$-body system is governed
by a set of uncoupled effective one-dimensional
hyperradial potential curves.
Thus, the strongly-interacting $N$-body problem reduces
to an effective one-dimensional problem.

Equation~(\ref{eq_energyunit})
predicts the existence of energy ladders
with spacing $2 \hbar \omega$~\cite{wern06,cast04}. Correspondingly, the 
energy spectrum contains breathing mode frequencies at integer multiples
of $2 \hbar \omega$, in addition to frequencies that are associated with
transitions between different hyperradial potential curves.
The latter class of frequencies
depends on 
$s_{\nu}-s_{\nu'}$, since frequencies in this class arise due to transitions
between two different  
hyperradial potential curves.
These predictions are a direct consequence of the
scale-invariance~\cite{tan04,wern06,cast04} and can be verified
experimentally.

The characterization of strongly-interacting
Fermi gases is non-trivial. Various approaches, including those discussed
at the end of the previous subsection, have been applied. 
Unlike in the three-body case, for which analytical
solutions to the hyperangular
Schr\"odinger equation are known
(see Subsec.~\ref{sec_hyperspherical}),
no analytical solutions are known for larger systems and
one generally has to resort to solving the hyperangular
equation or the full Schr\"odinger equation numerically.
For systems with $L=0$ and up to $N=4$
particles, the hyperangular Schr\"odinger equation
has been solved by adopting a hyperspherical correlated Gaussian
approach~\cite{stec09}, a variant of the stochastic variational approach
discussed earlier.
For larger systems, in contrast, obtaining the solution to
the hyperangular Schr\"odinger equation
(which would result in the $s_{\nu}$) is more
challenging than obtaining that to the full Schr\"odinger equation
(which results in $E_{\rm{tot}}^{\rm{unit}}$).
Using Eq.~(\ref{eq_energyunit}), which has been proven
to hold if the range $r_0$ of the underlying two-body potential
is sufficiently small~\cite{blum07,blum09a,wern08}, 
$E_{\rm{tot}}^{\rm{unit}}$ and $s_{\nu}$ can be 
converted into each other.

\begin{figure}
\vspace*{+1.5cm}
\includegraphics[angle=0,width=65mm]{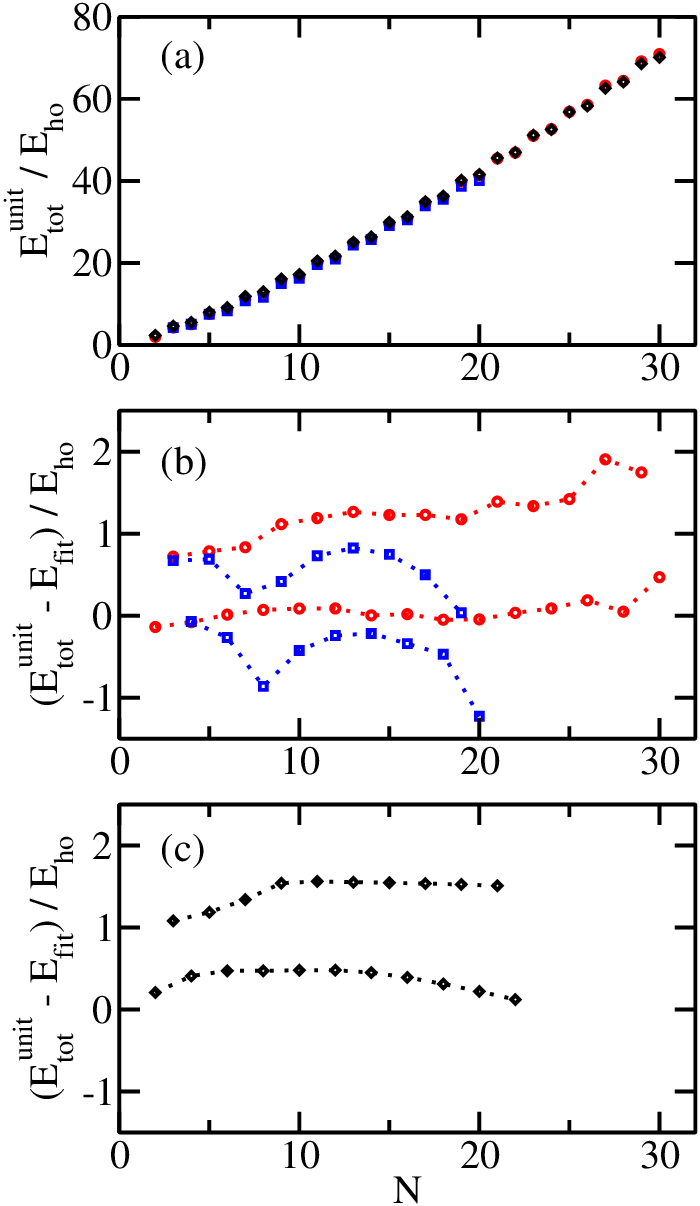}
\vspace*{-0.2cm}
\caption{(Color online)
Energetics of trapped two-component Fermi gas with $\kappa=1$ 
at unitarity as a function of $N$ for $N_1-N_2=0$ or 1.
Panel~(a) 
shows the
total energy $E_{\rm{tot}}^{\rm{unit}}$.
Panels~(b) and (c) show the energy difference
$E_{\rm{tot}}^{\rm{unit}}-E_{\rm{fit}}$ 
using $E_{\rm{fit}}=\sqrt{0.467} E_{\rm{tot}}^{{\rm{NI,ETF}}}$
(as a guide to the eye, dotted lines connect the data points for
even and odd $N$).
Circles, squares, and diamonds
show results based on the fixed node DMC energies from Ref.~\cite{blum07},
the lattice MC energies from Ref.~\cite{nich10},
and the DFT energies from the supplemental material
of Ref.~\cite{bulg07}.
}
\label{fig_energyunit}
\end{figure}

Figure~\ref{fig_energyunit}(a)
shows the total energy $E_{\rm{tot}}^{\rm{unit}}$ as a function of $N$ for 
two-component Fermi gases with $N_1-N_2=0$ or 1
at unitarity, calculated using three different 
approaches (i.e., a fixed node DMC approach~\cite{blum07}, 
a lattice MC approach~\cite{nich10} and a 
density functional theory (DFT)
approach~\cite{bulg07}).
These sets of energies have been chosen to illustrate
our current understanding (see Refs.~\cite{jaur07,chan07,zuba09} 
for other calculations).
The MC and DFT approaches are 
based on entirely different formulations
and the relatively good agreement displayed in 
Fig.~\ref{fig_energyunit}(a) is encouraging.
To interpret Fig.~\ref{fig_energyunit}(a), 
we relate the energies $E_{\rm{tot}}^{\rm{unit}}$
at unitarity
to the energies $E_{\rm{tot}}^{\rm{NI}}$ of the trapped non-interacting system.
Since the system at unitarity is characterized by the same number of
length scales as the non-interacting system, the energies
have to be directly proportional to each other~\cite{gior08}, 
\begin{eqnarray}
\label{eq_universalrelation}
E_{\rm{tot}}^{\rm{unit}}=\sqrt{\xi_{\rm{tr}}} E_{\rm{tot}}^{\rm{NI}};
\end{eqnarray}
the reason for writing the proportionality factor
as a 
square root will become clear below.
The total energy $E_{\rm{tot}}^{\rm{NI}}$ of the trapped non-interacting system
shows 
shell closures for 
$N_i=1,4,10,20,\cdots$, i.e., the energy
increases by $3 \hbar \omega/2$ when the first spin-up 
fermion and spin-down
fermion are added, by $5 \hbar \omega/2$ when the second through fourth
spin-up and spin-down fermions are added, and so on.
The energies at unitarity 
show---on the scale
chosen in Fig.~\ref{fig_energyunit}(a)---no 
signature of shell closures but instead odd-even oscillations.
Thus, it appears reasonable to determine the proportionality factor
$\sqrt{\xi_{\rm{tr}}}$
by ``smoothing'' the non-interacting energies, i.e.,
by approximating 
$E_{\rm{tot}}^{\rm{NI}}$ by
the semi-classical extended Thomas Fermi energies 
$E_{\rm{tot}}^{{\rm{NI,ETF}}}$~\cite{brac97},
\begin{eqnarray}
\label{eq_energynieft}
E_{\rm{tot}}^{{\rm{NI,ETF}}} = \frac{1}{4} (3 N)^{4/3} \left[
1 + c_{{\rm{ETF}}} \frac{1}{2} (3 N)^{-2/3} \right] \hbar \omega,
\end{eqnarray}
where $c_{\rm{ETF}}=1$ (in Subsec.~\ref{sec_fermigasunequalmasses},
we treat $c_{\rm{ETF}}$ as a fitting parameter).
We refer to 
the energy 
obtained by fitting the even $N$ energies 
to Eq.~(\ref{eq_universalrelation}) 
with $E_{\rm{tot}}^{\rm{NI}}$ replaced by
$E_{\rm{tot}}^{\rm{NI,ETF}}$ as $E_{\rm{fit}}$.
For the energies for $N=2-30$ of Ref.~\cite{blum07}, for example, 
$\xi_{\rm{tr}}$ is found to equal $0.465$.
In the large $N$ limit, the term proportional to $c_{\rm{ETF}}$
can be neglected and Eq.~(\ref{eq_energynieft}) reduces
to the local
density approximation (LDA) expression (see below)~\cite{gior08}.

Symbols in 
Figs.~\ref{fig_energyunit}(b) and (c)
show the quantity
$E_{\rm{tot}}^{\rm{unit}}-E_{\rm{fit}}$.
The energy difference
$E_{\rm{tot}}^{\rm{unit}}-E_{\rm{fit}}$ reveals clear odd-even oscillations
for all three data sets.
For the fixed node DMC data [circles in Fig.~\ref{fig_energyunit}(b)],
the energy difference
$E_{\rm{tot}}^{\rm{unit}}-E_{\rm{fit}}$ is approximately equal to zero
for even $N$ 
and 
of the order of 
$\hbar \omega$ (with a slight overall increase with increasing $N$)
for odd $N$.
For the lattice MC data set [squares in Fig.~\ref{fig_energyunit}(b)], 
in contrast, $E_{\rm{tot}}^{\rm{unit}}-E_{\rm{fit}}$
shows 
oscillations that follow the
shell structure of the non-interacting system for both even and odd $N$
(recall, the non-interacting system with $N_1=N_2$
possesses shell closures at $N=2,8,20,\cdots$),
with a roughly
constant offset between the even and odd $N$ data.
Thirdly, the DFT data [see Fig.~\ref{fig_energyunit}(c)]
show structure that appears to be
independent of
the shell structure of the non-interacting system.
It is currently not fully clear which of these behaviors can be attributed
to numerical artifacts and which reflect genuine physics.
We speculate that the data from Ref.~\cite{nich10}
overestimate the shell structure but note that
further studies are needed to fully understand the energetics
of trapped two-component systems at unitarity.
We note that modified versions of Eqs.~(\ref{eq_universalrelation})
and (\ref{eq_energynieft}),
which are based on a 
different interpretation of the underlying physics,
lead to somewhat different $\xi_{\rm{tr}}$ values~\cite{rupa09}.

To further interpret the results of the trapped system
at unitarity, 
we relate the energies of the
trapped and homogeneous systems via the 
LDA. The energy $E_{\rm{hom}}^{\rm{unit}}$ per particle
of the homogeneous system
at unitarity
is directly
proportional to the 
energy $E_{\rm{FG}}$ of the non-interacting
homogeneous gas,
$E_{\rm{hom}}^{\rm{unit}} = \xi_{\rm{hom}} E_{\rm{FG}}$.
The proportionality constant $\xi_{\rm{hom}}$, also
referred to as the Bertsch parameter~\cite{bertsch}, 
has been determined
theoretically and experimentally.
The most reliable estimate comes from very recent
fixed node DMC
calculations, which determine an upper bound for $\xi_{\rm{hom}}$,
$\xi_{\rm{hom}}=0.383(1)$~\cite{forb11}. 
The value of $\xi_{\rm{hom}}$ is of fundamental importance
and enters into 
effective theories that use
the equation of state of the homogeneous
system as input.
Application of the LDA (see, e.g., Ref.~\cite{meno02a}),
which treats the trapped Fermi gas as being locally (i.e., at each
distance from the trap center) characterized by a constant density,
predicts that
the energy of the trapped system at unitary is given by
$E_{\rm{tot}}^{\rm{unit}} = \sqrt{\xi_{\rm{hom}}} E_{\rm{tot}}^{\rm{NI}}$~\cite{gior08}.
Thus, $\xi_{\rm{tr}}$ (see above) provides an alternative estimate
of $\xi_{\rm{hom}}$.
The fact that the $\xi_{\rm{tr}}$ obtained by analyzing the energies of
the trapped system ($\xi_{\rm{tr}}=0.465$) is larger than the
value of $\xi_{\rm{hom}}=0.383(1)$~\cite{forb11} 
obtained from recent MC simulations for
the homogeneous system may be attributed to the fact that 
the calculations for the trapped system are
restricted
to
relatively small $N$ and/or that the energies of the trapped 
system may be slightly too high.
The latter may be a consequence of the fixed node approximation
employed in Ref.~\cite{blum07}.

Another quantity of interest is the excitation gap $\Delta E(N)$,
which
characterizes the odd-even oscillations
of two-component Fermi gases. 
\begin{figure}
\vspace*{+1.5cm}
\includegraphics[angle=0,width=65mm]{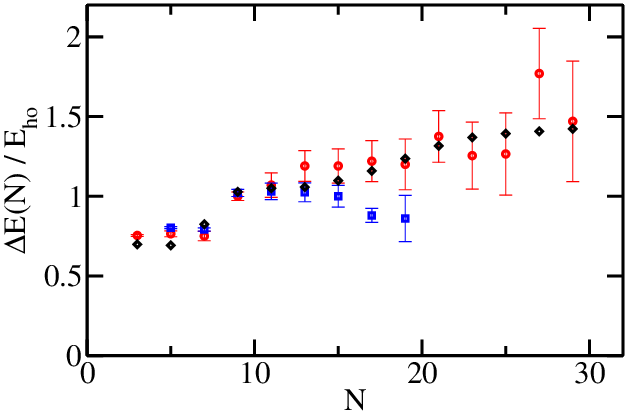}
\vspace*{-0.2cm}
\caption{(Color online)
Excitation gap $\Delta E(N)$
of trapped two-component Fermi gas with $\kappa=1$ 
at unitarity as a function of $N$ for $N_1-N_2=0$ or 1.
Circles, squares, and diamonds
show results based on the fixed node DMC energies from Ref.~\cite{blum07},
the lattice MC energies from Ref.~\cite{nich10},
and the DFT energies from the supplemental material
of Ref.~\cite{bulg07}.
For clarity, errorbars are only shown 
for the
fixed node DMC and lattice MC data sets.
}
\label{fig_gap}
\end{figure}
The excitation gap 
combines the energies of the imbalanced system
with $N_1-N_2=1$ (i.e., $N=2N_1-1$)
with the energies of the next smaller 
and next larger balanced systems~\cite{gior08},
\begin{eqnarray}
\label{eq_excitationgap}
\Delta E(N) = E_{\rm{tot}}(N_1,N_1-1) - 
\nonumber \\
\frac{E_{\rm{tot}}(N_1-1,N_1-1)+E_{\rm{tot}}(N_1,N_1)}{2}.
\end{eqnarray}
Figure~\ref{fig_gap} 
shows $\Delta E(N)$ as a function
of $N$ for two-component Fermi gases at unitarity
for the
same data sets as shown in Figs.~\ref{fig_energyunit}(a), 
\ref{fig_energyunit}(b) and \ref{fig_energyunit}(c).
While the energy difference $E_{\rm{tot}}^{\rm{unit}}-E_{\rm{fit}}$
shows a distinct dependence on the numerical approach
employed, the excitation gap determined by the three different
methods agrees fairly well.
The excitation gap $\Delta E(N)$ 
increases 
from about $0.7 \hbar \omega$ 
to about $1.5 \hbar \omega$ as $N$ increases from
3 to 29.
In the homogeneous system, the excitation gap equals half the 
energy it takes to brake a pair~\cite{gior08}.
In the trapped system, the interpretation is not
quite as straightforward because of the inhomogeneity 
induced by the confining potential.
In particular, the fixed node DMC calculations suggest that the extra
particle is located near the edge of the cloud as opposed to the center of
the cloud~\cite{blum07,stec08}. This finding supports the conclusion put
forward by Son~\cite{son07} that $\Delta E(N)$ scales as $N^{1/9}$ and not
as $N^{1/3}$ as would be expected based on the LDA.
The microscopic data available to date do not,
however, extend to
sufficiently large $N$ to unambiguously distinguish between the
$N^{1/9}$ and $N^{1/3}$ behaviors.
We note that the definition of the excitation
gap $\Delta E(N)$, Eq.~(\ref{eq_excitationgap}), 
applies to all interaction strengths, and not just the unitary regime.
The excitation gap's behavior in the weakly-attractive
regime has been discussed briefly in the context
of Fig.~\ref{fig_pertbcs}.

Lastly, we use the energies $E_{\rm{tot}}^{\rm{unit}}$ to arrive
at an interpretation of the unitary Fermi gas within
the hyperspherical framework.
The odd-even 
oscillations of the energies imply odd-even oscillations of the 
$s_{\nu}$ coefficients that
describe the ground state
and thus an odd-even
staggering of the hyperspherical potential curves $V_{\rm{eff}}(R)$
that govern the dynamics at unitarity~\cite{wern06,stec08}.
Application of the LDA predicts that
the $s_{\nu}$ coefficient
of the ground state
for even $N$ approaches $\sqrt{\xi_{\rm{hom}}} E_{\rm{tot}}^{\rm{NI}}/(\hbar \omega)$
in the large $N$ limit~\cite{blum07}.

The microscopic energies  as well as structural properties
have been used to
assess the accuracy of a number of effective approaches,
such as those based on fractional exclusion statistics~\cite{bhad08},
DFT~\cite{bulg07,zuba09,sala08,adhi08,zuba09a,bulg10}
and effective field theory.
Lastly, we note that 
effort has 
also
been devoted
to understanding population-imbalanced
Fermi gases with $N_1 \gg N_2$ at unitarity
within a microscopic framework~\cite{blum08f}.
In the limit of ``extreme'' population imbalance,
the trapped analog of the fermionic polaron problem
is realized~\cite{schi09,nasc10}.

\subsection{Mass-imbalanced Fermi gas}
\label{sec_fermigasunequalmasses}
This subsection considers $s$-wave interacting two-component
Fermi gases
with unequal masses. 
While equal-mass Fermi gases are, by now, relatively well understood,
our understanding of unequal-mass systems is much less complete.
One of the reasons is that experiments on unequal-mass systems are
just now becoming available~\cite{tagl06,tagl08,spie09,spie10,tren11}. 
Dual species experiments can be hampered by large three-body
loss rates, which reduce the system's lifetime
and thus make it challenging to cool dual-species
systems to degeneracy.
Trapped 
dual-species systems
with mass ratio $\kappa$ ($\kappa = m_1/m_2$, where
$m_i$ denotes the atom mass of the $i$$^{th}$ component) are,
in general,
characterized by two different angular trapping frequencies,
i.e., $\omega_1 \ne \omega_2$.
If $\omega_1$ and $\omega_2$ differ, the center of mass motion no
longer separates.
The 
behavior of the $(2,1)$ system changes with increasing $\kappa$
(see Sec.~\ref{sec_threebody}):
Three-body resonances may occur for 
$8.619< \kappa < 13.607$~\cite{wern06,nish08}
and Efimov physics plays a role 
for $\kappa >13.607$~\cite{braa06,braa07}.
The treatment of larger systems 
with $\kappa \gtrsim 8.619$ thus requires theoretical
tools that are capable of accounting for these three-body
phenomena and, at the same time, of capturing the
new physics introduced by the fourth, fifth, etc. particle.

\begin{figure}
\vspace*{+1.5cm}
\includegraphics[angle=0,width=65mm]{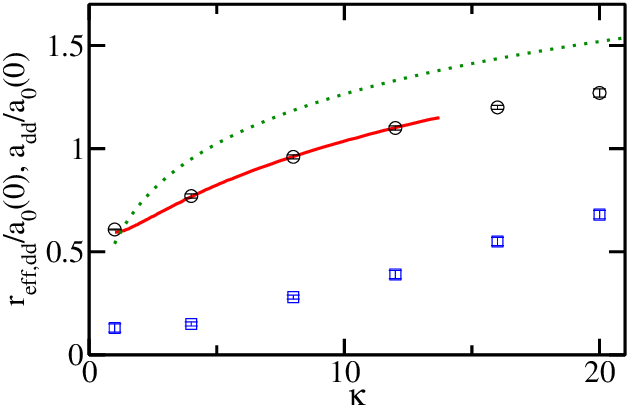}
\vspace*{-0.2cm}
\caption{(Color online)
Dimer-dimer scattering length $a_{\rm{dd}}$
and effective range $r_{{\rm{eff}},{\rm{dd}}}$
as a function of the 
mass ratio $\kappa$.
Solid lines show $a_{\rm{dd}}$
obtained by solving the free-space
Schr\"odinger equation with zero-range interactions (the results
are taken from Fig.~3 of Ref.~\protect\cite{petr05}) while dotted lines 
show $a_{\rm{dd}}$ obtained by applying the BO approximation
[see Eq.~(14) of Ref.~\protect\cite{marc08}].
Circles and squares show $a_{\rm{dd}}$ and $r_{{\rm{eff}},{\rm{dd}}}$
extracted from
the energy spectrum of the trapped $(2,2)$ system
(the values of $a_{\rm{dd}}$ and $r_{{\rm{eff}},{\rm{dd}}}$ are taken from
Table~II of Ref.~\protect\cite{stec08}).}
\label{fig_scattmass}
\end{figure}

We first consider unequal-mass Fermi gases with small, positive
$a_0(0)$.
Assuming that $a_0(0)$ 
fully characterizes the system dynamics,
the perturbative expression Eq.~(\ref{eq_pertbec})
may be generalized to unequal-mass systems~\cite{blum08e,stec07b}.
Thus, a natural question is how the atom-dimer and dimer-dimer
scattering lengths $a_{\rm{ad}}$ and $a_{\rm{dd}}$ vary with $\kappa$.
For the $(2,1)$ system, the atom-dimer $s$-wave scattering
length is associated with the $L^{\Pi}=0^+$ wave function;
for this channel, Efimov physics is absent.
The atom-dimer scattering length $a_{\rm{ad}}$ equals $1.18a_0(0)$ 
for $\kappa=1$ 
and increases with increasing $\kappa$~\cite{petr03}.
For the $(2,2)$ system, the 
$s$-wave dimer-dimer scattering length
was first determined via
a free-space calculation
that
employs a zero-range interaction
model and assumes the absence of three-body
bound states for $\kappa < 13.607$~\cite{petr05}.
$a_{\rm{dd}}$ changes monotonically from
$0.6a_0(0)$ to $1.14 a_0(0)$
as $\kappa$ increases from 1 to $13.607$~\cite{petr05}.
The solid line
in Fig.~\ref{fig_scattmass} shows the results for $a_{\rm{dd}}$.
The free-space calculations terminate at $\kappa = 13.607$, where
it is known that Efimov physics enters.
For $\kappa > 13.607$, the 
effective dimer-dimer interactions are
expected to depend on a three-body or Efimov parameter.

The dimer-dimer scattering length
$a_{\rm{dd}}$ for unequal-mass systems has also been 
extracted from the energy spectrum of the trapped $(2,2)$
system with equal frequencies and
finite-range interactions (see circles in 
Fig.~\ref{fig_scattmass})~\cite{stec08,stec07b}.
Figure~\ref{fig_scattmass} shows that the two
sets of calculations (solid line and circles)
are in good
agreement for $\kappa \lesssim 13.607$.
The latter calculation does not
terminate at $\kappa=13.607$ but goes beyond.
In particular, for $\kappa \gtrsim 13.607$, the scattering length $a_{\rm{dd}}$ 
is obtained by following the lowest state of the
dimer-dimer family,
which lies above a subset of states
whose energies show a pronounced dependence 
on
the three-body parameter. The dimer-dimer scattering lengths
for $\kappa \gtrsim 13.607$ have been confirmed independently
by an approximate
BO treatment (see 
dotted line in Fig.~\ref{fig_scattmass})~\cite{marc08}.
While the scattering lengths derived within the BO 
framework are larger than those obtained by the more exact treatment,
the overall behavior is similar. A more sophisticated BO treatment, 
referred to as
hybrid BO approach, leads to scattering lengths nearly indistinguishable 
from those shown by the solid line and the symbols in 
Fig.~\ref{fig_scattmass}~\cite{marc08}.
The energy spectrum
of the trapped four-fermion system additionally allows for the determination
of the dimer-dimer effective range $r_{{\rm{eff}},{\rm{dd}}}$ 
(squares in Fig.~\ref{fig_scattmass})~\cite{stec08,stec07b}.
Figure~\ref{fig_scattmass} shows that $r_{{\rm{eff}},{\rm{dd}}}$
increases more rapidly with $\kappa$ than 
$a_{\rm{dd}}$, indicating that finite-range effects 
for the dimer-dimer system become more
important with increasing $\kappa$.
This implies, e.g., that the validity regime of the 
dimer-dimer zero-range model 
decreases with increasing $\kappa$.
The positive dimer-dimer scattering length suggests that
two-component Fermi gases with large mass ratio $\kappa$ and 
small atom-atom $s$-wave scattering length 
form 
repulsively
interacting metastable molecular Bose gases. Losses
due to the formation of Efimov trimers in dimer-dimer collisions 
are predicted to decrease
exponentially with increasing mass ratio~\cite{marc08}.

\begin{figure}
\vspace*{+1.5cm}
\includegraphics[angle=0,width=65mm]{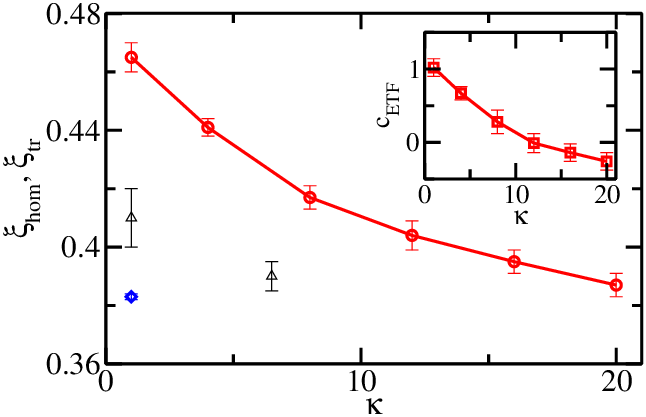}
\vspace*{-0.2cm}
\caption{(Color online)
Circles (main figure) and squares (inset) show
$\xi_{\rm{tr}}$ and $c_{\rm{ETF}}$, respectively, 
as a function of the mass ratio $\kappa$
(the values are taken from Fig.~6 of Ref.~\protect\cite{stec07b}).
For comparison, triangles show $\xi_{\rm{hom}}$
obtained by the fixed node DMC method~\cite{geze09};
the diamond shows a more recent value for $\xi_{\rm{hom}}$ 
for $\kappa=1$~\protect\cite{forb11}.}
\label{fig_unequalmassunit}
\end{figure}

Next, we consider the unitary regime
and assume that the angular trapping frequencies
$\omega_i$
are adjusted such that the harmonic oscillator lengths 
$a_{{\rm{ho}},i}$, where $a_{{\rm{ho}},i}=\sqrt{\hbar/(m_i \omega_i)}$,
are equal.
Small trapped
two-component systems with equal oscillator lengths,
$N \le 20$
and $1 < \kappa \le 20$ have been studied by the fixed node 
DMC method~\cite{blum08e,stec07b}.
These fixed node DMC calculations are performed using a 
guiding function that is constructed from the
solution of the two-body problem, a so-called ``pairing function''.
Since the densities 
of the two components 
overlap to a fairly large degree,
the fixed node DMC calculations can be used to
estimate the dependence of $\xi_{\rm{tr}}$
on $\kappa$. 
The fixed node DMC energies for
$\kappa>1$  suggest a slightly
modified analysis 
compared to that outlined 
around Eqs.~(\ref{eq_universalrelation})
and (\ref{eq_energynieft}) for $\kappa=1$.
In particular, $E_{\rm{fit}}$ is determined by treating the parameter
$c_{\rm{ETF}}$
in Eq.~(\ref{eq_energynieft})
as a fitting parameter
and not as a constant~\cite{stec07b}.
Figure~\ref{fig_unequalmassunit} shows that 
$\xi_{\rm{tr}}$ (circles in the main figure) 
and $c_{\rm{ETF}}$ (squares in the inset) decrease with increasing $\kappa$.
For comparison, triangles and a diamond show the values of $\xi_{\rm{hom}}$ 
for the homogeneous system obtained by performing 
fixed node DMC calculations~\cite{forb11,geze09}.
The slight decrease of $\xi_{\rm{hom}}$ from $\kappa=1$ to $6.5$ 
(triangles) has been interpreted
as providing a measure of non-$s$-wave contributions for 
the $\kappa=6.5$ system compared to the $\kappa=1$ system~\cite{geze09}.
The energy of the trapped system 
(see circles in Fig.~\ref{fig_unequalmassunit})
appears to be lowered more
than that of the homogeneous system (see triangles 
in Fig.~\ref{fig_unequalmassunit}),
which
might be a consequence
of the fact that the trapped system lowers its
energy if the densities are slightly mismatched.
Other causes for the discrepancy 
cannot be ruled out and further investigations are needed to
clarify this point.

Our discussion thus far has been based on the assumption that trimers 
are absent.
From the zero-range model with
infinite $a_0(0)$, however,
it is known that three-body bound states may
appear for $8.619 < \kappa < 13.607$~\cite{wern06,nish08}.
The possibility that three-body
physics,
and maybe even $N$-body physics, could
play a decisive
role in determining the
properties of two-component Fermi gases
for sufficiently large $\kappa$ is intriguing
and has recently stimulated interest among 
theorists~\cite{nish08,blum10,blum10a,gand10}.
In particular,
a three-body
bound state 
with $L^{\Pi}=1^{-}$
symmetry and a four-body bound state
with $L^{\Pi}=1^{+}$ symmetry have been predicted to
appear
for the FFX systems with $\kappa \approx 12.31$ and 
the FFFX system with $\kappa \approx 10.4$, 
respectively~\cite{blum10,blum10a}.
These results are obtained
by studying trapped systems with
purely attractive short-range
Gaussian model interactions
using
the stochastic variational approach.
An independent fixed node DMC study of the free-space system~\cite{gand10},
in which the F and X atoms
interact through different short-range model
potentials, 
arrives at similar conclusions, and furthermore
predicts the existence of a resonance for the $(4,1)$ system
for $\kappa$ slightly smaller than $10.4$.
Although 
the observed 
resonances depend, at least in principle, 
on the details of the underlying two-body
potential, it has been argued
that 
these
resonances 
occur most likely 
if $s_{\nu} \approx 1/2$~\cite{blum10,blum10a}.
The analysis that leads to this speculation is
based on the hyperspherical framework,
which allows for the
dependence of the eigenenergies 
on the hyperradial boundary condition to be studied.
In particular, it was found~\cite{blum10,blum10a} that
the energy-dependence of the transcendental 
eigenequation 
is weakest for $s_{\nu}=1/2$
and that the $N$-body resonances might thus be most amenable to 
tuning if $s_{\nu} \approx 1/2$.
Although the mass ratio of two-component Fermi gases is fixed by
nature (implying a 
particular $s_{\nu}$ value), 
the effective mass may be tuned by loading the system 
into an optical lattice~\cite{nish08}.
Future work needs to investigate to which extent 
$N$-body resonances can be utilized to study or engineer novel
many-body phenomena.
A first step in this direction was undertaken by
Nishida~{\em{et al.}}~\cite{nish08} 
who investigated the behavior of many-body systems
with two- and three-body resonances 
and $\kappa \approx 8.619$ and $13.607$.

Lastly, we consider the regime where $\kappa$ is greater than $13.607$,
i.e., the regime where three-body Efimov physics plays a role
(see Subsec.~\ref{sec_efimov}).
While the three-body
Efimov effect is quite well understood by now,
it is presently an open question to which extent
the behavior of the three-body system is
modified due
to the presence of other particles.
While such modifications due to the ``many-body background''
are expected to be relatively small 
for bosons, the situation may be
different for fermions 
due to
Pauli blocking.
In a lowest order semi-classical WKB treatment, the presence of the 
fermions that ``do not belong to''
the Efimov trimer have been accounted for by
constraining the momenta $\hbar k$ that contribute
to the three-body wave function 
to
$k>k_F$~\cite{macn10}. 
As a consequence,
the energy spacing between consecutive Efimov
states depends, in addition to $s_{\nu}$, on $k_F$.
While more studies are needed, the outlined example points toward a 
future frontier, namely the study of few-body phenomena in 
the presence of a
many-body background.
In passing, we also mention
that a four-body Efimov
effect has been predicted to exist for the FFFX system
with $13.384 < \kappa < 13.607$ and $1^+$ symmetry,
i.e., in a regime where the three-body Efimov effect is
absent~\cite{cast10}.

\subsection{Microscopic approach to thermodynamics
of equal-mass Fermi gas}
\label{sec_fermigasvirial}
So far, this review has focused on the zero-temperature
regime. An important question is, however, 
how the system properties depend on
temperature (see Refs.~\cite{nasc10,hori10} for very recent 
experiments).
Recently, the finite-temperature properties 
of macroscopic
two-component Fermi gases
have been
examined by employing a high temperature virial expansion approach that
uses the two- and three-body energies as 
input~\cite{liu09,liu10}.

For concreteness, we consider a spin- and mass-balanced
$s$-wave interacting
two-component Fermi gas consisting of $N$ mass $m$ point particles.
Since the chemical potential $\mu$
diverges logarithmically toward $-\infty$ as the temperature
approaches $\infty$, 
the fugacity $z$, $z=\exp[\mu/(k_B T)]$,
can be identified
as a small parameter~\cite{ho04a,rupa07,huan87}. 
It has been shown that a cluster expansion in terms of the
fugacity provides reliable
results for spin- and mass-balanced 
Fermi gases down to about 
half the Fermi
temperature  
if the expansion includes
the second and third order virial coefficients
(i.e., if the two- and three-body clusters are accounted 
for)~\cite{liu09,liu10}.
The high-temperature virial expansion approach is
equivalent to an expansion in terms of the ``diluteness
or degeneracy parameter''
$n \lambda_{\rm{dB}}^3$, where $n$ denotes the density
and $\lambda_{\rm{dB}}$ the de Broglie wave 
length~\cite{ho04a,rupa07,huan87}.
At high temperatures, the degeneracy parameter
is small.
However, near the transition temperature,
i.e., near $T/T_c \approx 1$, the degeneracy
parameter is of order one
(or $\lambda_{\rm{dB}} \approx \langle r \rangle$, where
$\langle r \rangle$ denotes the average interparticle
spacing). Thus, 
the applicability of 
the 
virial expansion approach is expected to break down in the 
low-temperature regime.

Following Liu {\em{et al.}}~\cite{liu09,liu10},
we work in the grand canonical ensemble
and write the thermodynamic potential
$\Omega$ in terms of the
partition function ${\cal{Z}}$, 
$\Omega=-k_B T \ln {\cal{Z}}$ with
${\cal{Z}}= \mbox{\rm{tr}} \exp[-(H- \mu N)/(k_B T)]$.
Once the thermodynamic potential $\Omega$ is known, other thermodynamic
observables such as the energy and pressure can be derived
from $\Omega$ by taking appropriate derivatives.
The idea behind the virial expansion is
to write the partition function
${\cal{Z}}$
in terms of  
the partition functions $Q_n$ of the $n$$^{th}$ cluster,
$Q_n=\mbox{\rm{tr}}_n \exp[-H_n/(k_B T)]$ and
${\cal{Z}}= 1+ z Q_1 + z^2 Q_2^2 + \cdots$.
Here, $H$ and $H_n$ denote the Hamiltonian of the full $N$-particle
system and the $n$-particle
subsystem or cluster, respectively.
The notation $\mbox{\rm{tr}}_n$ indicates that the 
trace is taken over the $n$-particle states.
If we denote the eigenenergies of the $n$$^{th}$ cluster
by $E_{n,j}$, then the partition function $Q_n$
becomes $Q_n=\sum_j \exp[-E_{n,j}/(k_B T)]$,
where $j$ collectively denotes the
set of quantum numbers
(i.e., the $Q_n$ are fully determined by the eigenenergies of the 
$n$$^{th}$ cluster).
Inserting the expansion for ${\cal{Z}}$ into the
thermodynamic potential $\Omega$ defines the virial
coefficients $b_n$,
$\Omega = -k_B T Q_1 ( z + b_2 z^2 + b_3 z^3+\cdots)$.
The second virial coefficient $b_2$, e.g., equals
$(Q_2-Q_1^2/2)/Q_1$ and depends on the $s$-wave scattering
length $a_0(0)$ as well as on the confining geometry and temperature.
The expansion of $\Omega$ in terms of $z$ shows that, if $|b_nz/b_{n-1}|$ 
is smaller than one, the $n$$^{th}$ contribution is suppressed by a factor
of $|b_nz/b_{n-1}|$ compared to the $(n-1)$$^{th}$ contribution.

\begin{figure}
\vspace*{+1.5cm}
\includegraphics[angle=0,width=65mm]{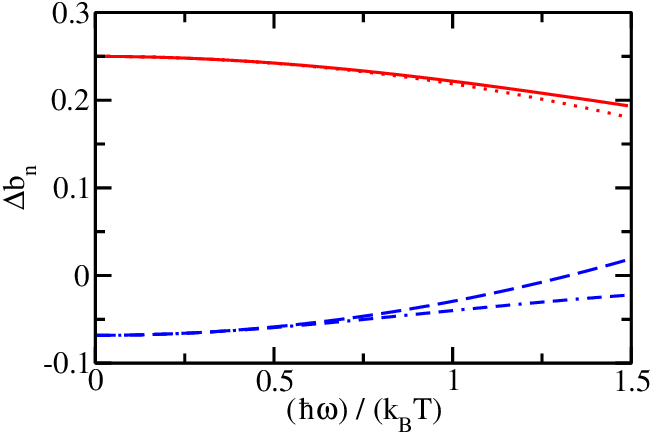}
\vspace*{-0.2cm}
\caption{(Color online)
Shifted virial coefficients 
$\Delta b_n$
as a function of the dimensionless inverse temperature $\tilde{\omega}$,
$\tilde{\omega}=\hbar \omega/(k_B T)$,
of the trapped equal-mass two-component Fermi gas at unitarity.
Solid and dotted lines show respectively the virial coefficient
$\Delta b_2$ 
and its Taylor
expansion up to order
$\tilde{\omega}^2$.
Dash-dash-dotted and
dashed lines show respectively the virial coefficient
$\Delta b_3$~\cite{dailprivate} 
and its Taylor
expansion up to order
$\tilde{\omega}^2$.
}
\label{fig_virial}
\end{figure}

We now consider an application to the 
harmonically trapped two-component
Fermi system
at unitarity, for which the two- and three-particle
energies are known with high precision~\cite{busc98,wern06a}.
It is convenient to
express the virial coefficients $b_n(a_0(0))$
of the interacting
system relative to the virial coefficients $b_n(a_0(0)=0)$
of the non-interacting system, i.e.,
we define $\Delta b_n = b_n(a_0(0)) - b_n(a_0(0)=0)$. 
At unitarity, the $s$-wave two-body energies 
equal $(2j+1/2)\hbar \omega +E_{\rm{cm}}$ ($j=0,1,\cdots$)
and $\Delta b_2$ can be calculated analytically by
summing over the quantum numbers
associated with the relative and center-of-mass degrees of freedom.
The result is
$\Delta b_2=\exp(-\tilde{\omega}/2) / [2(1+ \exp (-\tilde{\omega}))]$~\cite{liu09},
where $\tilde{\omega}$ denotes a
dimensionless ``inverse temperature'' or scaled frequency, 
$\tilde{\omega}= \hbar \omega / (k_B T)$.
The third, shifted virial coefficient 
$\Delta b_3$ 
requires summing the three-body energies,
which can be obtained by solving a
transcendental equation [see Subsec.~\ref{sec_threebodylimits}
and Eq.~(7) of Ref.~\cite{wern06a}], numerically.
The high temperature expansions of $\Delta b_2$ and $\Delta b_3$
at unitarity are
$1/4-\tilde{\omega}^2/32$ and
$-0.06833960+0.038867 \tilde{\omega}^2$, respectively~\cite{liu09}.
Figure~\ref{fig_virial} shows the shifted virial coefficients
$\Delta b_2$ and $\Delta b_3$ 
at unitarity
as a function of 
$\tilde{\omega}$.
As can be seen, the temperature dependence of the
virial coefficients is fairly weak in the high temperature (small
$\tilde{\omega}$) regime.
The Fermi energy $E_F$ of the harmonically trapped
non-interacting two-component Fermi gas, which can be written as
$E_F \approx (3N)^{1/3} \hbar \omega$, defines
the Fermi temperature $T_F$, 
$T_F = E_F / k_B$.
For $N=100$, e.g., one finds $k_B T_F \approx 7 \hbar \omega$, i.e., 
the
temperature dependence of the virial coefficients is 
fairly weak if the temperature $T$ is larger than the 
Fermi temperature $T_F$.

Knowing the virial coefficients
$\Delta b_2$ and $\Delta b_3$,
the thermodynamic properties of the trapped 
Fermi gas can be calculated.
As a first application, Liu {\em{et al.}}~\cite{liu09} 
determined the
interaction energy $E_{\rm{int}}$, $E_{\rm{int}}=E-E_{\rm{tot}}^{\rm{NI}}$ (here,
$E_{\rm{tot}}^{\rm{NI}}$ denotes the energy
of the corresponding non-interacting system),
of a trapped two-component lithium mixture
at unitarity in the temperature regime $T \gtrsim T_F$. 
The 
theoretical predictions connect smoothly to
experimental data from the Duke group~\cite{luo07}, 
which cover the temperature
regime $T \lesssim T_F$.
More recently, the virial expansion approach has been 
used to determine
other observables at and away from 
unitarity~\cite{hu10,liu10a}.

The virial expansion approach outlined can be extended
to the homogeneous system, thereby allowing for comparisons with
results obtained by
finite temperature MC simulations~\cite{bulg06,buro06,akki07} and effective 
many-body approaches.
To this end, Liu {\em{et al.}}~\cite{liu09} applied 
an ``inverse LDA'', which allowed for the determination of
the virial coefficients of the
homogeneous system from those of the inhomogeneous system.
The theoretically determined
high-temperature value of the
third order virial coefficient has been confirmed 
experimentally~\cite{nasc10}.
The examples discussed in this subsection underline
that microscopic bottom-up approaches provide,
in certain cases, 
powerful means to predict many-body properties.

\section{$s$-wave interacting Bose gas         
under confinement}
\label{sec_bosegas}
This section considers single-component
Bose gases 
under external confinement.
Subsection~\ref{sec_bosegaseqofstate} 
introduces the mean-field Gross-Pitaevskii (GP)
framework, which
assumes that the behavior
of weakly-interacting Bose gases is fully
governed by the $s$-wave scattering 
length $a_0(0)$.
The validity regimes of the GP equation and 
a modified GP equation are benchmarked for Bose gases under
spherically symmetric harmonic confinement.
Subsection~\ref{sec_bosegasinstability} discusses the mechanical
instability that arises for Bose gases with negative
$s$-wave scattering lengths.
Lastly, Subsec.~\ref{sec_bosegasthreebodyparam}
discusses the role of the effective
range and the three-body parameter for Bose systems.

\subsection{Equation of state of trapped Bose gas: The role
of the $s$-wave scattering length}
\label{sec_bosegaseqofstate}
Weakly-interacting Bose gases under external confinement 
at zero temperature are
most commonly described by the non-linear mean-field GP 
equation~\cite{ginz58,gros61,gros63,pita61,rupr95,baym96,dalf96},
which
can be derived by
approximating the many-body 
wave
function $\Psi(\vec{r}_1,\cdots,\vec{r}_N)$
by a product of single-particle orbitals $\Phi(\vec{r}_j)$,
$\Psi=\prod_{j=1}^N \Phi(\vec{r}_j)$,
where $\int |\Phi(\vec{r})|^2 d^3 \vec{r}=1$~\cite{esry97,esry97b}.
This product ansatz is fully symmetrized and geared toward 
the description of the energetically lowest lying gas-like 
state of the $N$-boson system.
Assuming that the interactions can be parameterized through
a sum of two-body potentials $V_{\rm{tb}}(\vec{r}_{jk})$
and minimizing the total energy $E_{\rm{tot}}(N)$ with respect to 
$\Phi^*$,
one obtains
the Hartree equation~\cite{esry97,esry97b}
\begin{eqnarray}
\label{eq_hartree}
\left[ \frac{-\hbar^2}{2m} \nabla_{\vec{r}}^2 +
V_{\rm{trap}}(\vec{r}) + (N-1)\int  V_{\rm{tb}}(\vec{r},\vec{r} \, ')
|\Phi(\vec{r} \, ')|^2 d^3 \vec{r} \, ' \right]
\Phi(\vec{r}) = \nonumber \\
\epsilon \Phi(\vec{r}).
\end{eqnarray}
In Eq.~(\ref{eq_hartree}), 
$m$ denotes the atom mass, $V_{\rm{trap}}$ the external confining
potential felt by each of the bosons, and $\epsilon$
the orbital energy
or chemical potential.
The integro-differential equation~(\ref{eq_hartree}),
which has to be solved
self-consistently for $\Phi(\vec{r})$ and $\epsilon$,
describes
a trapped ``test
particle''  that moves---just as in atomic structure 
calculations---in the mean-field
[the third term in square brackets on the left hand side
of Eq.~(\ref{eq_hartree})] created by the other
$(N-1)$ atoms.

For ultracold degenerate $s$-wave interacting Bose gases,
the two-body
interaction potential $V_{\rm{tb}}$
with $s$-wave scattering length $a_0(0)$
is frequently replaced by the
Fermi pseudopotential $V_F$, Eq.~(\ref{eq_fermipp}),
for which the integral in Eq.~(\ref{eq_hartree})
can be evaluated analytically,
\begin{eqnarray}
\label{eq_gp}
\left[ \frac{-\hbar^2}{2m} \nabla_{\vec{r}}^2 +
V_{\rm{trap}}(\vec{r}) + (N-1)\frac{4 \pi \hbar^2 a_0(0)}{m}
|\Phi(\vec{r})|^2 \right]
\Phi(\vec{r}) = 
\nonumber \\
\epsilon \Phi(\vec{r}).
\end{eqnarray}
Equation~(\ref{eq_gp}) is the
GP equation for $s$-wave interacting systems,
sometimes also referred to 
as
non-linear Schr\"odinger equation.
Although the GP equation has been derived by applying the variational
principle, 
the resulting total energy $E_{\rm{tot}}(N)$
does not constitute a variational upper bound to the energy
of the energetically lowest-lying gas-like state of the
many-body system since
the true atom-atom potential has been replaced by Fermi's
pseudopotential. 
The GP equation depends on
the product
$(N-1)a_0(0)$, and not on $(N-1)$ and $a_0(0)$ 
separately. This implies that the mean-field formulation is
only sensitive to the ``strength'' of the effective interaction
and not to its underlying microscopic origin, i.e.,
whether it is due to, say, large $N$ and small $a_0(0)$ or
small $N$ and large $a_0(0)$.
An alternative derivation
of Eq.~(\ref{eq_gp}) is based on a field operator description
and results in a 
factor of $N$ as opposed to $N-1$ (see, e.g., 
Refs.~\cite{dalf98,fett96,legg01}). 
While it is justified to replace $N-1$ by $N$ in the
large $N$ limit, 
the factor of $N-1$ should be used in the small 
$N$ limit where the difference between 
$N-1$ and $N$ is appreciable.
The outlined number conserving microscopic derivation of the GP equation
emphasizes that its application is not restricted to large number
of particles; in fact, as shown below, the GP equation provides a
good description even for dilute two-particle systems.

\begin{figure}
\vspace*{+1.5cm}
\includegraphics[angle=0,width=65mm]{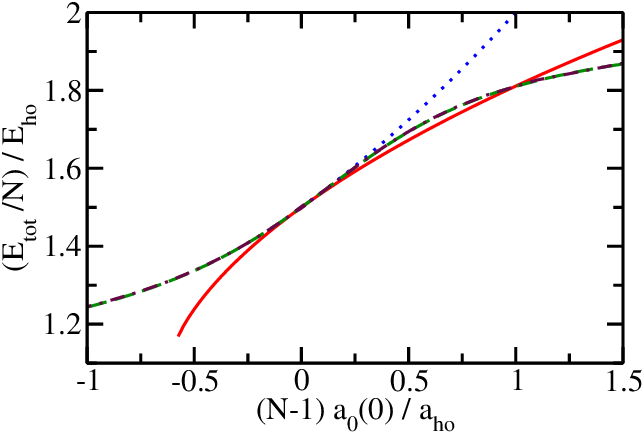}
\vspace*{-0.2cm}
\caption{(Color online)
Total energy $E_{\rm{tot}}/N$ per particle 
of a
dilute
Bose gas under
external spherically symmetric
confinement as a function of 
the ``interaction parameter'' $(N-1)a_{0}(0)/a_{\rm{ho}}$.
The solid line shows the mean-field GP energy.
Dotted, dashed and dash-dotted lines show the energies
obtained by solving the linear Schr\"odinger equation for two
particles interacting 
through a hardsphere potential, a square-well potential 
(with range $r_0$, $r_0=0.01a_{\rm{ho}}$,
and depth $V_0$ adjusted such that the free-space system
does 
support 
one 
bound state) and the zero-range pseudopotential $V_0^{\rm{pp}}$ 
[Eq.~(\ref{eq_pseudo3d})],
respectively.
Note that the dashed and dash-dotted lines are indistinguishable on
the scale shown.}
\label{fig_gp}
\end{figure}

The solid line in Fig.~\ref{fig_gp}
shows the total energy 
$E_{\rm{tot}}/N$ per particle, obtained by solving the GP equation
for a spherically symmetric harmonic confinement,
as a function of the interaction parameter $(N-1)a_0(0)/a_{\rm{ho}}$.
The solid line terminates at 
$(N-1)a_0(0)/a_{\rm{ho}} \approx -0.57497$~\cite{esry97b,dodd96}.
For more negative interaction parameters, the GP equation supports
a solution whose energy is unbounded from below (i.e., the energy
approaches $-\infty$ and the corresponding wave function
describes a high density state of size much smaller than $a_{\rm{ho}}$).
Generally speaking, the validity regime of the GP
equation is restricted to dilute systems,
i.e., to parameter combinations that fulfill the inequality
$n(0) |a_0(0)|^3  \ll 1$~\cite{bogo47}, 
where $n(0)$ denotes the peak density.
For $(N-1)a_0(0)/a_{\rm{ho}} \gtrsim -0.57497$, the diluteness parameter 
$n(0) |a_0(0)|^3 $ is much smaller than 1, assuming $N$ is not too
small.
This suggests that the
predicted drop of the GP energy 
(the termination of the solid line in Fig.~\ref{fig_gp}) lies
within the validity regime of the mean-field framework.
As discussed in Subsec.~\ref{sec_bosegasinstability}, the 
sudden drop of
the GP energy 
is
interpreted as a mechanical instability or collapse. 

We now compare the solutions of the GP
equation with those
obtained by solving the
linear Schr\"odinger equation.
Dotted, dashed and dash-dotted
lines in Fig.~\ref{fig_gp} show the 
total energy $E_{\rm{tot}}/N$
per particle for two particles under spherically symmetric 
harmonic confinement
interacting through 
a hardsphere potential with range $a_0(0)$
$[a_0(0)>0]$, 
a square well
potential with range $r_0$
and depth $V_0$ 
($r_0=0.01 a_{\rm{ho}}$ and $V_0$ are chosen
such that the potential supports 
one 
$s$-wave bound state in free space),
and the pseudopotential $V_0^{\rm{pp}}$
[see Eq.~(\ref{eq_pseudo3d})], respectively. 
The two-body energies for these potentials
are obtained by solving the 
linear
Schr\"odinger equation 
analytically (for the treatment of $V_0^{\rm{pp}}$, 
see Sec.~\ref{sec_twobody}).
For $(N-1)|a_0(0)|/a_{\rm{ho}} \lesssim 0.1$,
the agreement between the energies obtained by solving the linear
and non-linear  Schr\"odinger equations is excellent. However,
for larger
$(N-1)|a_0(0)|/a_{\rm{ho}}$ deviations are 
visible. 
The energy for two particles
interacting through
the hardsphere potential increases roughly linearly with $a_0(0)$
and lies above the other energies.
The comparatively large increase of the energy for the hardsphere
potential 
can be understood by realizing that
the excluded volume increases with increasing
$a_0(0)$, resulting in an infinite two-body energy as
$a_0(0)$ approaches infinity. This 
unphysical behavior suggests
that the applicability regime of the hardsphere potential 
is limited to small $a_0(0)$.
The GP energy lies below the energies for the zero-range and square-well
interaction potentials for $(N-1)a_0(0)/a_{\rm{ho}} \lesssim 1$
and above for $(N-1)a_0(0)/a_{\rm{ho}} \gtrsim 1$.
The discrepancy between the GP energy and the energies 
for the zero-range and square-well
interaction potentials
increases as $(N-1)a_0(0)/a_{\rm{ho}}$ increases further.
In the large interaction parameter regime, the diluteness
parameter $n(0)|a_0(0)|^3$ is no longer small 
compared to one and higher order corrections
need to be included in the mean-field description (see below).
For negative scattering lengths, the energies obtained by solving the
linear Schr\"odinger equation vary smoothly.
In particular, the mechanical instability predicted by the GP
equation has no two-body analog; at least three particles are 
needed to see an analog of the instability predicted
by the GP equation within the
framework of the linear Schr\"odinger equation
(see also Subsec.~\ref{sec_bosegasinstability}).

\begin{figure}
\vspace*{+1.5cm}
\includegraphics[angle=0,width=65mm]{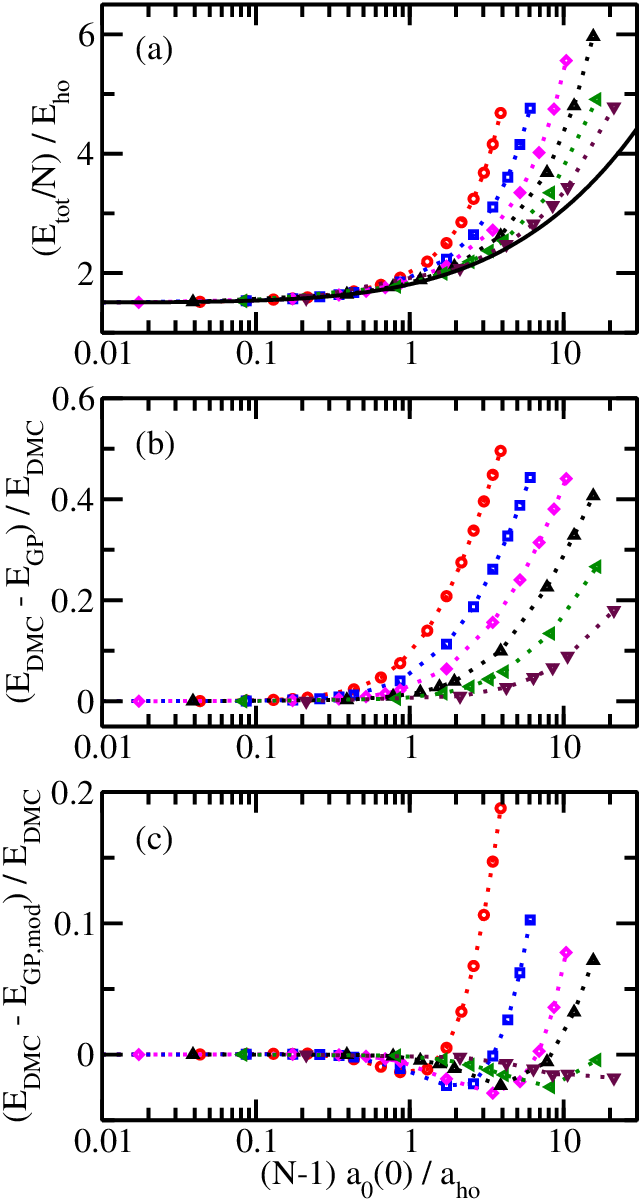}
\vspace*{-0.2cm}
\caption{(Color online)
Hardsphere equation of state for the Bose gas
under spherically symmetric confinement.
Symbols in panel~(a) show the total energy $E_{\rm{DMC}}/N$
per particle
for the $N$-particle Bose gas with hardsphere interactions
as a function of the
interaction parameter $(N-1)a_0(0)/a_{\rm{ho}}$
for 
$N=2$ (circles),
$N=3$ (squares),
$N=5$ (diamonds),
$N=10$ (up triangles),
$N=20$ (left triangles), and
$N=50$ (down triangles); the energies are obtained by solving the
linear Schr\"odinger equation by the DMC method.
For comparison, the solid line shows the total energy per particle $E_{\rm{GP}}/N$
obtained
by solving the mean-field GP equation.
Panels~(b) and (c) show the scaled energy
differences $(E_{\rm{DMC}}-E_{\rm{GP}})/E_{\rm{DMC}}$ and
$(E_{\rm{DMC}}-E_{{\rm{GP}},{\rm{mod}}})/E_{\rm{DMC}}$, respectively,
for the same $N$ as panel~(a);
to guide the eye,
dotted lines connect data points for the same $N$.
The figure has been adapted from Ref.~\protect\cite{blum01}.}
\label{fig_hardsphere}
\end{figure}

For larger number of particles, the determination of
the solutions
to the linear Schr\"odinger equation becomes more involved.
In addition to the increased number of degrees of freedom,
complications arise due to the fact that 
the $N$-boson system supports,
unlike $s$-wave interacting two-component Fermi systems, 
$N$-body ($N>2$) 
bound states
even if the underlying two-body potential does 
not support a two-body bound state. Thus, the 
energetically lowest lying gas-like state of the Bose gas is a 
metastable state and not the true ground state of the system.
Tightly bound $N$-body clusters,
whose properties are discussed further in 
Subsec.~\ref{sec_bosegasthreebodyparam},
are absent for systems interacting through
purely repulsive two-body potentials
such as the hardsphere potential. Although purely repulsive interaction models
exhibit, as discussed exemplarily above
for the hardsphere potential, unphysical behavior
for large $a_0(0)$, they provide a 
quantitatively correct description for small $a_0(0)$
and allow one
to estimate model dependent effects
for intermediate $a_0(0)$~\cite{blum01,holz99,dubo01,dubo03}.
Symbols in Fig.~\ref{fig_hardsphere}(a) show the energy $E_{\rm{tot}}/N$
per particle for $N$-boson systems, $N=2-50$,
with hardsphere interactions
under external spherically symmetric harmonic
confinement as a function of the interaction parameter
$(N-1)a_0(0)/a_{\rm{ho}}$~\cite{blum01}. 
It can be seen clearly that 
the solutions of the linear Schr\"odinger equation depend
on $N$ and $a_0(0)/a_{\rm{ho}}$ separately and not just the product
of these two parameters. 
For comparison, the solid line shows the GP energy per particle.
The DMC energies for different
number of particles (symbols), obtained by
solving
the linear Schr\"odinger by the DMC method~\cite{blum01}, 
lie above the GP energies.
Figure~\ref{fig_hardsphere}(b)
shows the energy difference $E_{\rm{DMC}}-E_{\rm{GP}}$.
For fixed $N$, the deviations between the DMC and GP
energies increase with increasing interaction
parameter.
For a fixed interaction parameter,
the deviations decrease with increasing $N$. 
An improved mean-field description of the hardsphere equation of state
is discussed below.

To investigate the small $|a_0(0)|/a_{\rm{ho}}$ behavior in more detail,
the 
$N$-boson system under external spherically
symmetric confinement with
two-body zero-range interactions can be treated perturbatively
within the framework of linear Schr\"odinger quantum mechanics.
Up to order $[a_0(0)/a_{\rm{ho}}]^2$,
the perturbative energy of the energetically
lowest-lying gas-like state of the $N$-boson system reads~\cite{john09}
\begin{eqnarray}
\label{eq_pertgp}
E_{\rm{tot}} \approx
\frac{3N}{2} \hbar \omega +
\left( \begin{array}{c} N \\ 2 \end{array} \right)
\sqrt{\frac{2}{\pi}} \frac{a_0(0)}{a_{\rm{ho}}} 
\hbar \omega
+ \nonumber \\
\left[
\left( \begin{array}{c} N \\ 2 \end{array} \right)
\frac{2(1-\mbox{ln} (2))}{\pi}-
\left( \begin{array}{c} N \\ 3 \end{array} \right)
\beta \right]
\left[ \frac{a_0(0)}{a_{\rm{ho}}}  \right]^2
\hbar \omega,
\end{eqnarray}
where 
\begin{eqnarray}
\beta = \frac{2}{\pi} \left[
4\sqrt{3}-6+6 \mbox{ln} \left( \frac{4}{2+\sqrt{3}}\right) 
\right]
\approx 0.8557583 
\end{eqnarray}
and
\begin{eqnarray}
\left( \begin{array}{c} N \\ n \end{array} \right) = \frac{N!}{n! (N-n)!}.
\end{eqnarray}
The first term on the right hand side is
the energy of the non-interacting (unperturbed) $N$-boson system.
The 
leading order 
energy shift of the 
$N$-particle system is equal
to the 
energy shift that $N(N-1)/2$ independent two-body systems would experience.
The next order shift contains a positive term that
scales with the number of pairs and a negative term that
scales with the number of trimers.
The latter contribution has been interpreted as an effective
attractive three-body interaction~\cite{jons02,john09}. 
This interpretation
emerges naturally in an effective field
theory treatment~\cite{john09}, which derives
an effective low-energy
Hamiltonian that contains 
one-body, two-body and three-body terms.
We stress that the effective attractive
three-body interaction arises from the pairwise zero-range interactions
and not from an explicit three-body potential. The 
need for and role of an
actual three-body potential is discussed in 
Subsec.~\ref{sec_bosegasthreebodyparam}.
The effective three-body interaction 
for the harmonically trapped Bose gas was first
predicted by Jonsell {\em{et al.}}~\cite{jons02},
who employed  the BO and adiabatic approximations
within the hyperspherical 
framework to
place bounds on the 
three-body coefficient.
For a Bose gas confined to a box with periodic boundary conditions,
the effective three-body interaction has a positive 
coefficient~\cite{huan57} (see also Refs.~\cite{wu59,lee57a,bean07,tan08d}),
which illustrates that the effective $N$-body interactions
can be tuned by varying the external confinement.
The second term on the right hand side
of Eq.~(\ref{eq_pertgp}) can also be derived
within the GP framework
by 
approximating $\Phi$
in Eq.~(\ref{eq_gp}) by the non-interacting wave function.

Equation~(\ref{eq_pertgp}) can be viewed
as the few-particle trap analog of the low-density expansion of
the equation of state of the homogeneous Bose gas
with density $n$~\cite{lee57a,lee57},
\begin{eqnarray}
\label{eq_homexpand}
\frac{E_{\rm{hom}}/N}{\frac{\hbar^2}{2ma_0^2(0)}} \approx 
4 \pi n a_0^3(0) 
\left[
1+ \frac{128}{15 \sqrt{\pi}} \sqrt{n a_0^3(0)} 
\right].
\end{eqnarray}
Application of the LDA to the first term
on the right hand side of Eq.~(\ref{eq_homexpand})
gives the second term on the right hand side of
Eq.~(\ref{eq_pertgp}).
The next order correction in Eq.~(\ref{eq_pertgp}), in contrast,
cannot be determined by applying the LDA to the $\sqrt{n a_0^3(0)}$
term
in Eq.~(\ref{eq_homexpand})~\cite{jons02}.

Lastly, we comment on the 
large scattering length regime.
The experimental realization
of cold atomic gases with large scattering length
and sufficiently long lifetimes would open the possibility to
study strongly-correlated Bose systems with unprecedented
control. Historically~\cite{lond38,lond38a}, the study of liquid helium,
which is characterized by a gas parameter $n a_0^3(0)$ of
about $0.21$ 
and a condensate fraction of about 7\%~\cite{cepe95,moro97,glyd00}, 
is tremendously important.
Liquid helium has 
an $s$-wave scattering length
that is roughly 10 times larger than
the effective range, which in turn is of the order of 
the interparticle
spacing.
Although the $s$-wave scattering length of Bose
gases can, for a subset of species such as $^{85}$Rb~\cite{robe98},
be tuned over a wide range through the application
of an external magnetic field in the vicinity of a Fano-Feshbach
resonance,
Bose gases with large scattering length exhibit
detremental losses 
since the three-body recombination rate shows
an overall $|a_0(0)|^4$ scaling~\cite{esry99b,niel99,beda00,braa01,fedi96}.
This is in contrast to two-component Fermi gases with equal masses, which
are stabilized by the ``Pauli pressure'' that results
from the Pauli exclusion principle.
While it is not clear at present
to which extent the equilibrium
properties of Bose gases with large scattering length
can be probed experimentally~\cite{papp08},
several theoretical studies have investigated
the properties of Bose gases with large scattering length
based on the assumption that the system's lifetime
would be sufficiently long to reach equilibrium.
The first studies in this direction were performed for the
trapped three-boson system by
Jonsell {\em{et al.}}~\cite{jons02} and 
Blume {\em{et al.}}~\cite{blum02b}.
More recent studies have estimated the lifetime of the
trapped three-boson system~\cite{wern06a,port11}.
Extensions to larger systems have also been 
considered~\cite{adhi08,cowe02,heis04,thog07,song08,lee10}.
The work by Song and Zhou~\cite{song08}, 
e.g., predicts the
existence of a fermionized three-dimensional Bose gas
with large scattering length (see also Ref.~\cite{ho04}).
This is an intriguing prospect 
since the fermionization of Bose gases has been 
studied extensively in one-dimensional systems.

If the interaction parameter $(N-1)a_0(0)/a_{\rm{ho}}$
becomes appreciable but not too large,
corrections to the GP equation can be accounted for
by a modified mean-field GP 
equation~\cite{blum01,braa97,timm97,fabr99},
\begin{eqnarray}
\label{eq_gpmod}
[ \frac{-\hbar^2}{2m} \nabla_{\vec{r}}^2 +
V_{\rm{trap}}(\vec{r}) + (N-1)\frac{4 \pi \hbar^2 a_0(0)}{m}
|\Phi(\vec{r})|^2 \times  \nonumber \\
\left( 
1 + \frac{32}{3 \sqrt{\pi}} 
[a_0^3(0)(N-1)]^{1/2} 
\Phi(\vec{r})
\right)
]
\Phi(\vec{r}) = \epsilon \Phi(\vec{r}).
\end{eqnarray}
The second term in the big round brackets
in the second line of Eq.~(\ref{eq_gpmod})
can be derived from the ``quantum fluctuation
term''
of the
homogeneous system, i.e., the second term on
the right hand side of Eq.~(\ref{eq_homexpand}).
Figure~\ref{fig_hardsphere}(c) shows
the energy difference $E_{\rm{DMC}}-E_{{\rm{GP}},{\rm{mod}}}$,
where the energy $E_{{\rm{GP}},{\rm{mod}}}$ has been derived from the orbital 
energy $\epsilon$
that results when solving Eq.~(\ref{eq_gpmod}).
A comparison of the vertical scales of Figs.~\ref{fig_hardsphere}(b)
and \ref{fig_hardsphere}(c) shows that the modified GP equation
extends the validity regime of the mean-field treatment.

\subsection{Bose gas with negative $s$-wave
scattering length}
\label{sec_bosegasinstability}
\begin{figure}
\vspace*{+1.5cm}
\includegraphics[angle=0,width=65mm]{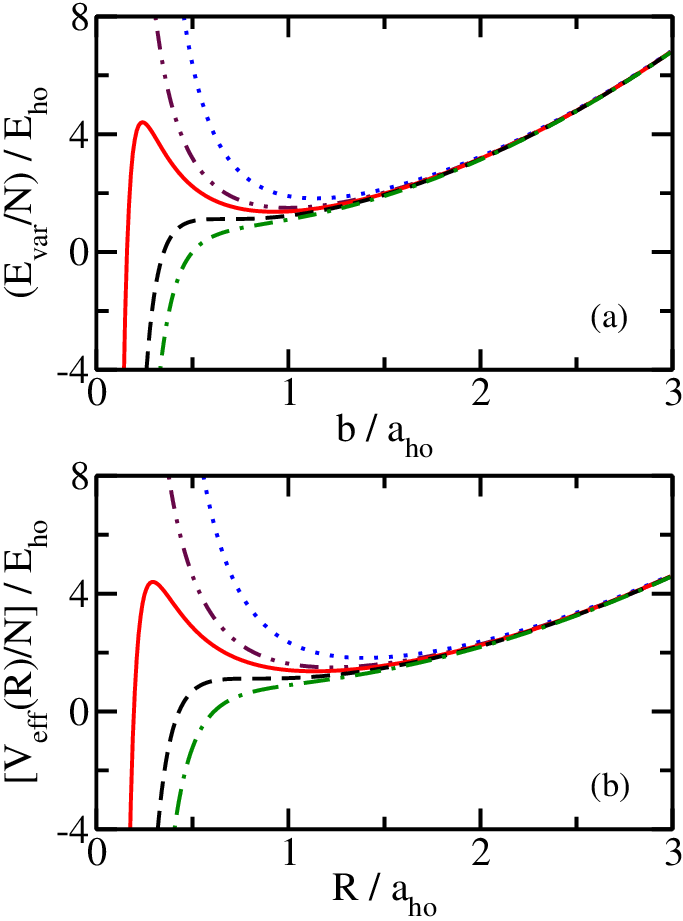}
\vspace*{-0.2cm}
\caption{(Color online)
Mechanical instability of Bose gases with negative scattering length.
(a) Dotted, dash-dot-dotted, solid, dashed and dash-dotted lines show
the variational energy $E_{\rm{var}}/N$ per
particle, Eq.~(\ref{eq_var}), as a function of $b/a_{\rm{ho}}$
for $(N-1)a_0(0)/a_{\rm{ho}}=1,0,-0.3,-0.67$ and $-1$, respectively.
(b) Dotted, dash-dot-dotted, solid, dashed and dash-dotted lines show
the effective potential curve $V_{\rm{eff}}(R)/N$, 
Eq.~(\ref{eq_effbose}), as a function of $R/a_{\rm{ho}}$
for $(N-1)a_0(0)/a_{\rm{ho}}=1,0,-0.3,-0.67$ and $-1$, respectively
($N$ is fixed, $N=1000$).
}
\label{fig_attractive}
\end{figure}

To shed light on the mechanical instability
for Bose gases with negative scattering length
(i.e., the termination of the solid line in Fig.~\ref{fig_gp}
for negative scattering length),
we solve the GP equation for 
a spherically symmetric harmonic confinement
using a variational Gaussian wave function
with width $b$.
Writing $\Phi(\vec{r})= b^{-3/2} \pi^{-3/4} \exp[-r^2/(2b^2)]$,
the
variational energy becomes~\cite{pere97}
\begin{eqnarray}
\label{eq_var}
\frac{E_{\rm{var}}}{N} \approx
\left[ 
\frac{3a_{\rm{ho}}^2}{4b^2} + \frac{3b^2}{4a_{\rm{ho}}^2}+
\frac{(N-1)a_0(0)a_{\rm{ho}}^2}{\sqrt{\pi} b^3}
\right] \hbar \omega
,
\end{eqnarray}
where the first, second and third terms on the right hand side
are the kinetic, trap and 
interaction energies, respectively.
Since the interaction energy 
varies as $b^{-3}$ while the 
kinetic and potential energies vary
as $b^{-2}$ and $b^2$, respectively,
it is clear that there exists a
critical negative interaction parameter 
beyond which the 
variational energy  possesses a global minimum for $b=0$ but not a local 
minimum~\cite{pere97}.
Similar conclusions have been drawn 
in related studies~\cite{kaga96,shur96,stoo97}.
Lines in Fig.~\ref{fig_attractive}(a)
show the variational energy per particle as a function 
of $b/a_{\rm{ho}}$ for various
interaction parameters $(N-1)|a_0(0)|/a_{\rm{ho}}$. 
For small $(N-1)|a_0(0)|/a_{\rm{ho}}$,
the variational energy shows a minimum
around $b/a_{\rm{ho}} \approx 1$.
As $(N-1)|a_0(0)|/a_{\rm{ho}}$ increases
[$a_0(0)$ negative], the energy minimum moves to smaller 
$b/a_{\rm{ho}}$ and the barrier that separates the local 
and global energy minima
decreases.
At the critical value
of $(N-1)a_0(0)/a_{\rm{ho}} \approx -0.67051$
[see dashed line in Fig.~\ref{fig_attractive}(a)], 
the barrier disappears, suggesting that the Bose gas
can lower its energy---without having to ``tunnel'' through
an ``energy barrier''---by shrinking toward (or collapsing into)
a so-called high density
``snowflake state''.
The variational Gaussian approach describes the onset
of the mechanical instability quite well, i.e.,
it predicts a slightly more negative critical interaction
parameter 
than the full solution to the GP equation 
[$(N-1)a_0(0)/a_{\rm{ho}} \approx -0.67051$ compared to $-0.57497$].
Furthermore, it highlights that the trapped Bose gas is,
for sufficiently small $(N-1)|a_{0}(0)|/a_{\rm{ho}}$ [$a_0(0)<0$],
stabilized by the positive energy contributions that arise  
due to the presence of the trap. If the system was homogeneous,
an infinitesimally small $|a_0(0)|$ [$a_0(0)<0$]
would induce collapse.
While it is suggestive to interpret the energy curves shown
in Fig.~\ref{fig_attractive}(a) as effective potentials, it must
be
kept in mind that the width $b$ corresponds to a variational
parameter and not to an actual spatial coordinate.

The collapse dynamics of Bose gases with negative $a_0(0)$
has also been investigated 
by solving 
the linear Schr\"odinger
equation 
with zero-range interactions within the 
BO hyperspherical 
approximation~\cite{bohn98}.
For a variational
wave function that is written as the product of a constant
hyperangular piece and a hyperradial function $R^{-(3N-1)/2}F(R)$,
the effective
hyperradial potential curve
$V_{\rm{eff}}(R)$ takes the form~\cite{bohn98}
\begin{eqnarray}
\label{eq_effbose}
V_{\rm{eff}}(R)= \frac{\hbar^2 (3N-1)(3N-3)}{8 Nm R^2}
+ \frac{1}{2} Nm \omega^2 R^2 + \nonumber \\
\sqrt{\frac{1}{2 \pi}} \frac{\hbar^2 (N-1)a_0(0)}{m} 
\frac{\Gamma(3N/2)}{\Gamma((3N-3)/2) N^{3/2}}
\frac{N}{R^3},
\end{eqnarray}
where $NR^2=\sum_i  \vec{r}_i^2$.
The effective potential $V_{\rm{eff}}(R)$
enters into the hyperradial equation $-\hbar^2/(2Nm)F''+V_{\rm{eff}}F=E_{\rm{tot}}F$,
which can be solved readily.
In Eq.~(\ref{eq_effbose}), 
the first, second and third terms on the right hand side
arise due to the kinetic energy operator, the external confinement, 
and the pairwise interactions, respectively.
Lines in Fig.~\ref{fig_attractive}(b) show 
$V_{\rm{eff}}(R)/N$ for various interaction
parameters $(N-1)a_0(0)/a_{\rm{ho}}$ and $N=1000$. We note that
the curves $V_{\rm{eff}}(R)/N$ depend only weakly on 
$N$~\cite{bohn98}.
A comparison of Figs.~\ref{fig_attractive}(a) and \ref{fig_attractive}(b)
[and, equivalently, of Eqs.~(\ref{eq_var}) and (\ref{eq_effbose})]
shows that $b$ plays a role similar
to that played by $R$. 
The key difference is that 
$b$ is a variational parameter
while 
the hyperradius $R$ is a spatial coordinate.
Physically, the small $R$ region describes a system in which the
atom-atom distances are small, i.e., where the system
is self-bound.
Using the effective potential curves $V_{\rm{eff}}$,
tunneling from
the region where the potential exhibits a local minimum
to the small $R$ region where the effective potential is unbounded 
from below has been estimated within the WKB approximation~\cite{bohn98}.
The conclusions discussed
here
have been confirmed 
for the three-body system
by performing an analysis that goes beyond the variational 
treatment and furthermore accounts
for the coupling between channels~\cite{blum02b}.
For larger systems, confirmation comes from 
treatments that explicitly account for two-body 
correlations~\cite{sore03} and experiments~\cite{brad97,robe01a}.

\subsection{Beyond the $s$-wave scattering length:
Effective range and three-body parameter}
\label{sec_bosegasthreebodyparam}

The previous two subsections assumed that the equation of state
of trapped Bose gases is fully determined by the $s$-wave scattering
length $a_0(0)$.
However,
atom-atom
potentials possess a finite range, which induces range-dependent
corrections to the equation of state.
Furthermore, the properties of Bose gases
may additionally depend on a three-body parameter whose
role for
the three-boson
system with large $s$-wave scattering length has already been discussed
in Subsec.~\ref{sec_efimov}.
In the following, we present simple arguments that 
illustrate at which order these corrections appear
in the equation of state. 

To account for the finite range of the underlying atom-atom
potential,
we resort
to the effective range expansion, Eq.~(\ref{eq_effrange}).
The effective range $r_{\rm{eff}}$ is determined by the shape
and range of the two-body potential. For the hardsphere potential,
e.g., the effective range $r_{\rm{eff}}$ is directly
proportional to $a_0(0)$ [i.e.,  $r_{\rm{eff}}=2a_0(0)/3$].
For the van der Waals potential with $r^{-6}$ tail, the
effective range 
$r_{\rm{eff}}$ is, in the low-energy regime,
also determined by the scattering length
$a_0(0)$~\cite{gao98}.
To estimate the order at which finite range
effects occur, we replace the
$s$-wave scattering length $a_0(0)$ in the leading order 
term of Eq.~(\ref{eq_pertgp}) by $a_0(E)$ and, in a 
second step, approximate
$a_0(E)$ by Eq.~(\ref{eq_effrange}).
A simple calculation  shows that finite range 
corrections enter, according to this estimate, at order 
$r_{\rm{eff}}a^2_0(0)/a_{\rm{ho}}^3$. Thus, finite range effects
are, assuming that $|r_{\rm{eff}}|$ is
smaller than $a_{\rm{ho}}$, suppressed compared to
the effective three-body interaction
[see Eq.~(\ref{eq_pertgp}) in Subsec.~\ref{sec_bosegaseqofstate}], 
which scales as $[a_{0}(0)/a_{\rm{ho}}]^2$,
by a factor of $r_{\rm{eff}}/a_{\rm{ho}}$.
The calculation could be made rigorous by carrying out a
perturbative analysis for a sum of zero-range
two-body potentials that account for 
finite-range effects through derivative operators~\cite{huan57}
[i.e., that depend on $a_0(0)$ and $r_{\rm{eff}}$].
Within the mean-field framework, effective range corrections have been
accounted for by adding a gradient term that is proportional
to $g_2 (N-1)\nabla^2 |\Phi(\vec{r})|^2$ to the 
mean-field potential of the GP equation~\cite{fu03}.
To determine this modified mean-field term and the
constant $g_2$, 
Fu~{\em{et al.}} derived a pseudopotential
that reproduces the real part of the scattering amplitude up to order
$k^2$~\cite{fu03} and
then used this revised pseudopotential in the Hartree
framework (see Subsec.~\ref{sec_bosegaseqofstate}).

The role of the three-body parameter
has 
previously been discussed in Subsec.~\ref{sec_efimov} for the
 three-boson system with infinitely large two-body
scattering length $a_0(0)$.
We now consider the small $a_0(0)$ regime
and estimate at which order the
three-body parameter
enters into the equation of state.
As already discussed, an imaginary $s_{\nu}$ gives rise for the need of
a three-body parameter. For small $|a_0(0)|$, Fig.~\ref{fig_snuroot}
shows that there exists an imaginary root for sufficiently
small $R$.
This implies that the sum of zero-range two-body potentials has to be 
complemented by a hyperradial three-body potential $V_3$
or,
equivalently, a three-body boundary condition in the hyperradial
coordinate.
While this boundary condition does not, as in the 
large $|a_0(0)|$ limit, lead to the appearance of
an infinite ladder of geometrically spaced
finite energy states, it does
have an effect on
the energy spectrum.

To estimate the effect of $V_3$ on the energy spectrum
of the trapped three-boson system,
we perform a perturbative analysis.
We assume that $V_3$ is proportional
to $(\hbar^2 A^4(k)/m) \delta(\vec{r}_1 - \vec{r}_2)
\delta(\vec{r}_2 - \vec{r}_3)$, where
$A^4(k)$ 
denotes the generalized energy-dependent
three-body
scattering length. 
Dimensional analysis shows that $A^4(k)$ has
units of (length)$^4$.
This three-body pseudopotential can alternatively be written in
terms of a $\delta$-function in the hyperradius. 
The quantity $A^4(0)$ conveniently 
parameterizes the three-body parameter in the 
weakly-interacting regime and is related to 
the hypervolume $D$ defined by Tan~\cite{tan08d}.
In 
first order perturbation theory,
this pseudopotential leads to
an energy 
shift proportional to
$[A(k)/a_{\rm{ho}}]^4 \hbar \omega$.
For the $N$-boson system, 
this energy correction gets multiplied by the number of trimers.
This analysis reveals that the three-body potential introduces
an energy shift of higher order 
in the inverse harmonic oscillator length
than both the effective
three-body interaction and the two-body effective range correction.
We note
that the effect of the three-body parameter on the 
energy spectrum of the weakly-interacting $N$-boson system in a
box
with periodic boundary conditions has been discussed in 
Refs.~\cite{bean07,tan08d}.

We now turn to the large $a_0(0)$ regime and ask 
two questions: {\em{(i)}} Does the description of successively
larger Bose systems require a new parameter 
for each successively larger weakly-bound Bose system?
{\em{(ii)}} What are the imprints of the three-body parameter 
on the $N$-body ($N>3$) spectrum?
The first question has been studied extensively 
ever since Efimov's work from the early 
seventies~\cite{efim70,efim71,efim73,amad73,naus87,plat04}.
It is
fairly well established
by now that the treatment of 
the low-energy physics of $N$-body ($N>3$)
Bose systems does not require additional parameters, i.e.,
the low-energy properties of the $N$-body system
are believed to be determined by $a_0(0)$ and a three-body
parameter~\cite{plat04,hamm07,stec09a}.
A consequence of the absence of four- and higher-body 
parameters is that there exists no four- or higher-body
Efimov effect for Bose systems with $0^+$ symmetry, i.e.,
there exists no infinite sequence of $N$-body bound states
whose geometric spacing is determined by an $N$-body parameter ($N>3$).
Other interpretations do, however, exist~\cite{yama06}. 
The second question has been studied extensively
over the past five years or so, mostly but
not exclusively in the context of the four-body system in 
free space~\cite{hamm07,stec09a}.
For each Efimov trimer, there exist two universal four-body
states whose properties are determined by those of the Efimov trimer
that the four-body states are ``attached to''.
These four-body states are, in fact, resonance states
with finite lifetime~\cite{delt11}.
The energies of the tetramer ``ground state'' and
``first excited state'' 
in free space are about 4.58
and 1.01 times larger than the binding energy of the 
respective trimer~\cite{hamm07,stec09a}.
The factor of 4.58 can be rationalized
by realizing that the tetramer can be thought of as consisting of four 
trimers.
In addition to the energy spectrum, other observables---such 
as the scattering lengths for which the tetramers become unbound
and dissociate into four free atoms---obey universal 
relations~\cite{dinc09,delt10,delt11,ferl09,poll09}.
At present, the study of larger systems is still in its
infancy. Initial predictions that relate the $N$-body
energies to those of the trimer have been 
made~\cite{plat04,hann06,stec10,yama10} 
and it is expected that
this research area will florish in the years to come.

\section{Summary and Outlook}
\label{sec_summary}
Trapped ultracold
atomic and molecular gases 
present a 
research area that is blossoming and whose future looks bright.
This area
has attracted the attention of researchers from diverse areas,
including those with backgrounds in
atomic physics, nuclear physics, few-body physics, quantum
optics, condensed matter physics and high energy physics.
One attractive aspect of
this research area is that experimental and theoretical 
efforts go hand in hand, with experiments being motivated
by theoretical work and theoretical work being motivated by
experimental progress.

This review provides an introduction
to trapped atomic and molecular physics following a bottom-up
approach. 
Building on a detailed discussion of
two-body and three-body systems,
the properties of trapped fermionic and bosonic systems 
with varying number of particles have been investigated.
The discussions and examples presented 
are meant to give the reader a
flavor of
this rich
and rapidly developing interdisciplinary field.
In the following, we 
comment on
selected present and future frontiers that 
have only been touched upon 
or not been covered at all in this review.

{\em{Other mixtures:}} This review focused primarily
on the properties of two-component Fermi gases and 
single-component Bose gases. Extensions of these 
studies to other mixtures, such as Bose-Fermi or Bose-Bose
mixtures, introduce additional degrees of
freedom (i.e., another scattering length and possibly
another confining length and/or mass ratio) and will allow one to
investigate the interplay between interactions and symmetry
in more depth. Studies along these lines
may, e.g., lead to a bottom-up
understanding of induced interactions
(see, e.g., Ref.~\cite{heis00}).

{\em{Multi-component Fermi gases:}}
While two-component Fermi gases are stable for
essentially any interaction strength, Fermi gases with more than
two components are not necessarily stable against 
collapse.
In fact, 
the phase diagram of multi-component Fermi gases
is currently being debated~\cite{hone04b,zhai07,paan07,cher07,blum08}. 
Bosenova-type collapse,
BCS-type phases and phases whose properties
are governed by competing dimer and trimer formation have been
suggested. Studies of trapped multi-component
Fermi gases may shed light on the effective
interactions between two trimers and its connections to Efimov
physics. 

{\em{$p$-wave Fermi gases:}} This review focused
almost exclusively on $s$-wave interacting
systems in the vicinity of
broad Fano-Feshbach resonances. While present experiments on 
higher partial wave resonances are plagued by detrimental
losses, it seems feasable that these losses could be suppressed
by loading the gas into an optical lattice
or by utilizing atomic species for which dipolar relaxation 
rates can be suppressed~\cite{gaeb07}.
Some theoretical few-body studies on 
$p$-wave interacting systems 
have been
performed~\cite{levi07,levi08,jona08,dinc08},
and it is anticipated that additional 
studies
will 
further clarify questions related to the system's lifetime
and address
which aspects of higher partial wave systems
are universal and which ones are not.

{\em{Dipolar systems:}}
Much progress has been made in trapping and cooling 
systems with magnetic and electric dipoles.
While weakly-interacting dipolar Bose gases have been
modeled quite successfully 
using a mean-field 
description, the study of strongly-interacting
dipolar systems is still in its infancy.
The three-body system has been investigated for
a few limiting scenarios within a microscopic 
framework~\cite{tick10,wang11} but little has been done yet
for larger systems. One possible approach is to
pursue a description that is based on 
an effective Hamiltonian, which
incorporates our understanding gained from
microscopic two- and three-body studies.

{\em{Extended Efimov scenario and $N$-body parameter:}}
The Efimov effect, in its original formulation,
applies to three-body systems in free-space.
Ever since Efimov's papers from the early 1970s,
researchers have asked 
whether or not the Efimov effect generalizes to larger systems.
Although much progress has been made, there are still 
open questions. Future studies of free-space 
and trapped four-, five-
and higher-body systems will address
in more detail how the energy spectrum
of the $N$-body system ($N>3$) depends on the three-body
parameter and how finite-range corrections enter.
It is anticipated that answers to these  
questions will be found by combining the insights 
gained from microscopic
and effective approaches applied to systems of varying 
symmetry and size.

{\em{Systems with two- and/or three-body resonances:}}
This review discussed how two- and
three-body interactions can be tuned by varying the
trappping frequency of the harmonic confinement, by
taking advantage of the optical lattice structure,
or by inducing tunable effective
three-body interactions that correspond
to varying short-range hyperradial boundary conditions.
For example, the scattering properties of three- and four-body
systems in effectively one- and two-dimensional confining geometries
have been analyzed in some detail~\cite{mora04,mora05}.
Another intriguing example is the appearance of resonances 
in mixed dimensional systems~\cite{nish08a,lamp10}.
It is expected that future work will uncover other means 
to manipulate the interactions.
The resulting effective interactions are expected to
open unique opportunities for designing larger
systems with exotic properties. This is a research branch
that appears to be equally intriguing to few- and many-body
physicists.

{\em{Input for Hubbard Hamiltonian:}}
Any type of effective Hamiltonian such as the Bose or Fermi 
Hubbard Hamiltonian depends on effective parameters
such as the tunneling amplitude and on-site interactions.
As experiments are being refined and many-body simulation
techniques advance, it is paramount to 
refine the determination of the effective parameters
that define a given effective
Hamiltonian~\cite{will10,will11}. 
Studies of trapped few-body systems
are expected to contribute to this endeaver.

{\em{Beyond the LDA:}}
Connections between inhomogenous and homogeneous systems are
often made through the LDA. While
this approach has been tremendously successful, 
several limitations of this approach are 
known. A future goal of trapped atom and molecule studies
should be to develop alternative approaches. These could
involve pushing microscopic calculations of trapped atom systems
to larger number of particles and extending
condensed matter techniques to trapped systems.

{\em{Time-dependent dynamics of Bose gases:}}
Although this review focused primarily on time-independent
studies, it should be noted that
much progress has been made over the past few years
in solving the time-dependent many-body Schr\"odinger equation.
One-dimensional Bose gases, e.g., 
have been treated by
various
variants of the multi-configurational 
approach~\cite{alon08}
(borrowed from quantum chemistry) and the time-evolving block
decimation or density-matrix renormalization group
approaches~\cite{vida03a,scho05,baue11} 
(borrowed from condensed matter physics).
From 
a fundamental as well as an experimental point of view,
the emergence of fragmented 
states of Bose gases is particularly interesting. 
Another intriguing set of studies addresses the tunneling
dynamics in double-well geometries. In particular, the interplay between
single particle and pair tunneling has been
investigated~\cite{zoll08}.
It is expected that studies along these lines will be extended
to higher-dimensional systems in the future.

{\em{Finite-temperature physics:}}
Except for Subsec.~\ref{sec_fermigasvirial}, this review
considered system properties at zero-temperature.
Many features, such as the shell structure of
two-component Fermi gases~\cite{dyke11} and peaks in
the loss rate due to Efimov resonances~\cite{dinc04}, become
smeared out as the temperature increases. At the microscopic
level, few studies to date have considered thermally averaged
quantities, despite the fact that thermometry in optical
lattice systems is currently a hot topic~\cite{mcka10}. 
It is expected that the
few-body community will push to develop the machinery
that allows for the treatment of
trapped systems at finite-temperature.

{\em{Non-equilibrium few-body dynamics:}}
One of the key advantages of small trapped systems
is that their entire excitation
spectrum is, in certain cases, amenable to analytical or numerical
treatments. This opens the possibility to prepare a 
system in a given initial state and to then follow the dynamics
in response to a slow variation of a system parameter or a 
rapid quench.
Initial studies along these lines have been 
performed~\cite{mies00,stec07c}.
It is expected that this area of research will further intensify
as studies of the non-equilibrium dynamics of optical lattice systems
are progressing dramatically.
Answers related to equilibration times, entanglement dynamics,
and critical behavior are expected to emerge.
For example, it is anticipated that few-body studies
will address what happens when two initially separated 
optical lattice sites or plaquettes are being merged.

\ack
The author has benefitted greatly from close collaborations 
with M. Asad-uz-Zaman, Kevin Daily, Krittika Goyal, Chris Greene, 
Gabriel Hanna, Debraj Rakshit,
Seth Rittenhouse, and Javier von Stecher
on topics covered in or related to this review.
The author thanks K. Daily for comments on the manuscript,
and for providing the full temperature dependent
$\Delta b_3$ data shown in Fig.~\ref{fig_virial}.
Support by the NSF through
grant PHY-0855332
is gratefully acknowledged.


\begin{thebibliography}{100}
\bibitem{dail10explain}
The lowest $s_{\nu}$ value for $L=0$ derived within the ``atom-dimer'' picture
  differs from that derived using the semi-classical framework; the atom-dimer
  picture gives $2$ in this case, which should be compared with the
  semi-classically derived result of $3$ and the exact result of $2.166$.
\end{thebibliography}
\end{document}